\documentclass[aps, prc, twocolumn, superscriptaddress]{revtex4-1}
\usepackage{graphicx}
\usepackage[super]{nth}
\usepackage{cancel}
\usepackage{xcolor}
\begin{document}

\title{Measurement of differential cross sections for deuteron-proton breakup reaction at 160~MeV}


\author{W.~Parol}
\email[]{wiktor.parol@ifj.edu.pl}
\email[]{wiktor.parol@gmail.com}
\author{A.~Kozela}
\affiliation{Institute of Nuclear Physics Polish Academy of Sciences, PL-31342 Krak{\'o}w}
\author{K.~Bodek}
\affiliation{M.~Smoluchowski Institute of Physics, Jagiellonian University, PL-30348 Krak{\'o}w, Poland}
\author{A.~Deltuva}
\affiliation{Institute of Theoretical Physics and Astronomy, Vilnius University, Vilnius, Lithuania}
\author{M.~Eslami-Kalantari}
\affiliation{Department of Physics, School of Science, Yazd University, Yazd, Iran}
\author{J.~Golak}
\affiliation{M.~Smoluchowski Institute of Physics, Jagiellonian University, PL-30348 Krak{\'o}w, Poland}
\author{N.~Kalantar-Nayestanaki}
\affiliation{KVI-CART, University of Groningen, NL-9747 AA Groningen, The Netherlands}
\author{G.~Khatri}
\affiliation{CERN, CH-1211 Geneva 23, Switzerland}
\author{St.~Kistryn}
\affiliation{M.~Smoluchowski Institute of Physics, Jagiellonian University, PL-30348 Krak{\'o}w, Poland}
\author{B.~K{\l}os}
\affiliation{Institute of Physics, University of Silesia, PL-41500 Chorz{\'o}w, Poland}
\author{J.~Kubo{\'s}}
\affiliation{Institute of Nuclear Physics Polish Academy of Sciences, PL-31342 Krak{\'o}w}
\author{P.~Kulessa}
\affiliation{Forschungszentrum Juelich IKP, DE-52425 Juelich, Germany}
\author{A.~{\L}obejko}
\affiliation{Institute of Physics, University of Silesia, PL-41500 Chorz{\'o}w, Poland}
\author{A.~Magiera}
\affiliation{M.~Smoluchowski Institute of Physics, Jagiellonian University, PL-30348 Krak{\'o}w, Poland}
\author{H.~Mardanpour}
\affiliation{KVI-CART, University of Groningen, NL-9747 AA Groningen, The Netherlands}
\author{J.G.~Messchendorp}
\affiliation{KVI-CART, University of Groningen, NL-9747 AA Groningen, The Netherlands}
\author{I.~Mazumdar}
\affiliation{Tata Institute of Fundamental Research, Mumbai 400 005, India}
\author{R.~Skibi{\'n}ski}
\affiliation{M.~Smoluchowski Institute of Physics, Jagiellonian University, PL-30348 Krak{\'o}w, Poland}
\author{I.~Skwira-Chalot}
\affiliation{Faculty of Physics, University of Warsaw, PL-02093 Warsaw, Poland}
\author{E.~Stephan}
\affiliation{Institute of Physics, University of Silesia, PL-41500 Chorz{\'o}w, Poland}
\author{A.~Ramazani-Moghaddam-Arani}
\affiliation{Departments of Physics, Faculty of Science, University of Kashan, Kashan, Iran}
\author{D.~Rozp{\c e}dzik}
\affiliation{M.~Smoluchowski Institute of Physics, Jagiellonian University, PL-30348 Krak{\'o}w, Poland}
\author{A.~Wilczek}
\affiliation{Institute of Physics, University of Silesia, PL-41500 Chorz{\'o}w, Poland}
\author{H.~Wita{\l}a}
\affiliation{M.~Smoluchowski Institute of Physics, Jagiellonian University, PL-30348 Krak{\'o}w, Poland}
\author{B.~W{\l}och}
\affiliation{Institute of Nuclear Physics Polish Academy of Sciences, PL-31342 Krak{\'o}w}
\author{A.~Wro{\'n}ska}
\affiliation{M.~Smoluchowski Institute of Physics, Jagiellonian University, PL-30348 Krak{\'o}w, Poland}
\author{J.~Zejma}
\affiliation{M.~Smoluchowski Institute of Physics, Jagiellonian University, PL-30348 Krak{\'o}w, Poland}


\date{\today}

\begin{abstract}
Differential cross sections for deuteron breakup $^{1}H(d, pp)n$ reaction were measured for a large set of 243 geometrical configurations at the beam energy of 80~MeV/nucleon. The cross section data are normalized by the luminosity factor obtained on the basis of simultaneous measurement of elastic scattering channel and the existing cross section data for this process. The results are compared to the theoretical calculations modeling nuclear interaction with and without taking into account the three--nucleon force (3NF) and Coulomb interaction. In the validated region of the phase space both the Coulomb force and 3NF  play an important role in a good description of the data. There are also regions, where the improvements of description due to including 3NF are not sufficient.
\end{abstract}


\maketitle

\section{Introduction}

One of the most basic topics in modern nuclear physics is the nature of the forces acting between nucleons. Exact knowledge of all features of the two--nucleon (NN) system dynamics should provide a basis for understanding of properties and interactions in heavier systems. This presumption has been verified by applying models of the NN interaction to describe systems composed of three nucleons (3N). Theoretical predictions of observables are obtained by means of the rigorous solution of Faddeev equations \cite{faddev_1961, witala_1988, huber_1993, glockle_1996}, including NN interaction as so-called realistic potential models, based on the meson exchange theory, originally  proposed by Yukawa \cite{yukawa_1935} and confirmed by Occhialini and Powell \cite{powell_1947}. Early stage of experimental studies of the deuteron-proton elastic scattering in the range of intermediate energies and theoretical efforts \cite{witala_2001} have proven the dominant, but not sufficient, role of the pairwise NN interaction. The missing piece of the dynamics, referred to as three--nucleon force (3NF), also contributes. The effects of this force, much smaller than parwise NN contribution, arise in systems consisting of at least three nucleons. Modern NN potentials like Argonne V18 (AV18) \cite{wiring_1995}, CD Bonn (CDB) \cite{machleidt_2001}, and Nijmegen I and II \cite{stoks_1994} have yielded a remarkably good agreement (with a $\chi^{2}$ of around 1) between the predictions of the calculations with the experimental data for two-nucleon systems. To describe three--nucleon systems these realistic NN potentials are used in Faddeev equations together with present models of 3NF like Urbana IX \cite{pudliner_1997} or Tucson-Melbourne \cite{coon_2001}. In another approach, three--nucleon interaction can be introduced within the coupled-channel (CC) framework by an explicit treatment of the $\Delta$-isobar excitation \cite{deltuva_2003, deltuva1_2003, deltuva_2008}. Alternatively, contributions of NN and 3NF to the potential energy of a 3N system can be calculated within the Chiral Perturbation Theory \cite{machleidt_2011, epelbaum_2012}. Here, the many-body interactions appear naturally at higher orders (non-vanishing 3NF at next-to-next-to leading order). Modern calculations include also other ingredients of few--nucleon dynamics such as Coulomb interactions \cite{ishikawa_2003, deltuva_2006} or relativistic effects \cite{witala_2005, skibinski_2006}. Predicted effects in differential cross sections emerge in various parts of the phase space of the deuteron-proton breakup reaction with different magnitude. Existing experimental data \cite{kistryn_2003, kistryn_2005, sekiguchi_2005, eslami_2009, mardanpour_2010, kistryn_2011, ciepal_2015} demonstrate quite sizable 3NF and Coulomb effects, and confirm their importance for the correct description of differential cross sections for the deuteron breakup reaction at energies above 65 MeV/nucleon and below 400~MeV/nucleon.

The present work is a continuation of experimental campaign focusing on the investigation of contributions from various dynamical ingredients (3NF, Coulomb force and relativistic component) of nuclear interaction via measuring various observables in few-nucleon systems for large parts of the phase space. Measured differential cross sections at 80~MeV/nucleon enlarges the systematic database for the deuteron--proton breakup reaction at intermediate energies. Produced feedback allows for further validation of available and future theoretical models of nuclear interaction.

In Section~\ref{sec:exp} the experimental setup is described. Section~\ref{sec:data_anal} gives an overview of the data analysis, while in section \ref{sec:results} the obtained results are presented. Section \ref{sec:summary} summarizes the main outcome of the presented studies.

\section{Experiment}
\label{sec:exp}
The experiment was performed at Kernfysisch Versneller Instituut (KVI) in Groningen, the Netherlands (currently KVI-CART). The deuteron breakup reaction, $^{1}H(d,pp)n$, was measured simultaneously with elastic scattering of deuterons on liquid hydrogen target. Deuteron beam of 80~MeV/nucleon energy was provided by the cyclotron AGOR (Accelerateur Groningen-ORsay) \cite{agor_1987}, while charged reaction products were detected by the BINA (Big Instrument for Nuclear-polarization Analysis, \cite{bina_manual}) setup. The BINA detection system is characterized by: high angular acceptance (almost 4$\pi$), good (in forward region) and moderate (in backward part) angular resolution, the ability to identify and to provide complete kinematical information for two or more charged particles in the final state. All these features make the BINA detector an excellent tool for studying the systems of few nucleons in the intermediate energy range.

\begin{figure}
  \includegraphics[width=7.cm]{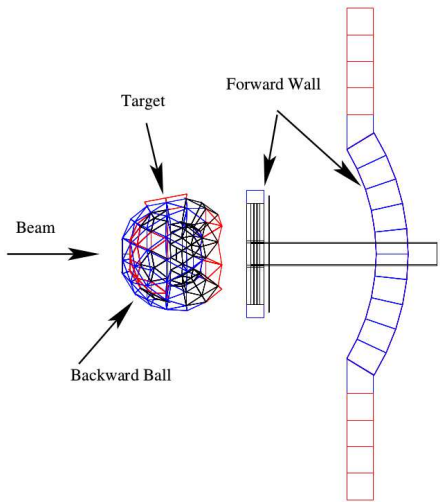}
 \caption{\label{fig:bina} Schematic side view of BINA detection system}
\end{figure}

The BINA detection system, Fig.~\ref{fig:bina}, consist of two main parts, forward Wall and backward Ball. The liquid hydrogen target cell is positioned in the center of the Ball, which served in this experiment as the scattering chamber only.
The front part, Wall, consists of three detector elements positioned in planes perpendicular to the beam line: a multi-wire proportional chamber (MWPC) and two scintillator hodoscopes, forming a set of 120 $\Delta$E--E virtual telescopes. The Wall is optimized for detecting protons and deuterons in the energy ranges of 20--130 MeV and 25--200 MeV, respectively.

Precise measurement of scattering angles is accomplished by MWPC \cite{volkerts_1999} positioned directly behind the thin vacuum window as the first detector intersected by the reaction products. It consists of three active planes: a plane measuring x-coordinate with vertical wires, a plane measuring y-coordinate with horizontal wires, and a diagonal plane U with wires inclined by 45 degrees. All wires within a plane are spaced by 2~mm and combined in pairs to form 118, 118 and 148 separate detector channels for X, Y and U planes, respectively. The active area of the MWPC is 38$\times$38~cm$^{2}$. It forms a pixel system allowing to precisely determine the crossing point of a charged particle, and thus to reconstruct the emission angles of the outgoing reaction products. The angular acceptance of the detector in polar angles is $\vartheta\in(10^{\circ}, 40^{\circ})$, with the full azimuthal coverage up to $30^{\circ}$.

The $\Delta$E transmission detector consists of 24 vertical strips of a 2~mm thick plastic scintillator (BICRON type BC-408 \cite{bicron_2015}). The signals from each $\Delta$E stripe are read by one photomultiplier tube (PMT) coupled through a light guide modeled for optimal light collection. As the signals are proportional to the specific energy loss of charged particles they play a crucial role in particle identification.

The E detector is made of horizontally--arranged 120~mm thick scintillator slabs (BICRON type BC-408 \cite{bicron_2015}). In order to minimize particle cross-overs between neighboring scintillators central ten elements of E detector follow a cylindrical symmetry, with the cylinder center at the target position (see Fig.~\ref{fig:bina}). The additional ten elements attached from the top and bottom to the cylindrical part were not used in the present experiment. Energy deposited by particles in the E slab was converted to scintillation light registered by two PMTs, attached to both ends of each detector. This allows to compensate for the light attenuation along the scintillator resulting in a position independent output.

Other details concerning the setup as well as electronic and read-out systems used in the experiment can be found in Ref. \cite{stephan_2013}.  Data acquisition system was based on GSI Multi-Branch System (MBS) \cite{essel_2000}.

\section{Data Analysis}
\label{sec:data_anal}
The data analysis started with the selection of time periods characterized by stable operation of the cyclotron and all elements of the detector. The selection was based on the scaler rates recorded for all individual channels of the detector. In order to minimize random coincidences, additional time gates rejecting particles not correlated with the trigger signal had been set.

\subsection{Tracking of individual particles}
\label{sec:cal}
\subsubsection{Track Reconstruction}
\label{sec:tracking}
Charged particles passing through the Wall detector deposit their energy successively in MWPC, $\Delta$E and E. In the simplest case, for a single particle only three wires (one wire per plane) in MWPC give a signal while in reality, clusters of two or more wires are observed. As no signal amplitude from MWPC was collected, in such cases the hit position is represented by the center of the cluster.

\begin{figure}
  \includegraphics[width=4.1cm]{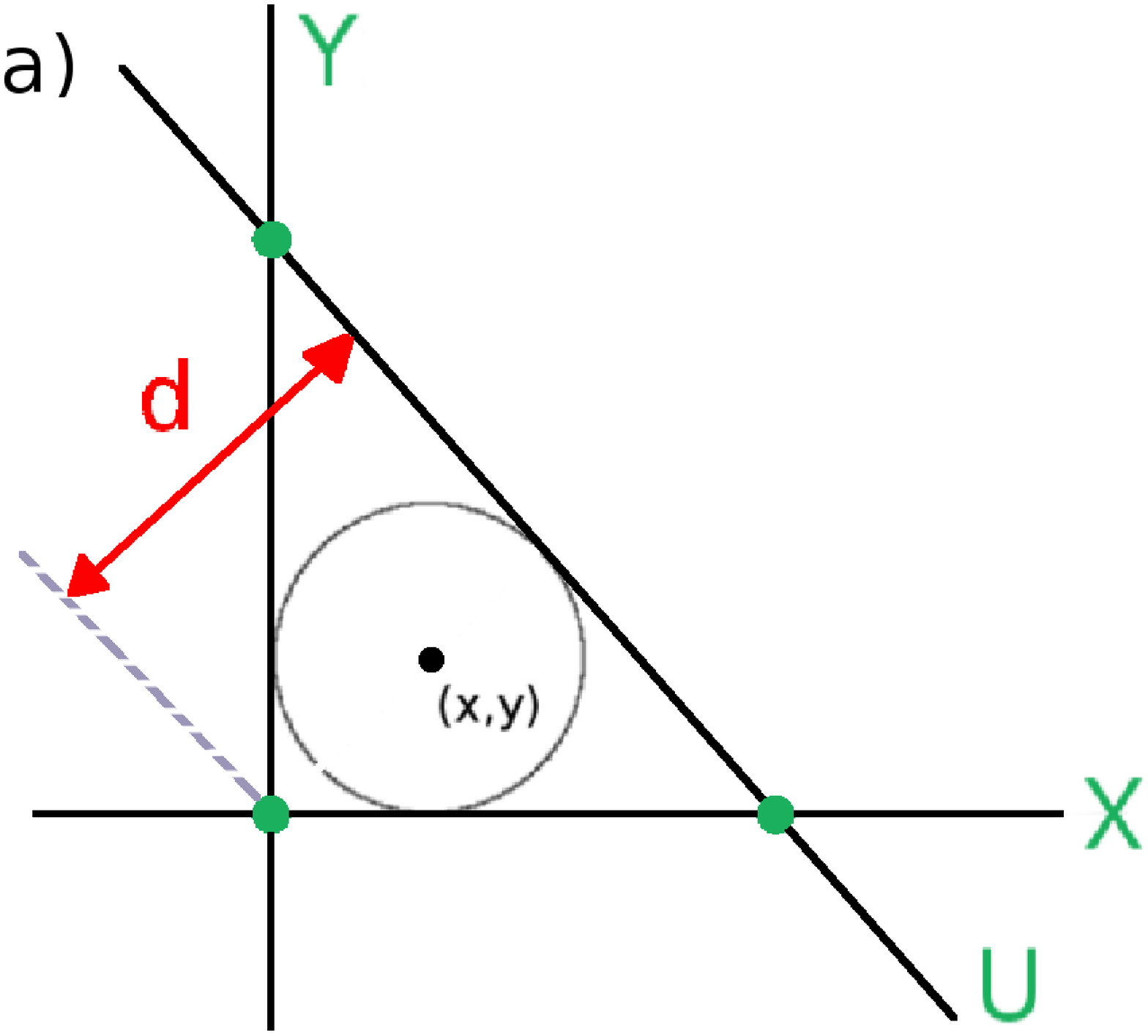}
  \includegraphics[width=4.1cm]{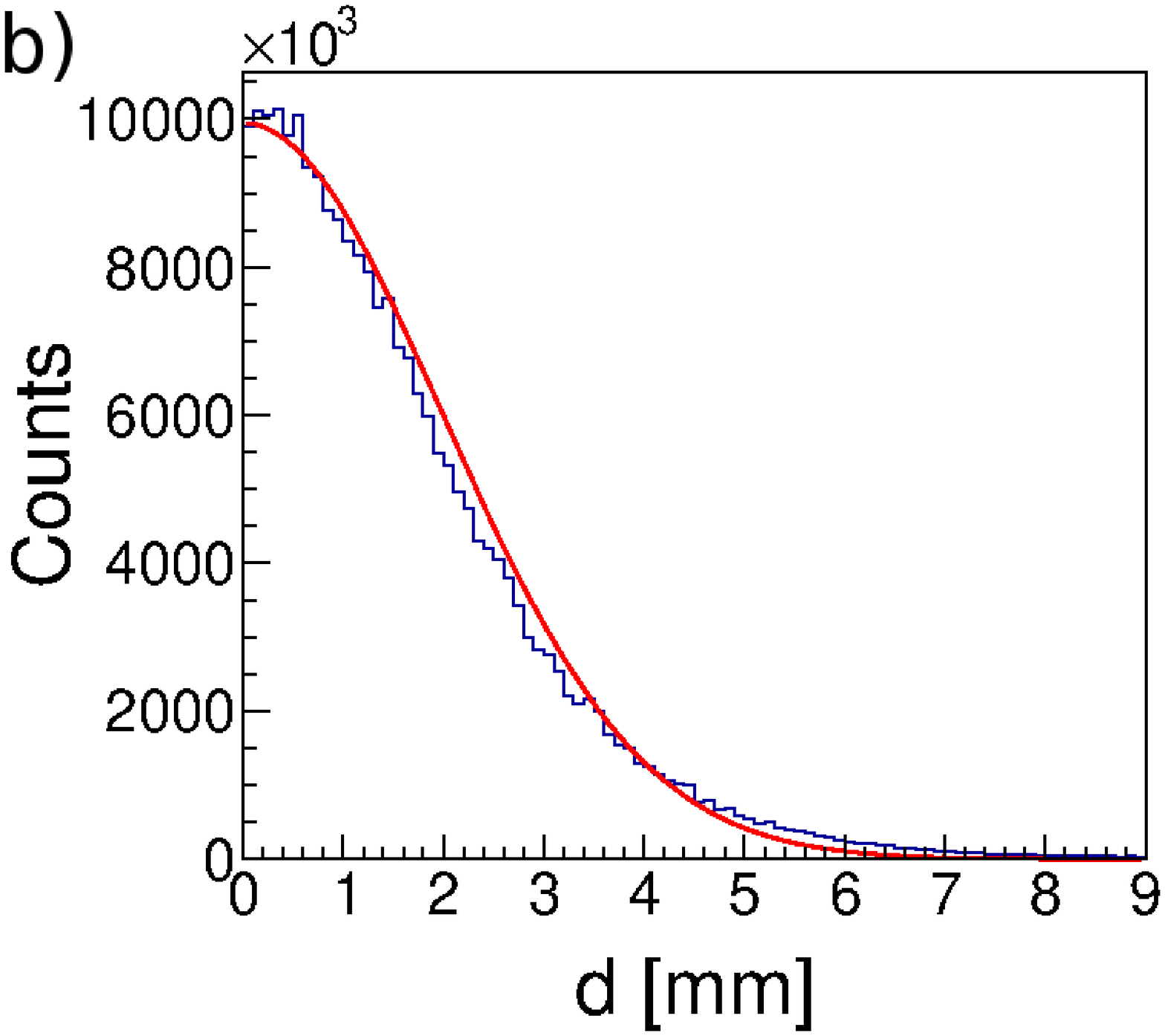}
\caption{\label{fig:xyu} \emph{Panel a:} Geometrical reconstruction of the $(x,y)$ coordinates of the full--track intersection with the Y-plane. Lines X, Y and U represent centroids of clusters in the respective planes projected onto Y-plane. Reconstructed $(x,y)$ coordinates for all three types of weak--tracks are shown as green dots. \emph{Panel b:} Distribution of $d$ (defined in the panel a) for the whole dataset. Limit of 7~mm for $d$ corresponds to the $\sim 3\sigma$ of the fitted gaussian distribution (red solid line).}
\end{figure}

In order to accept the track information from the MWPC several different strategies may be used. In the analysis published in several earlier papers e.g. in Ref.~\cite{ciepal_fbs_2019}, a coincidence of all three planes was required, with the X and Y planes defining $(x,y)$ coordinates of the intersection of the track with the MWPC and the U-plane was used to validate this intersection. In the data analysis of deuteron-deuteron scattering presented in Ref.~\cite{ciepal_prc_2019} we accepted also events with only two planes hit, with the condition that no other hits are present in MWPC and the resulting position information is correlated with hits in $\Delta$E and E detectors. This kind of tracks, which are further referred to as ''weak--tracks'' (as opposed to ''full--tracks'' indicating 3-plane coincidences), is important for consideration of systematic effects such as e.g. energy and position dependent MWPC efficiency. The analysis presented in this paper is based on full--tracks and, in addition, we take advantage from the position information supplied by the U-plane. In such a case, the final position is given by the center of a circle inscribed in the triangle defined by corresponding cluster centroids projected (from the target center direction) onto a common plane (see Fig.~\ref{fig:xyu}).  Assuming equal position resolution of all planes, this algorithm improves the final angular resolution for the polar angle $\vartheta$ up to $0.4^{\circ}$ and the azimuthal angle $\varphi$ to $0.67^{\circ}-1.39^{\circ}$ (depending on the polar angle).  It is also clear that weak--tracks involving U-plane feature lower position resolution in one direction than those defined by X and Y plane.
It is important to mention that 3-dimensional track parameters, in this case polar and azimuthal angles, can be obtained by the following formulas:
\begin{eqnarray}
\label{eqn:defangles1}
\vartheta &=& \arctan{\left(\frac{\sqrt{x^{2}+y^{2}}}{Z_{Y}}\right)},\\
\label{eqn:defangles2}
\varphi &=&  \text{atan2}{\left(y, x\right)},
\end{eqnarray}
under the assumption that the corresponding particle was emitted from the target center ($Z_{Y}$ is the distance of the projection plane from this center). The $\text{atan2}()$ function calculates the principal value of the $\arctan\left(\frac{y}{x}\right)$, using the signs of the two arguments to determine the quadrant of the result \cite{cfunction_manual}. Smearing of the reaction point due to target thickness and beam size is included into systematic uncertainty of the final results.
Having those parameters one may check if the track coincides, within given position resolution, with hits in the E and $\Delta$E detector elements -- only such events are considered in further analysis. In order to combine  the hits in the individual planes into the full--track event, a cut has been imposed on the distance between the centroid of the cluster reconstructed in the U-plane and the cross-point between centroids in the X and Y planes ($d$-variable in Fig.~\ref{fig:xyu}).

\subsubsection{Particle identification}
\label{subsec:pid}
\begin{figure}
  \includegraphics[width=8.6cm]{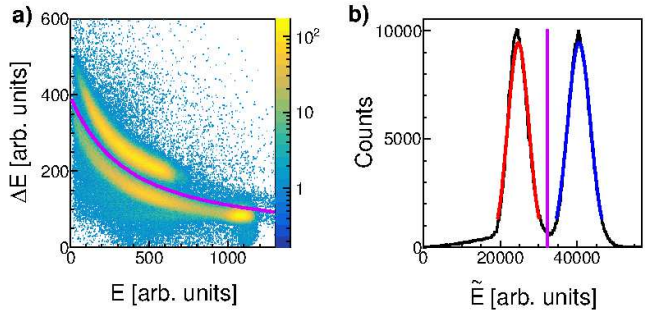}
  \caption{\label{fig:edeLin} Example of the identification spectra for a chosen virtual telescope ($\Delta$E=13, E=8). \emph{Panel a:} $\Delta$E-E signal distribution. The violet line separates the proton and deuteron bands and corresponds to the vertical line indicated in the distribution on the right side. \emph{Panel b:} Projection of the linearized spectrum onto the $\widetilde{E}$ variable. The two peaks correspond to the proton and deuteron bands. The 2$\sigma$ ranges of the fitted Gaussian functions are shown as red and blue lines for protons and deuterons, respectively.}
\end{figure}

Neglecting traces of heavier ions from beam interactions with the target frames, the particle identification can be reduced in this experiment to simple distinction between protons and deuterons, while the later ones come exclusively from the elastic scattering. The identification is based on the linearization technique applied to the $\Delta$E-E spectra \cite{ploskonka_1975}. It allows for identification of reaction products by analytically-determined conditions. Following simplified consideration based on Bethe-Bloch formula one introduces a new variable $\widetilde{E} = (E + \Delta E)^{\kappa} - E^{\kappa}$, the value of which is constant for each type of particles in wide energy range \cite{parol_lin_2014}. The index $\kappa$ characterizes the detector material, its internal structure (variations of transparency, quality of the surface) and geometry and is determined for each virtual telescope separately.
As a result, one-dimensional distribution of $\widetilde{E}$ variable is obtained in which protons and deuterons are visible as distinct peaks (Fig.~\ref{fig:edeLin}). In order to improve the sensitivity of this method, the fine tuning of $\kappa$ (as well as of the parameters: $\mu_{p}$, $\sigma_{p}$, $\mu_{d}$, $\sigma_{d}$  corresponding to the centroids and widths of proton and deuteron peaks, respectively) was performed for each virtual $\Delta$E-E telescope. For this purpose a sample of the data with well balanced number of protons and deuterons (from \emph{dd} scattering experiment at the same beam energy) has been used. The $\kappa$ index has been varied to get maximal separation between  proton and deuteron peaks. The obtained final values of $\kappa$ range from 1.63 to 1.85, while according to Bethe-Bloch rule $\kappa = 1.73$ is expected for an ideal scintillator. This method allows for controllable selection of different event samples not biased by subjective cuts.

\subsubsection{Energy Reconstruction}
Energy calibration gives a relation between the registered ADC channel and the deposited energy ($E^{D}$) in a given scintillator element. Since 2~mm thick $\Delta$E stripes remove a relatively small fraction of particle energy and, furthermore, this information is strongly biased by light attenuation along the scintillator and the light guide, only E detector was used for reconstruction of particles energies. The calibration was carried out using protons from elastic scattering and Monte-Carlo simulations including full detector geometry implemented in the GEANT4 simulation package \cite{wilczek-phd, parol_acta_2014}.
The detector is characterized by noticeable variation of the PMT signal amplitude depending on the point of the interaction along the scintillator. This dependence, caused by light attenuation and losses, can be significantly suppressed by applying geometrical mean of responses of the left and right PMTs \mbox{($C = \sqrt{c_{L}\cdot c_{R}}$)}. For the two middle E slabs, partially cut in the center in order to accommodate an opening for the beam pipe, plain sum of the signals (\mbox{$C = c_{L} + c_{R}$}) was applied. The remaining small dependence of the signal on position is taken into account in the position dependent energy calibration.

\begin{figure}
  \includegraphics[width=8.3cm]{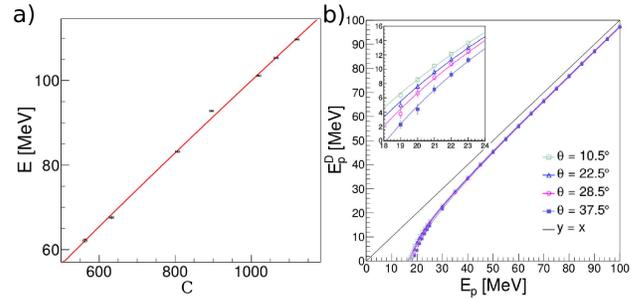}
 \caption{\label{fig:dep2vex} Energy calibration. \emph{Panel a:} Example of the correlation between experimentally obtained centroids of the distribution of the variable {\emph C} and of the corresponding distribution of the simulated energy deposited in E detector together with the fitted function defined in Eq.~(\ref{eqn:cal}).
 \emph{Panel b:} Set of polynomials transforming energy deposited by proton, $E^{D}(\vartheta)$, to its initial kinetic energy at the reaction point.}
\end{figure}

In order to extend the calibration over energies of protons from the breakup reaction, a dedicated measurement has been performed using energy degraders, placed between the $\Delta$E and E detectors. Degraders were made of steel plates of precisely-defined thicknesses, which were mounted in several configurations allowing for satisfactory coverage of the energy range. The elastically scattered protons were selected according to kinematical conditions: co-planarity and $\vartheta_{p}\text{~vs.~}\vartheta_{d}$ relations. The detector plane was divided into 180 sectors, each of them labeled by the side ($s = left, right$), E-scintillator element number ($N = 0,1,\dots,9$) and the polar angle bin number ($\bar{\vartheta}=0,\dots, 8$). All the sectors were calibrated separately using the information from GEANT4 simulations and the following two-parameter function: 
\begin{eqnarray}
\label{eqn:cal}
E^{D}_{s, N, \bar{\vartheta}}(C) &=& a_{s, N, \bar{\vartheta}}C + b_{s, N, \bar{\vartheta}}\sqrt{C},
\end{eqnarray}
where take out: $E^{D}_{s, N,\bar{\vartheta}}(C)$ stands for the energy deposited in this particular detector element as a function of variable $C$, i.e. the combination of signals from left and right PMTs defined above. An example of the fit is shown in Fig.~\ref{fig:dep2vex}a.

Deuteron energy calibration was based on that for protons and corrected for different light output corresponding to the same energy deposited by particles of different mass. Particle-dependent light output for the known scintillator material has been taken from Bicron data sheets \cite{bicron_2015}, and additionally validated in the dedicated studies (see Ref.~\cite{khatri_phd_2015} for the details).

To reproduce the initial kinetic energy of a particle at the reaction point ($E_{i}$), the conversion formula has been found based on the energy loss of simulated mono-energetic protons and deuterons on their way to and inside the E detector:
\begin{eqnarray}
\label{eqn:dep2vex}
E_{i}(\vartheta) &=& P^{8}_{i, \vartheta}(E^{D}_{i}(\vartheta)),
\end{eqnarray}
where subscript \emph{i} stands for particle type (proton or deuteron) and \{$P^{8}_{i, \vartheta}$\} is a set of \nth{8}--order polynomials with factors calculated from deposited-to-initial energy relations obtained from GEANT4 simulations of the experiment (see Fig.~\ref{fig:dep2vex}b).

Final energy resolution reaches 2.1\% for 123~MeV protons, and deteriorates with energy in accordance with photon statistics.

\subsection{\label{sec:pid} Process identification}
\begin{figure}
  \includegraphics[width=4.25cm]{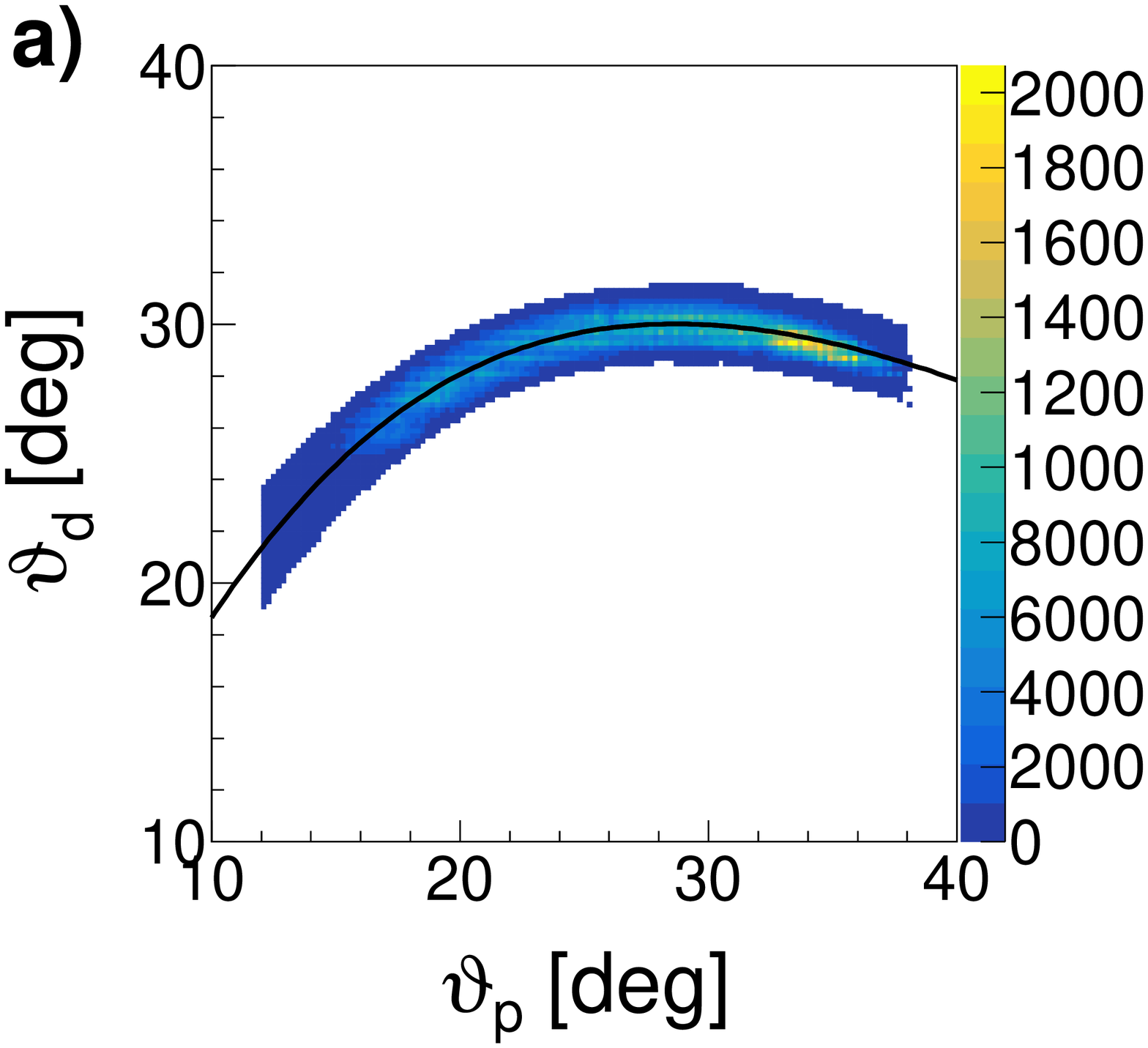}
  \includegraphics[width=4.25cm]{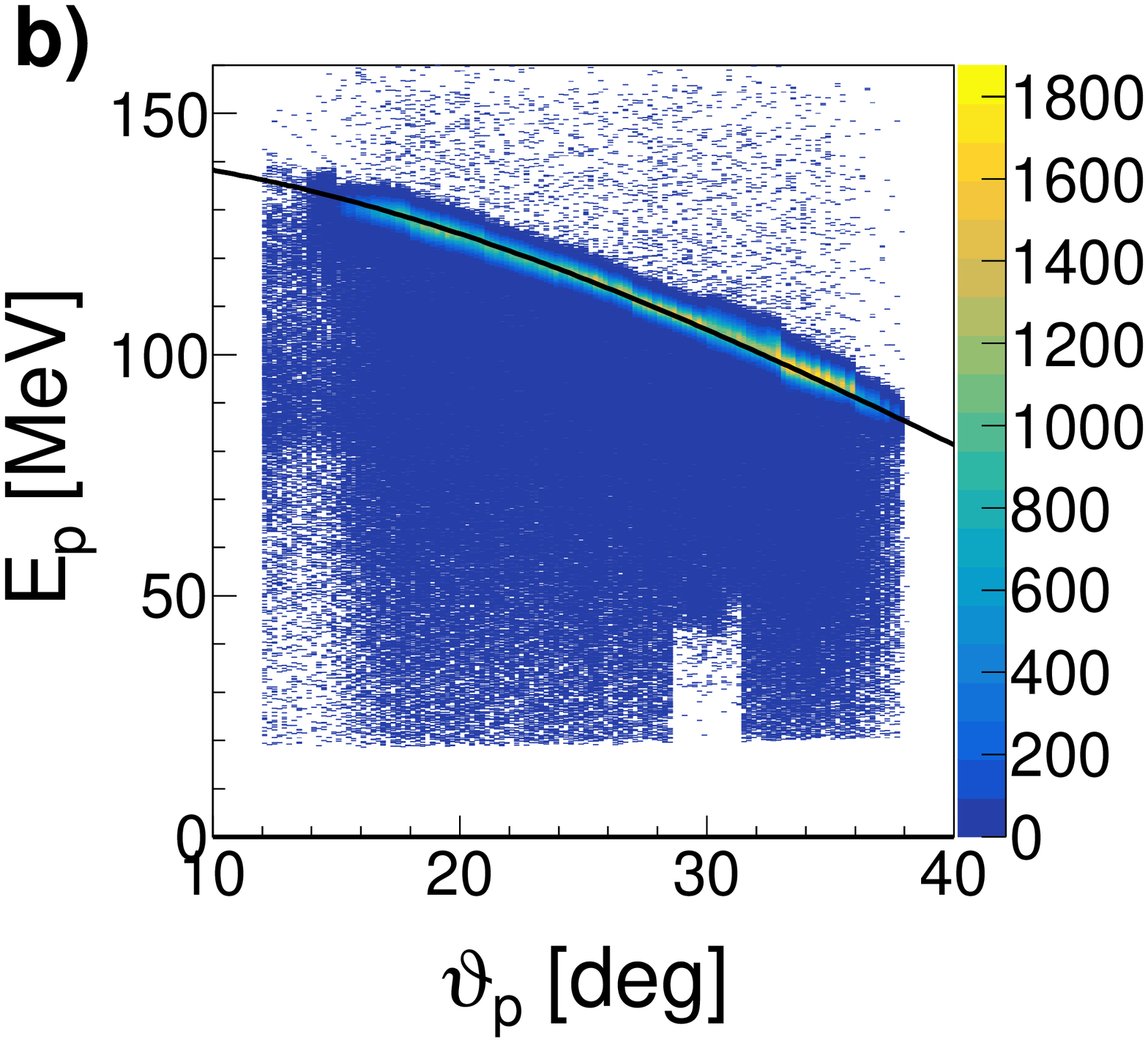}
 \caption{\label{fig:kin2} \emph{Panel a:} Kinematic relation between polar angles of coincident coplanar particles with $\pm3\sigma$ cut around the theoretical kinematics of \emph{dp} elastic scattering (black line). \emph{Panel b:} Energy distribution of particles identified on the basis of coplanarity and polar angle cut (shown in the left panel) as elastically scattered protons.}
\end{figure}

\begin{figure}
  \includegraphics[width=4.25cm]{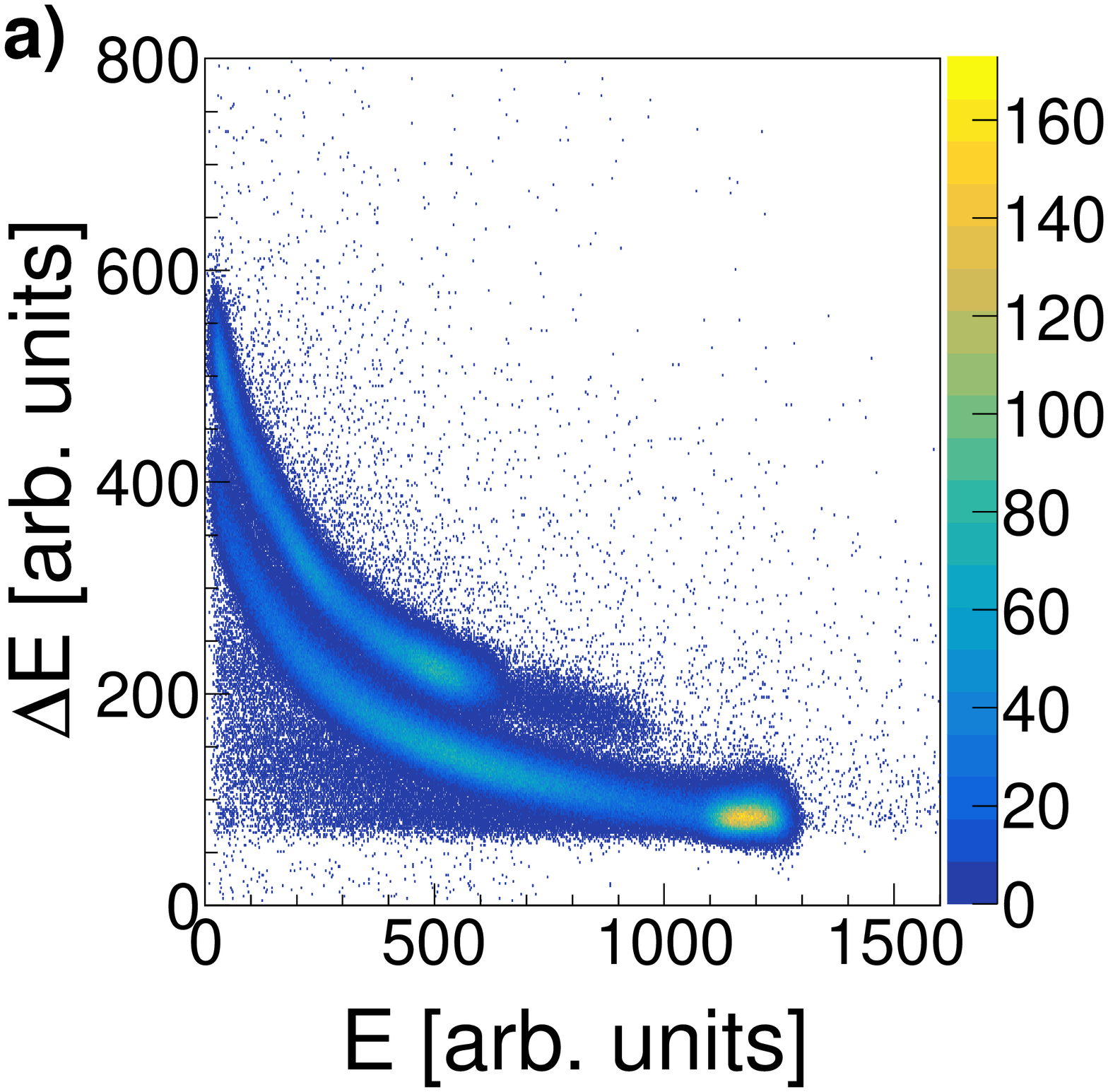}
  \includegraphics[width=4.25cm]{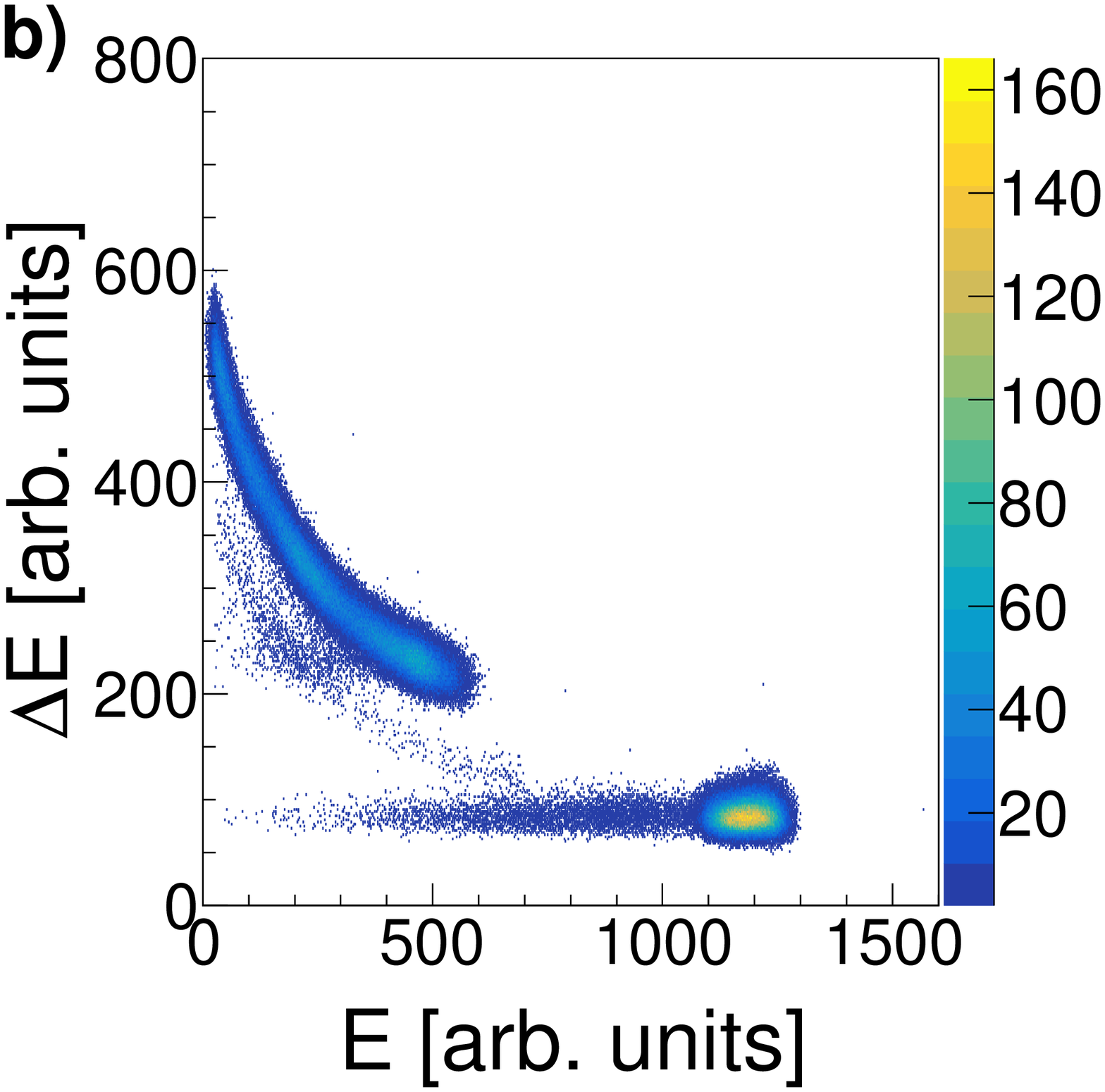}
 \caption{\label{fig:ede0} Relation of particle energies deposited in chosen $\Delta$E-E telescope (E=1, $\Delta$E=9). \emph{Panel a:} spectrum for particles registered in coincidence with any other particle in Wall; \emph{Panel b:} the same with an additional condition that coincident particles meet angular kinematical relations of the elastic scattering.}
\end{figure}

Neglecting small admixtures of electromagnetic processes, deuteron--proton interaction in the studied energy range may result in elastic scattering or breakup. A deuteron in the exit channel uniquely identifies the reaction as elastic scattering, while two protons have to be identified in order to identify the breakup reaction. This goal can be accomplished for most of the recorded events using the PID procedure described in Sec.~\ref{subsec:pid}.
On the other hand, precise measurement of angles and strict kinematic relations of the scattering angles and energies in the elastic deuteron-proton scattering (Fig.~\ref{fig:kin2}) allow for correct identification of the process even when the regular PID method fails.
As a consequence two--track events can be identified as deuteron-proton elastic scattering even when one, or both particles underwent hadronic interaction in the thick scintillator, which influences the energy measurement and prevents successful application of $\Delta$E-E--based the PID method (Fig.~\ref{fig:ede0}).

\subsection{\label{sec:eff} Detector efficiency}
Determination of a true number of events of a given type from the number of events registered by the detector requires knowledge of the detector efficiency.  Since successful registration of a single particle is conditioned with complete information from three detectors (MWPC, $\Delta$E and E), the total detection efficiency, $\varepsilon(x, y)$, can be considered as a product of individual efficiencies of those detectors.
\begin{eqnarray}
\label{eqn:eff_part}
\varepsilon(x, y) &=& \varepsilon_{\text{MWPC}}(x, y)\cdot\varepsilon_{\Delta E}(x, y)\cdot\varepsilon_{E}
\end{eqnarray}

The efficiency of the E detector ($\varepsilon_{E}$) has been assumed to be 100\%. This is justified by a very tight fitting of the detector slabs. The gap between adjacent elements is of the order of 50 microns  as compared to 10 cm width of the front face of each element. Therefore, the problem of efficiency reduces to particle-type dependent energy threshold. 

Since the main sources of inefficiency for both  $\Delta$E and MWPC detectors are well localized, proper accounting for them required construction of position-dependent efficiency maps. This was possible using the position information from MWPC.

\begin{figure}
  \includegraphics[width=8.6cm]{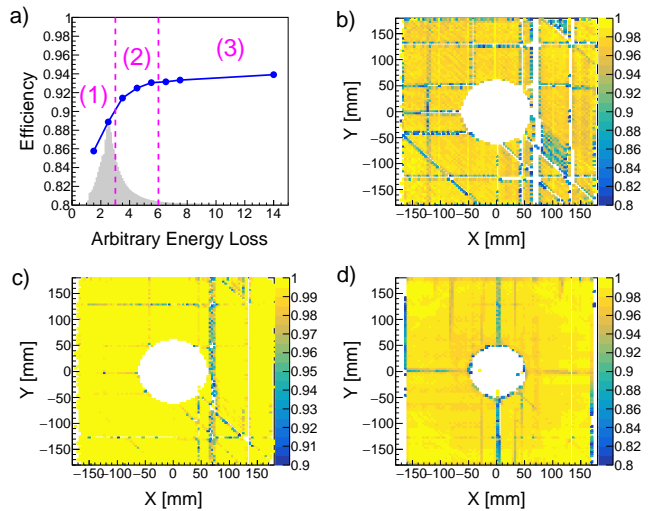}
 \caption{\label{fig:hardware} \emph{Panel a:} The average MWPC efficiency (blue dots connected by line) as a function of the energy loss in the gas mixture, Eq.~\ref{eqn:enloss}. The distribution of events is shown in gray. \emph{Panels b and c:} Position dependence of the MWPC efficiency shown only for the range (1), separately for full- (b) and weak--tracks (c). \emph{Panel d:} Analogous map of the $\Delta$E efficiency.}
\end{figure}

The efficiency of the $\Delta$E detector, $\varepsilon_{\Delta E}(x, y)$ (Fig.~\ref{fig:hardware}d), is calculated directly on the basis of a single particle events according to the following formula:
\begin{eqnarray}
\label{eqn:eff_de}
\varepsilon_{\Delta E}(x, y) &=& \frac{N_{\text{ref}+\Delta E}(x, y)}{N_{\text{ref}}(x, y)},
\end{eqnarray}
where the reference number of events, $N_{\text{ref}}$, corresponds to the number of all particles registered by MWPC with a correlated hit in the E scintillator, regardless of the $\Delta$E information, while for $N_{\text{ref}+\Delta E}$, additional matching with information from the $\Delta$E detector was required.

Position dependent efficiency maps of the multi-wire proportional chamber have been obtained from single plane efficiencies ($\varepsilon_{X},\varepsilon_{Y}, \varepsilon_{U}$) according to the following formulas:
\begin{eqnarray}
\label{eqn:eff_mwpc}
\varepsilon_{\text{MWPC}}(x, y) &=& \varepsilon_{X}(x, y) \cdot \varepsilon_{Y}(x, y) \cdot \varepsilon_{U}(x, y),\\
\label{eqn:eff_mwpc_weak}
\varepsilon^{\text{weak}}_{\text{MWPC}}(x, y) &=& \varepsilon_{\text{MWPC}}(x, y)\\  
\nonumber &+& \varepsilon_{X}(x, y) \cdot \varepsilon_{Y}(x, y) \cdot (1-\varepsilon_{U}(x, y))\\
\nonumber &+& \varepsilon_{X}(x, y) \cdot (1-\varepsilon_{Y}(x, y)) \cdot \varepsilon_{U}(x, y)\\
\nonumber &+& (1-\varepsilon_{X}(x, y)) \cdot \varepsilon_{Y}(x, y) \cdot \varepsilon_{U}(x, y).
\end{eqnarray}
The first one (Eq.~(\ref{eqn:eff_mwpc})) corresponds to full--tracks, when the coincidence of all 3 planes was required, and the second (Eq.~(\ref{eqn:eff_mwpc_weak})) to the analysis in which also weak--tracks were accepted (see section \ref{sec:tracking}). Efficiencies of individual planes have been calculated using position information from the remaining two planes, and requiring information from both scintillator hodoscopes, in a way analogous to the one already introduced for $\Delta$E. It is clear that the efficiency in the approach allowing for weak--tracks is much less sensitive to local defects (dead channels) present in each plane. Average efficiency in this case was as high as 98\%  (see Fig.~\ref{fig:hardware}c), as compared to 85\% for full--tracks (see Fig.~\ref{fig:hardware}b). In order to better understand systematics associated with MWPC efficiency, full analysis of cross section has been performed with and without accepting weak--tracks. The good agreement obtained strengthens our confidence in the final results of efficiency calculations \cite{ciepal_prc2_2019}. The data analysis is based on the full--tracks due to better angular resolution achieved in this way.

The MWPC efficiency depends also on the particle type and energy and this effect is visible in efficiency maps for full--tracks. This in fact can be traced back to the dependence on the relative energy loss of a given particle within the detector gas-mixture. In order to account for this effect, a new variable was introduced:
\begin{eqnarray}
\label{eqn:enloss}
E_{\text{loss}} &\sim& q^{2}\frac{m}{E_{k}},
\end{eqnarray}
where $q$ and $m$ are the particles' charge and mass, while $E_{k}$ denotes their kinetic energy. This variable was normalized in a way that for the most energetic among all the registered particles, elastically-scattered protons, it equals to one. Fig.~\ref{fig:hardware}a  presents  distribution of the $E_{\text{loss}}$ (regardless the particle type) and the average efficiency obtained for various bins in $E_{\text{loss}}$. In the final analysis, in order to retain acceptable statistics in each bin of the efficiency map, only three satisfactory populated ranges of $E_{\text{loss}}$ have been defined, as shown in Fig.~\ref{fig:hardware}a. The final efficiency map for minimum ionizing particles  (Fig.~\ref{fig:hardware}b) registered in this experiment (region marked as (1) in $E_{\text{loss}}$ distribution) is compared to the efficiency maps constructed for $\varepsilon^{\text{weak}}_{\text{MWPC}}$ (Fig.~\ref{fig:hardware}c). Because of contact problems at hardly accessible places, certain electronics channels of MWPC did not work. There are quite numerous inefficient regions in MWPC, especially for full--tracks corresponding to region (1) of $E_{\text{loss}}$. The correction defined in Eq.~(\ref{eqn:eff_mwpc}) is not effective for crossing inefficient wires in two or more planes. In any such cases or in more general cases of low final detector efficiency $\varepsilon(x,y)<0.5$ (Eq.~(\ref{eqn:eff_part})) the affected detector region was rejected from the analysis. The acceptance loss was calculated with Monte Carlo simulation and corrected for, as described in the next section.

\subsubsection{Configurational Efficiency}
When two particles enter the same detector element the reconstructed information is distorted. This leads to false energy reconstruction and may corrupt particle identification. Such effect, further referred to as configurational efficiency, depends strongly on geometry of the final state and of the detector, and can be accounted for by Monte--Carlo simulations.

\begin{figure}
  \includegraphics[width=8.3cm]{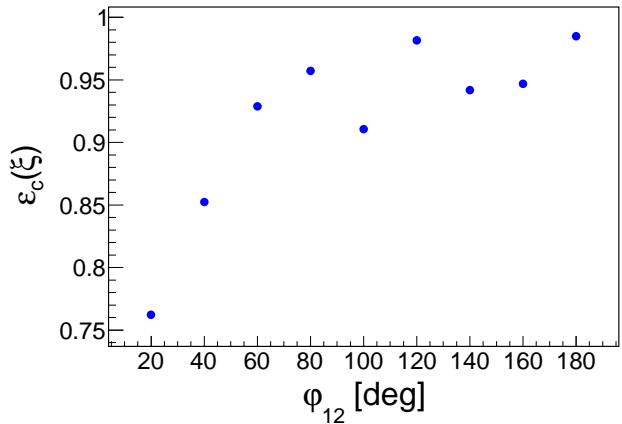}
 \caption{\label{fig:geoeff} Configurational efficiency for a set of breakup proton-proton configurations characterized with \mbox{$\vartheta_{1}=27^{\circ},\vartheta_{2}=21^{\circ}$} and $\Delta\varphi=10^{\circ}$. General trend, observed also for other combinations of polar angles, shows the decrease of the efficiency with decreasing relative azimuthal angle between protons, $\varphi_{12}$, with local minima determined by the geometry of the E detector.}
\end{figure}

Due to coplanarity condition, the loss of events corresponding to elastic scattering due to configurational efficiency is practically negligible. For the breakup reaction the efficiency strongly depends on breakup kinematics, defined by polar angles of emission of both protons, $\vartheta_{1}$ and $\vartheta_{2}$, and relative azimuthal angle, $\varphi_{12}$, between them. Due to axial symmetry of the cross section, the so-defined configuration is rotated around the beam axis. The configurational efficiency is determined by the analysis of a set of breakup events simulated with the use of Geant4 framework with the Wall detector geometry included. Since a good statistical accuracy of such correction has been ensured, the only significant uncertainty may originate from the inaccuracies of the experimental setup (detector or beam geometry) and the applied model of an event generator. In the following, the uniform 3-body breakup phase space distribution has been used, which is well justified in the case of narrow angular ranges applied in defining the configuration. The configurational efficiency for a given geometry $(\vartheta_{1},\vartheta_{2},\varphi_{12})$ is defined as the ratio of the number of events for which both particles were registered by separate detector elements, to the number of all simulated events. As expected, the configuration efficiency rises with increase of $\varphi_{12}$, with pronounced local minima reflecting the structure of the E detector (Fig.~\ref{fig:geoeff}). Due to much finer granularity of the MWPC as compared to the hodoscopes, the contribution of this detector to the configurational efficiency is very small. In the simulation, the distribution of cluster sizes observed in the experiment was used to account for hit losses due to coalescence of clusters produced by different particles. The final correction for acceptance losses takes into account losses due to two particles registered in the same element and, earlier-discussed regions of low efficiency. It is calculated as follows:
\begin{eqnarray}
\label{eqn:geoeff}
\varepsilon_{c}(\xi) &=& \frac{N_{\text{rec}}(\xi)}{N_{\text{tot}}(\xi)},
\end{eqnarray}
where $\xi$ defines the geometry of the reaction products: \mbox{$\xi = \{\vartheta_{1},\vartheta_{2}, \varphi_{12}\}$} for the breakup reaction and \mbox{$\xi=\{\vartheta_{p}\}$} for the elastic-scattering channel, $N_{\text{tot}}$ is the total number of coincidences generated for this configuration and $N_{\text{rec}}$ counts only those events which are successfully registered by the virtual BINA detector.

The total correction factor related to the efficiencies for registering of a number ($N$) of coincident events in the chosen configuration $\xi$ can be written as:
\begin{eqnarray}
\label{eqn:eff_tot}
\varepsilon^{\xi}_{N} &=& N\varepsilon_{c}(\xi)\left(\sum_{<i,j>}\frac{1}{\varepsilon(x_{i}, y_{i}) \cdot \varepsilon(x_{j}, y_{j})}\right)^{-1},
\end{eqnarray}
where $\epsilon(x_{i},y_{i})$ is the single particle efficiency defined in Eq.~\ref{eqn:eff_tot} and $<i,j>$ symbolizes the set of $N$ coincident pairs.

\subsubsection{Hadronic reactions}
\label{sec:hadro}
\begin{figure}
  \includegraphics[width=4.2cm]{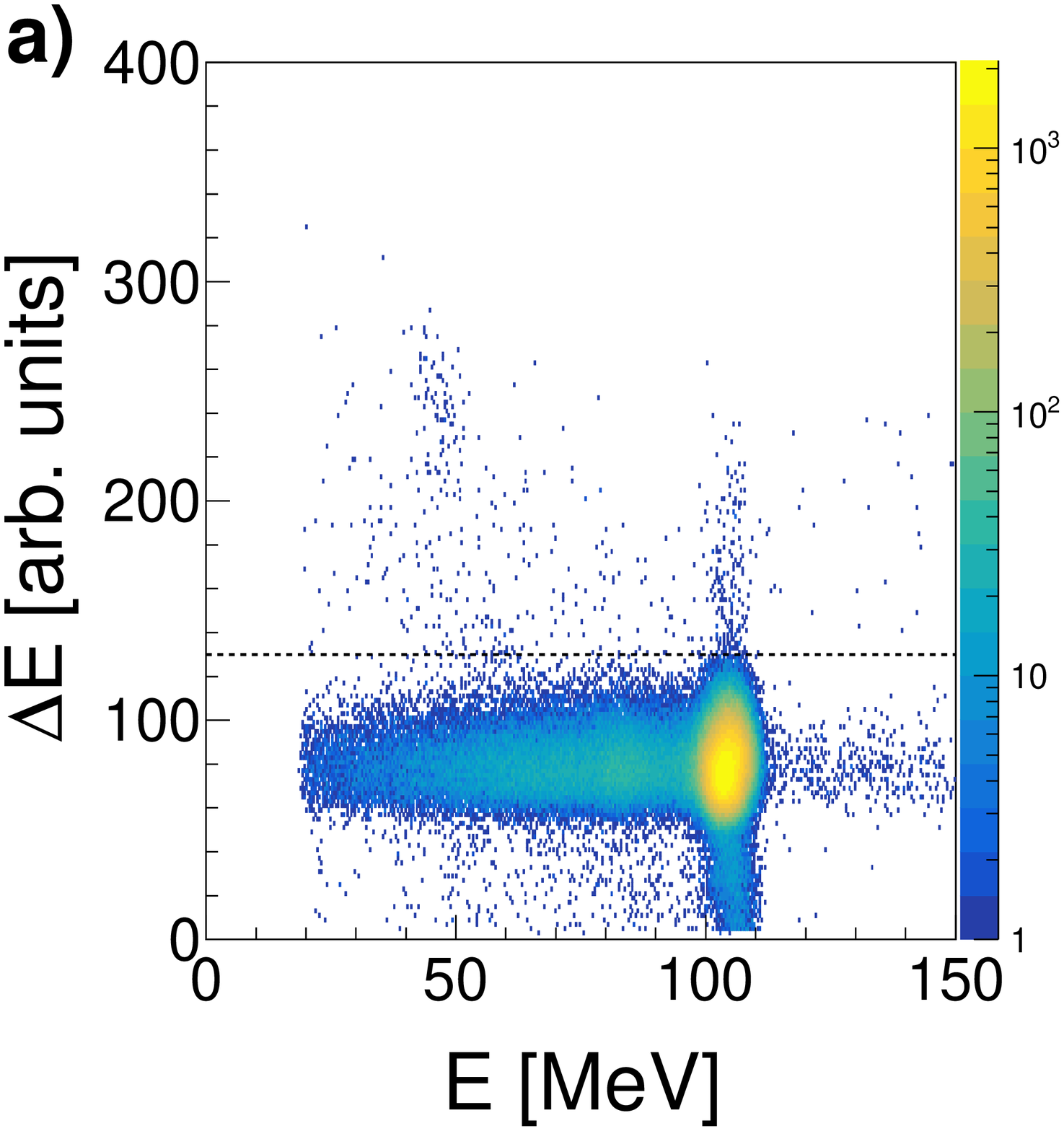}
  \includegraphics[width=4.2cm]{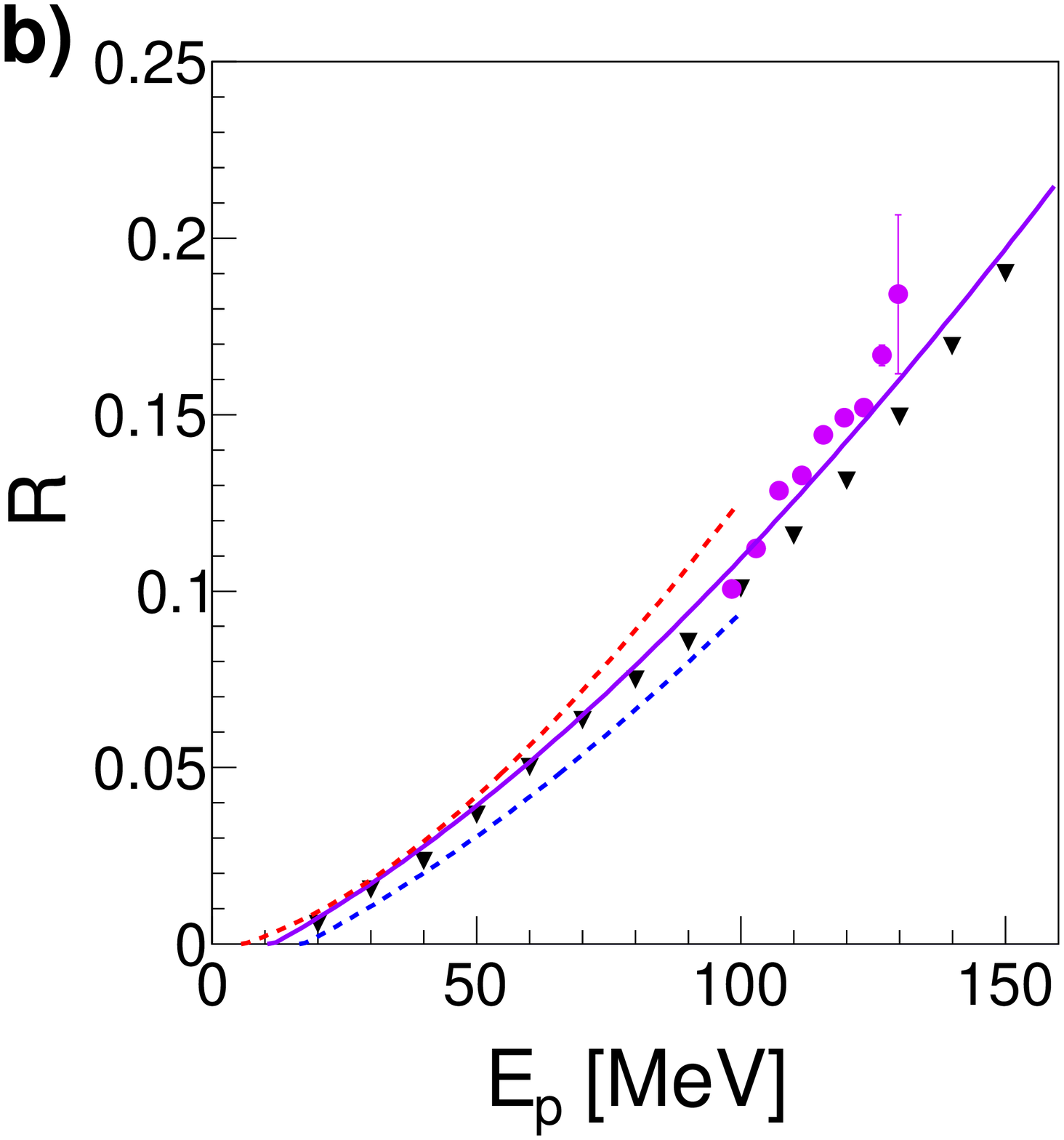}
 \caption{\label{fig:hadronic}  \emph{Panel a:} sample of $\Delta$E-E spectrum for chosen polar angle of elastically-scattered protons $\vartheta_{p} = 31^{\circ}\pm1^{\circ}$. Horizontal band extending on the left from elastic spot corresponds to hadronic interactions lowering the registered energy. Events below dashed line are used in the calculation of tail-to-peak ratio $R$. \emph{Panel b:} Obtained tail-to-peak ratio (points) compared to theoretical calculations at lower energies \cite{measday_1969} (lines) and to simulations \cite{ciepal_fbs_2019} (triangles). Solid line represents predictions of a simple model based on the material composition of the scintillator \cite{measday_1969} extrapolated to energies above 100~MeV.}
 \end{figure}

The calibration and particle identification procedures fail when the particle undergoes a hadronic reaction with large momentum transfer on its way to- or inside the E detector. In such cases a part of particle energy is lost, less light is produced in the scintillator, and as a consequence, the reconstructed kinetic energy is underestimated, leading to event rejection due to the PID cut.
The amount of affected events has been estimated based on experimental $\Delta$E--E spectra gated by kinematical conditions defining \emph{dp} elastic scattering (Fig.~\ref{fig:hadronic}a) in order to reject the breakup band. In these spectra hadronic interactions are visible as horizontal band protruding on the low-energy side from the elastic-scattering spot. Number of events integrated within this band has been normalized to the number of events inside the elastic peak (tail-to-peak ratio $R$).
Results obtained for several energies are in satisfactory agreement with theoretical predictions based on the effective inelastic cross section model for protons in a combination of materials building the plastic scintillator \cite{measday_1969} extrapolated to energies above 100 MeV (Fig.~\ref{fig:hadronic}b).
Therefore, the $R$ values corresponding to the solid line in Fig.~\ref{fig:hadronic}, considered as validated, were used in further analysis with up to 3.8\% systematic uncertainty. In order to account for the loss of breakup--originated protons due to hadronic interactions, a dedicated correction factor ($\eta(E_{p})$) has been introduced on the basis of $R$: 
\begin{eqnarray}
\label{eqn:eff_had}
\eta(E_{p}) = 1 + R.
\end{eqnarray}
In the case of elastic scattering protons and deuterons are identified on the basis of kinematical relations of angles. As a consequence, events were lost only if hadronic interactions had occurred before the particle reached scintillators and the corresponding correction was negligible.

\subsection{Luminosity}
\begin{figure}
  \includegraphics[width=8.4cm]{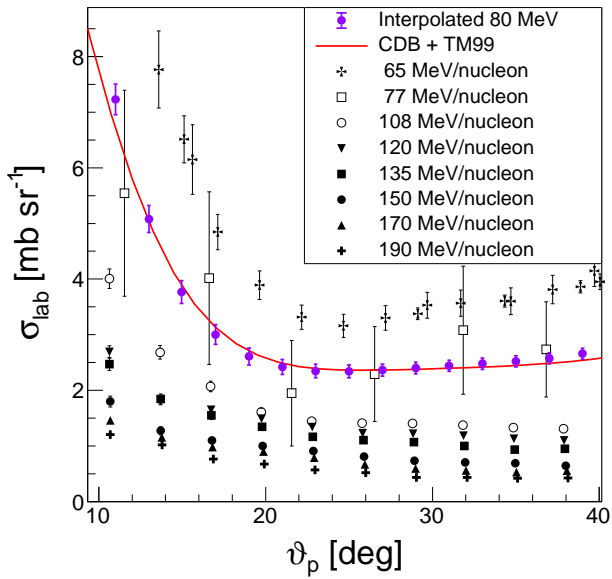}	
 \caption{\label{fig:interpol} Interpolated cross section for deuteron-proton elastic scattering at the energy of 80~MeV/nucleon (violet) in the laboratory frame in comparison to the theoretical calculations based on CDB+TM99 potential \cite{witala_1998}. The experimental data used for this interpolation are shown in black.}
\end{figure}

Direct measurement of absolute differential cross sections requires precise knowledge of the beam current, target thickness and scattering angles of the reaction products. In the present experiment, neither target cell surface density nor very low beam current (in the range of few pA) were known sufficiently precisely. The normalization factor had to be obtained on the basis of simultaneously-measured elastic \emph{dp} scattering and the corresponding cross section derived from previous experiments. Since no published cross section data for the \emph{dp} elastic scattering process at 80~MeV/nucleon exist, model independent interpolation has been done based on all existing experimental data in the range of 65--190~MeV/nucleon \cite{shimizu_1982, davidson_1963, sekiguchi_2002, ermisch_2003}.
The obtained absolute values of the differential  cross section, $\sigma_{\text{lab}}$, agree very well with theoretical calculations based on the Charge Dependent Bonn potential supplemented with the TM99 three--nucleon force \cite{witala_1998} (Fig.~\ref{fig:interpol}). The reference data were subsequently used to calculate experimental integrated luminosity according to:
\begin{eqnarray}
\label{eqn:lumi}
L (\vartheta_{p}) &=& \frac{N_{\text{pd}}(\vartheta_{p})}{\sigma_{\text{lab}}(\vartheta_{p})\cdot\Delta\Omega \cdot\varepsilon(\vartheta_{p})},
\end{eqnarray}
where $\Delta\Omega$ is the solid angle and $N_{\text{pd}}(\vartheta_{p})$ is the number of elastically-scattered protons registered in certain bin in polar angle of proton emission. To reduce the influence of acceptance losses resulting from reaching edges of a square-like detector, the elastically-scattered particles were collected from limited space close to diagonals of MWPC defined by azimuthal angle $\varphi_{12} = \{45^{\circ}, 135^{\circ}, -45^{\circ}, -135^{\circ}\}$ with tolerance $\pm 15^{\circ}$.
\begin{figure}
  \includegraphics[width=8.4cm]{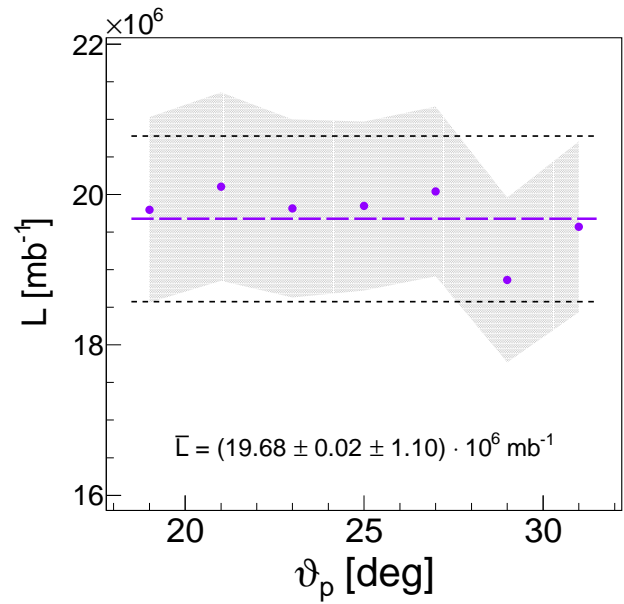}
 \caption{\label{fig:lume} Luminosity integrated over time determined on the basis of elastic scattering, independently for each proton polar angle. Statistical errors are negligible, while  gray shade corresponds to the range of systematic errors of individual data points. The average value of the luminosity (violet dashed line) is known with accuracy dominated by systematic uncertainty (dashed black lines).}
\end{figure}

Obtained values are, as expected, consistent with one another within experimental uncertainties (Fig.~\ref{fig:lume}). As the final integrated luminosity value, the average \mbox{$\bar{L} = \left(19.68\pm0.02^{stat}\pm1.10^{syst}\right)\cdot10^{6} [\text{mb}^{-1}]$} was taken.

\subsection{Differential Cross Sections for Breakup Process}
\begin{figure}
  \includegraphics[width=8.6cm]{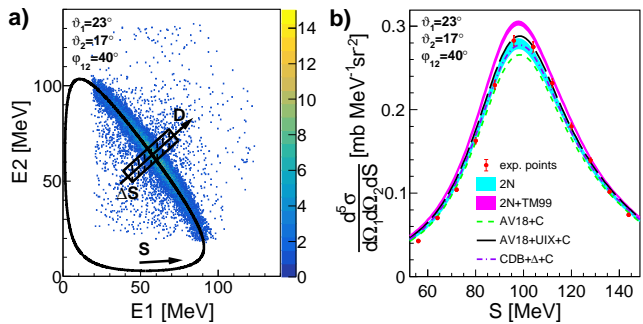}
  \caption{\label{fig:ee} \emph{Panel a:} Distribution of events for a selected breakup configuration together with the corresponding kinematic curve; definitions of variables $S$ and $D$ are presented in a graphic way. \emph{Panel b:} Measured cross section distribution as a function of $S$-variable compared to the theoretical predictions for this configuration. See Sec.~\ref{sec:results} for details of theoretical models specified in the legend.}
\end{figure}

The breakup cross section has been determined for 243 angular configurations. The angular range for integration of events has been chosen to $2^{\circ}$ in polar and $20^{\circ}$ in azimuthal angles. For each configuration, breakup events are placed on the $(E_{1},E_{2})$ plane, in which they group along the corresponding kinematical curve (Fig.~\ref{fig:ee}a).
The width of the distribution in the direction perpendicular to the curve, $D$--coordinate, depends on the energy resolution and spread of kinematics corresponding to the size of the angular bin. Arc-length measured along the kinematic curve is used to define $S$-coordinate \cite{parol_phd_2016}.

The five-fold differential cross section for the deuteron breakup reaction has been calculated according to the following formula:
\begin{eqnarray}
\label{eqn:crosssec}
\frac{d^{5}\sigma(\xi, S)}{d\Omega_{1}d\Omega_{2}dS} &=& \frac{N^{\text{BR}}(\xi, S)\cdot\eta(E_{1})\cdot\eta(E_{2})}{\varepsilon^{\xi}_{N^{\text{BR}}}\cdot L\cdot \Delta\Omega_{1}\cdot\Delta\Omega_{2}\cdot\Delta S},
\end{eqnarray}
where $L$ is the luminosity integrated over time and $N^{\text{BR}}$ corresponds to the number of events falling into the chosen geometry $\xi$ within the ranges of integration defined below. $\varepsilon^{\xi}$ is the total detector efficiency as defined in Eq.~(\ref{eqn:eff_tot}), and $\eta(E_{1}),\ \eta(E_{2})$  account for the energy-dependent hadronic interaction corrections. Ordering of protons in the case of analysis of symmetric configurations $(\vartheta_{1} = \vartheta_{2})$ is random, while in the case of asymmetric configurations $(\vartheta_{1} \neq \vartheta_{2})$, the proton scattered at a larger polar angle is marked as the first one ($\vartheta_{1} > \vartheta_{2}$).
 
Determining the cross section starts from the measured number of events $(N^{\text{BR}})$ from the deuteron breakup channel. All the accepted events are classified into kinematic configurations defined by scattering angles $(\vartheta_{1}, \vartheta_{2}, \varphi_{12})$. The adopted grid assumes 9 intervals in the relative azimuthal angle (with width of $\Delta\varphi_{12}=20^{\circ}$) and 27 combinations of intervals in $\vartheta_{1}$ and $\vartheta_{2}$ angles, $2^{\circ}$ wide. Centers of these intervals are given by the formula:
\begin{eqnarray}
\label{eqn:configurations}
\vartheta_{1, 2}  &=& 17^{\circ} + 2^{\circ}k,\ \ k=0, 1, 2,\dots,6 \\
\nonumber \varphi_{1, 2}  &=& 20^{\circ}j,\ \ j=1, 2,\dots,9
\end{eqnarray}
For each geometry, the two--dimensional $E_{2}~\text{vs.}~E_{1}$ distribution is constructed and events falling within a single bin in $S$ with $\Delta S=8$~MeV (see e.g. a hatched rectangle in Fig.~\ref{fig:ee}) are projected onto the axis locally perpendicular to the $S$-curve ($D$-coordinate). For each $S$ bin, a gaussian function is fitted to the distribution of events along $D$ and integrated over $\pm3\sigma$. The resulting $N^{BR}$ value is normalized according to Eq.~(\ref{eqn:crosssec}) and the cross section distribution as a function of $S$ is obtained (see Fig.~\ref{fig:ee}).

\subsection{Experimental Uncertainties}
The elastic scattering and breakup reactions were measured simultaneously, by the same detector and under the same experimental conditions, like beam current, triggers, dead--time etc.. Though some of the systematic effects are the same, the clear differences between both processes (coplanarity of elastic scattering kinematics and different particle types in the exit channels) lead to different balance of systematic uncertainties which will be discussed separately for each reaction channel.

For the elastic scattering, systematic errors were calculated in bins of $\vartheta_{p}$. These systematic factors bias the luminosity and, as a consequence, the global normalization of the breakup cross section. Estimation of systematic errors of differential cross sections for deuteron breakup was performed separately for each configuration defined by $\xi$ or for an individual data point. Global results are presented in Table~\ref{tab:errors}, while the individual uncertainties of data points are shown as bands in the Figs.~\ref{fig:cs1}--\ref{fig:cs9}.

The methods adopted to reconstruct physical parameters of the registered particles were studied as one of the potential sources of systematic errors. Among them, uncertainties associated with the reconstruction of angles based on the assumption of a point-like target were determined. For that purpose, reaction point was varied within the volume of beam--target intersection and a corresponding range of variation of angles was determined. The analysis was repeated with all angles shifted within their uncertainty and the resulting change of cross section was included into the systematic uncertainty.

\begin{figure}
  \includegraphics[width=6cm]{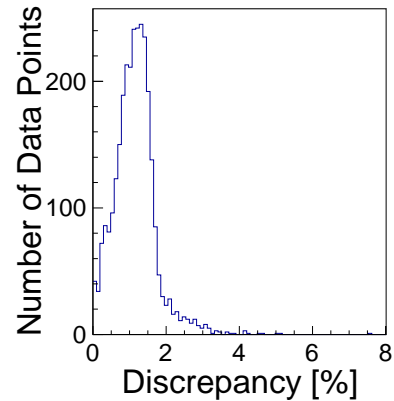}
  \caption{\label{fig:pid_discrepancy} Distribution of relative difference between breakup cross section values obtained with 2$\sigma$ and 3$\sigma$ cut on proton PID peak.}
\end{figure}

The effect of the particle-identification method based on linearization of the E--$\Delta$E spectra has been estimated by data analysis for different ranges of accepted protons around the corresponding peaks in the $\widetilde{E}$ variable (See Sec.~\ref{sec:pid}). 
The shape of PID peaks is not exactly gaussian, and corresponding factors have been calculated from the real distribution. In an ideal case applying these correction factors should result in the same cross section values for any range of PID peak used in analysis, with only statistical uncertainties affected. As the PID-related uncertainty, the maximal deviation from this behavior was taken, while the proton acceptance range was varied between 2$\sigma$ and 3$\sigma$ of the corresponding peaks. Observed percentage discrepancy for most measured cross section points lie below 2\% (see Fig.~\ref{fig:pid_discrepancy}) which is adopted as PID-related systematic uncertainty. In the case of elastic scattering the effect of PID was limited to only one particle, since the coincident deuteron was identified based on the strict kinematical relation of elastic scattering. Here, PID affects additionally the value of luminosity by about 0.7\%.

The systematic uncertainty related to the configurational efficiency has been studied and presented in details in a former publication \cite{ciepal_prc_2019}. Discrepancies between different methods of calculations of configurational efficiency presented in that work allow to state that this uncertainty is small for most of the configurations and it rises with decreasing relative azimuthal angles, $\varphi_{12}$. In particular, the largest contribution corresponds to a few selected configurations characterized by the smallest $\varphi_{12}$, while elastic scattering channel ($\varphi_{12}=180^{\circ}$) is practically immune to this component.
\begin{table}
 \caption{\label{tab:errors} Evaluation of total systematic errors.}
 \begin{ruledtabular}
 \begin{tabular}{lc}
  \textbf{Normalization factor (luminosity):} & \\
  Reconstruction of angles & 0.24\% \\
  Particle identification  & 0.7\% \\
  Cross-section interpolation & 4.1\% \\
  \hline
  \textbf{Breakup cross section:} & \\
  Reconstruction of angles & 0.15--0.24\% \\
  Particle identification &  0.12--7.6\% \\
  Hadronic reactions & 3.8\% \\
  Configurational efficiency & 0.1--16.3\% \\
 \end{tabular}
 \end{ruledtabular}
 \end{table}

\section{Results and discussion}
\label{sec:results}
The measured \textbf{2944} points of differential cross sections for 243 geometrical configurations of the deuteron breakup at 160~MeV were used to validate modern theoretical calculations.

\begin{figure}
  \includegraphics[width=4.275cm]{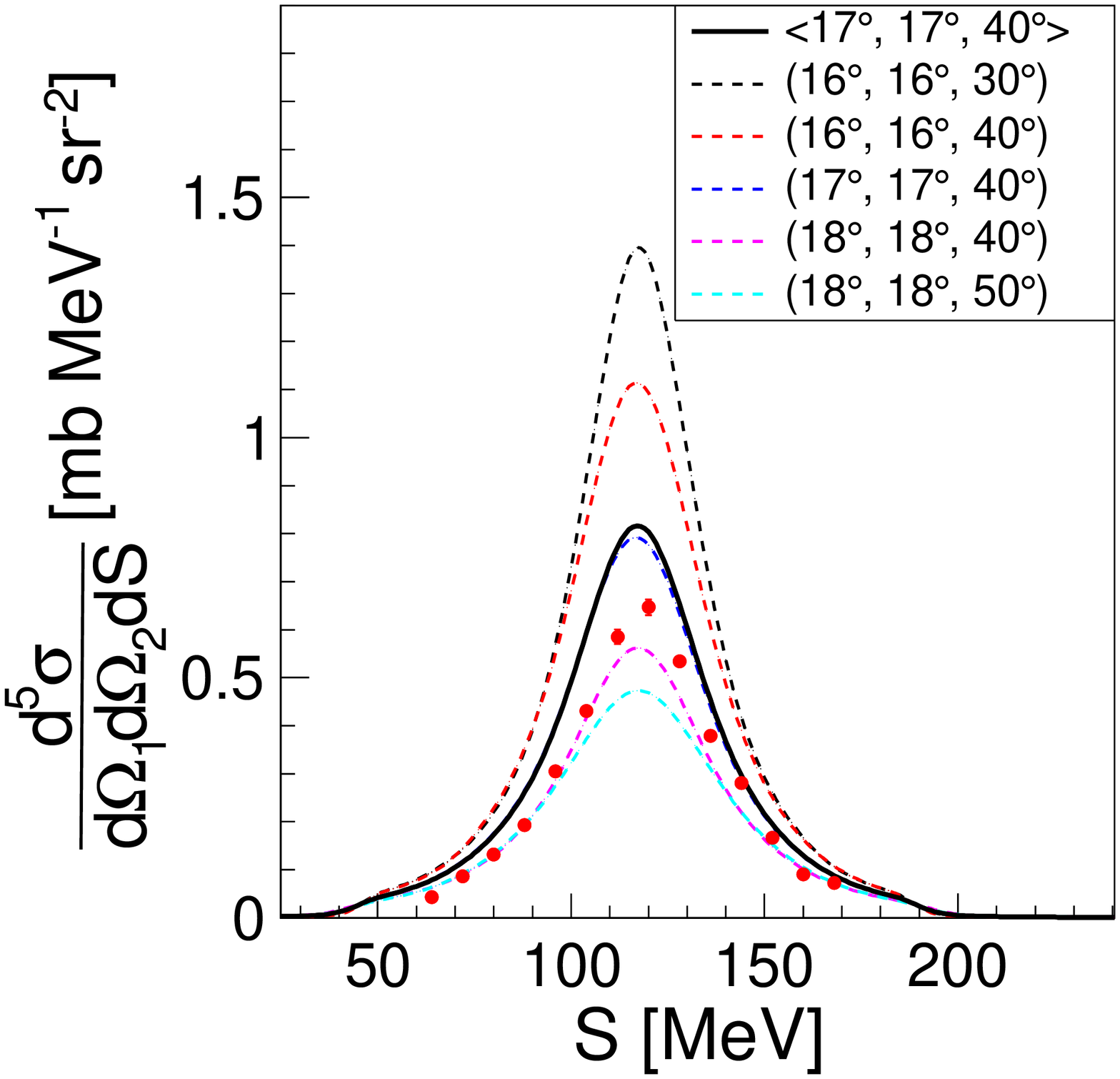}
  \includegraphics[width=4.275cm]{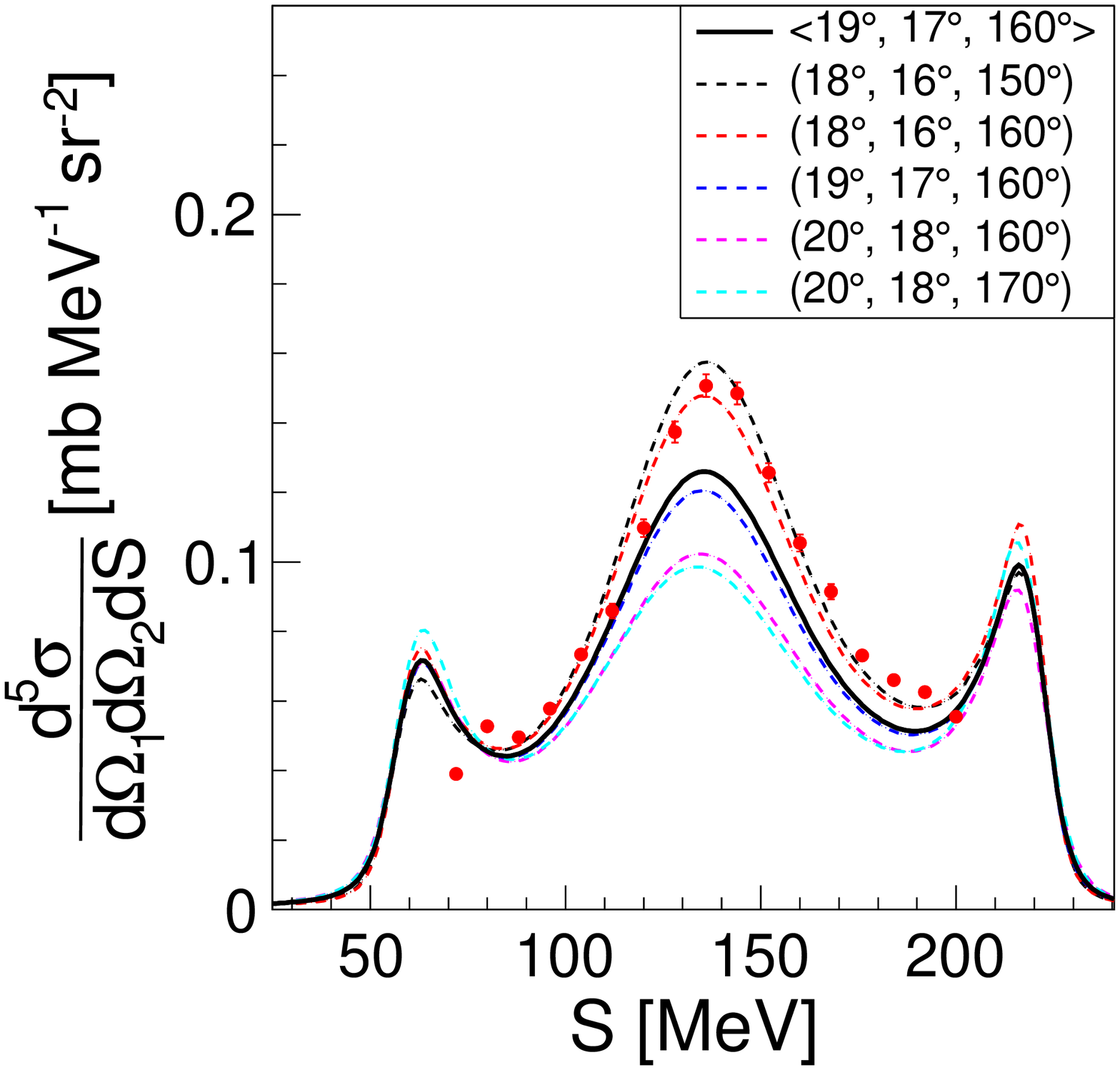}
  \caption{\label{fig:ateo} Effect of averaging (CDB+$\Delta$+C) calculations for a configuration given in region of large variations of the cross sections. Black solid line presents the average, dashed blue line is the prediction for central geometry while other dashed lines represent predictions for the limits of accepted angular range, each as a function of $S$ along the central kinematic. For comparison, the experimental data are also shown (red points).}
\end{figure}

In order to account for possibly large variations of the theoretical calculations within the finite bin size, the comparison must include values ​​of the cross section at the same angular range as in the experiment: $(\vartheta_{1}\pm\Delta\vartheta,\vartheta_{2}\pm\Delta\vartheta, \varphi_{12}\pm\Delta\varphi)$, where $\Delta\vartheta=1^{\circ}$ and $\Delta\varphi=10^{\circ}$. The final value representing the theoretical cross section for a given configuration includes, beside the value at the central point, also 26 points enclosing the corresponding bin, all of them projected onto a common relativistic
kinematics calculated for the central geometry. The $S$--coordinate is defined individually for each configuration, but the same step width, $\Delta S$ of 8~MeV, has been set for all. As an example, the data compared to the sample of raw and averaged predictions are presented in the Fig.~\ref{fig:ateo}.
\begin{figure}
  \includegraphics[width=8.cm]{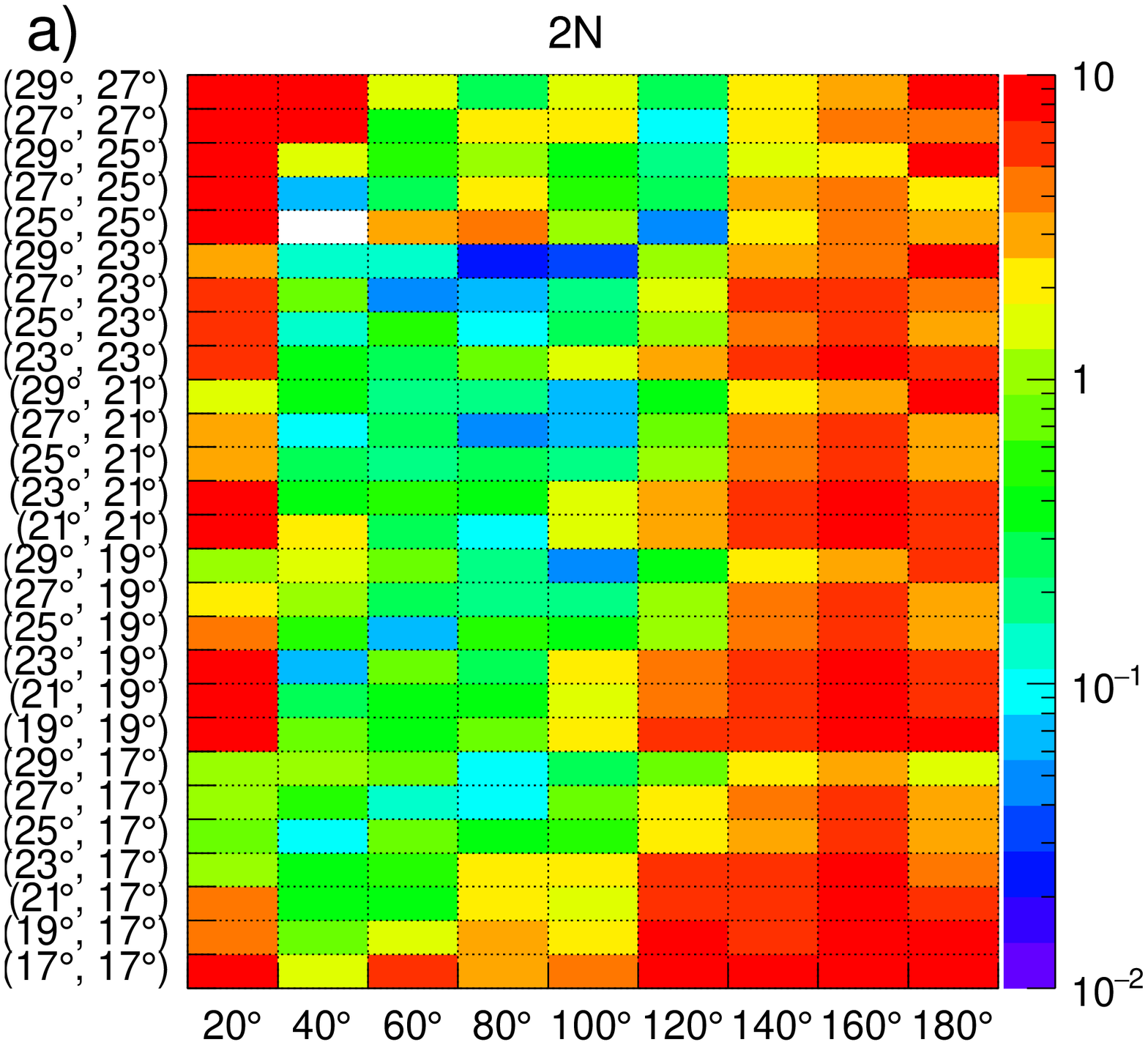}
  \includegraphics[width=8.cm]{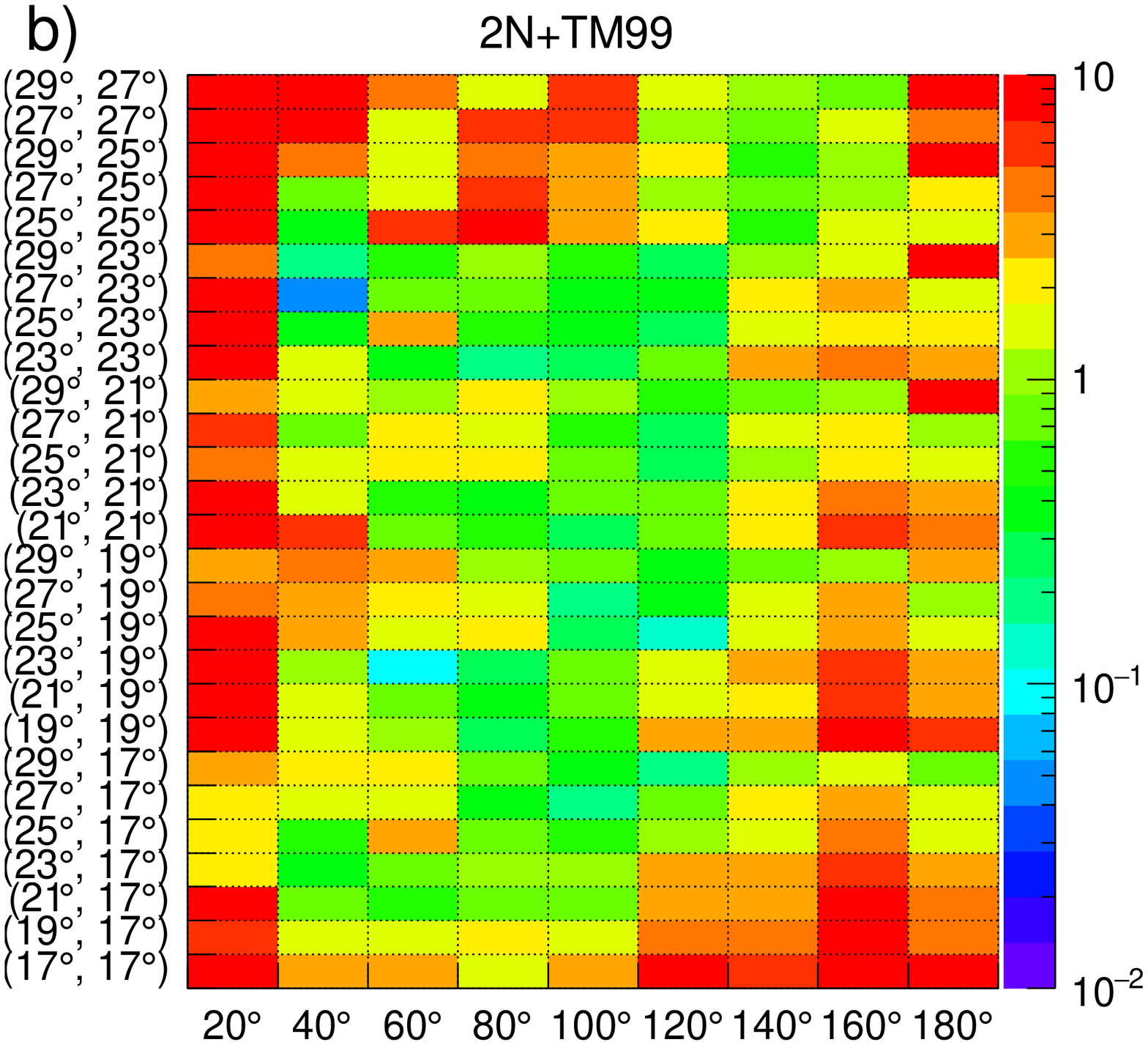}
  \caption{\label{fig:chi2_0} Maps of $\chi^{2}$ for individual geometries of breakup reaction defined by polar ($\vartheta_{1}, \vartheta_{2}$ on vertical axis) and azimuthal ($\varphi_{12}$ on horizontal axis) angles. Data are compared to the center of the bands corresponding to 2N (panel a) and  2N+TM99 (panel b) calculations. For details see the text.}
\end{figure}

The predicted values ​​of cross sections were calculated for (by H.~Wita\l a et al.) the set of nucleon--nucleon (2N) phenomenological potentials (CD Bonn \cite{machleidt_2001}, Argonne V18 \cite{wiring_1995}, Nijmegen I \cite{stoks_1994}, Nijmegen II \cite{stoks_1994}) and for these potentials supplemented with the Tucson-Melbourne (TM99) \cite{coon_2001} three--nucleon force (2N+TM99).
The next group of calculations (by A.~Deltuva) is based on the Argonne V18 potential in the variants with the added 3N force model of Urbana IX (AV18 + UIX) \cite{pudliner_1997} and taking into account the Coulomb force (AV18+C, AV18+UIX+C). Another set of calculations (by A.Deltuva) is based on the coupled channel formalism with  Charge Dependent Bonn potential with intermediate $\Delta$ creation (CDB+$\Delta$) \cite{deltuva_2008}, also taking into account the electromagnetic interaction (CDB+$\Delta$+C).
Calculation for 2N phenomenological potentials are presented in the form of bands, the width of which reflects the range of predictions obtained with individual potentials. The calculations for 2N+TM99 are presented in a similar way, while all other calculations are presented as individual lines. The complete set of experimental results is shown in Figs.~\ref{fig:cs1}--\ref{fig:cs9}. Each figure consists of three parts corresponding to three combinations of polar angles, as specified in the legends. The data are shown as red dots (full circles) surrounded by gray bands of systematic errors. Statistical errors are usually smaller than data points. The most striking observation is that the theories overestimate the cross section values for the configurations with small relative angles $\varphi_{12}$, and underestimate them for $\varphi_{12}$ lager than $120^{\circ}$. This effect is visible for all investigated polar angle combinations and is consistent with the observations of \emph{dp} breakup measurement at 130~MeV (65~MeV/nucleon) \cite{kistryn_2005}. The studies at 65~MeV/nucleon showed that including Coulomb interaction practically solved the problem of discrepancy, provided 3NF effects were also taken into account. In the present data, the effect is significantly reduced, but not removed for models including Coulomb force. In the certain areas of phase space the disagreement between the experiment and theory is in general large and cannot be accounted for by estimated systematical uncertainties. These uncertainties are in the most cases comparable or even larger than the differences between theoretical predictions given by the different calculations with or without three--nucleon force.
\begin{figure}
  \includegraphics[width=8.cm]{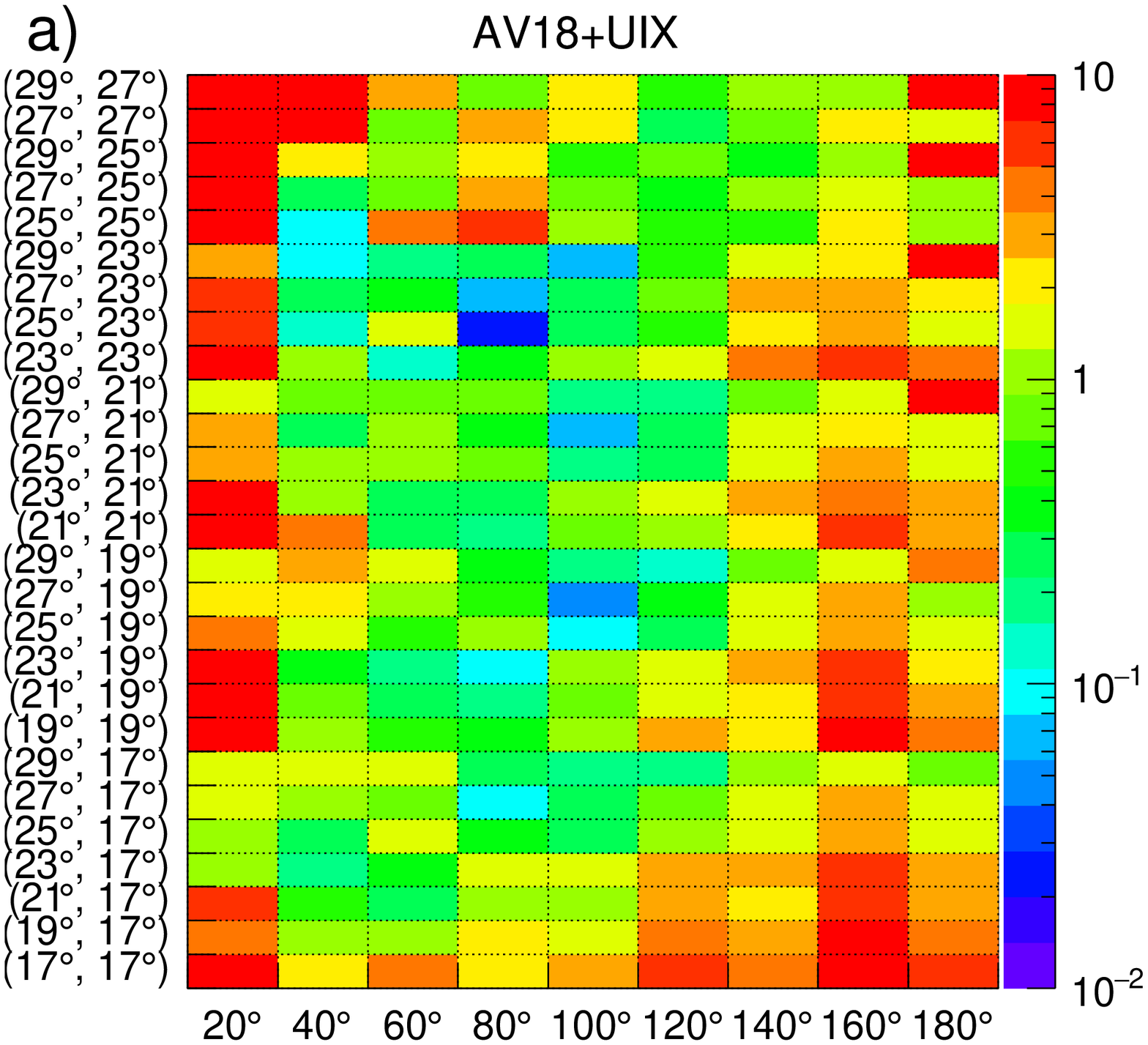}
  \includegraphics[width=8.cm]{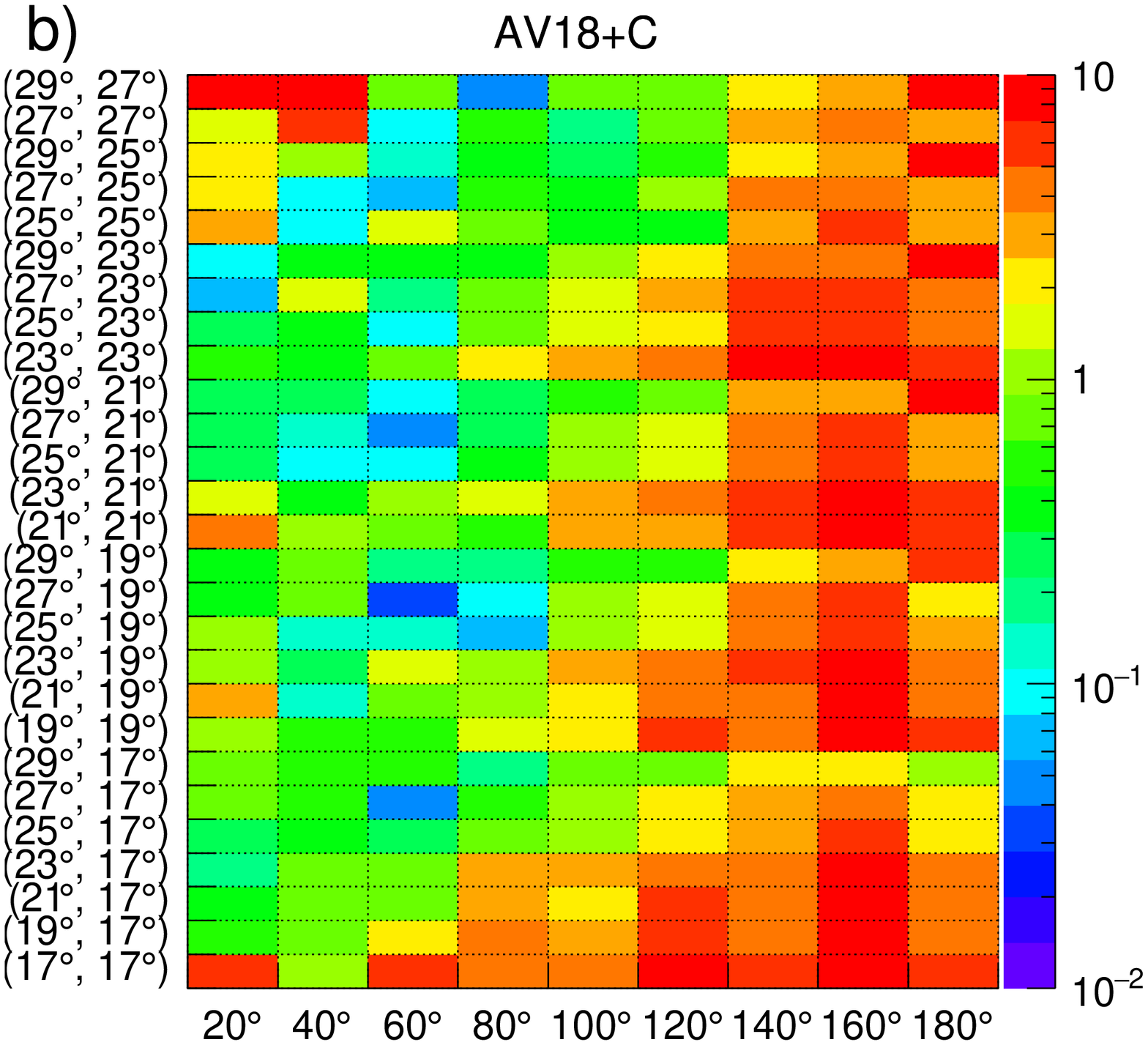}
  \includegraphics[width=8.cm]{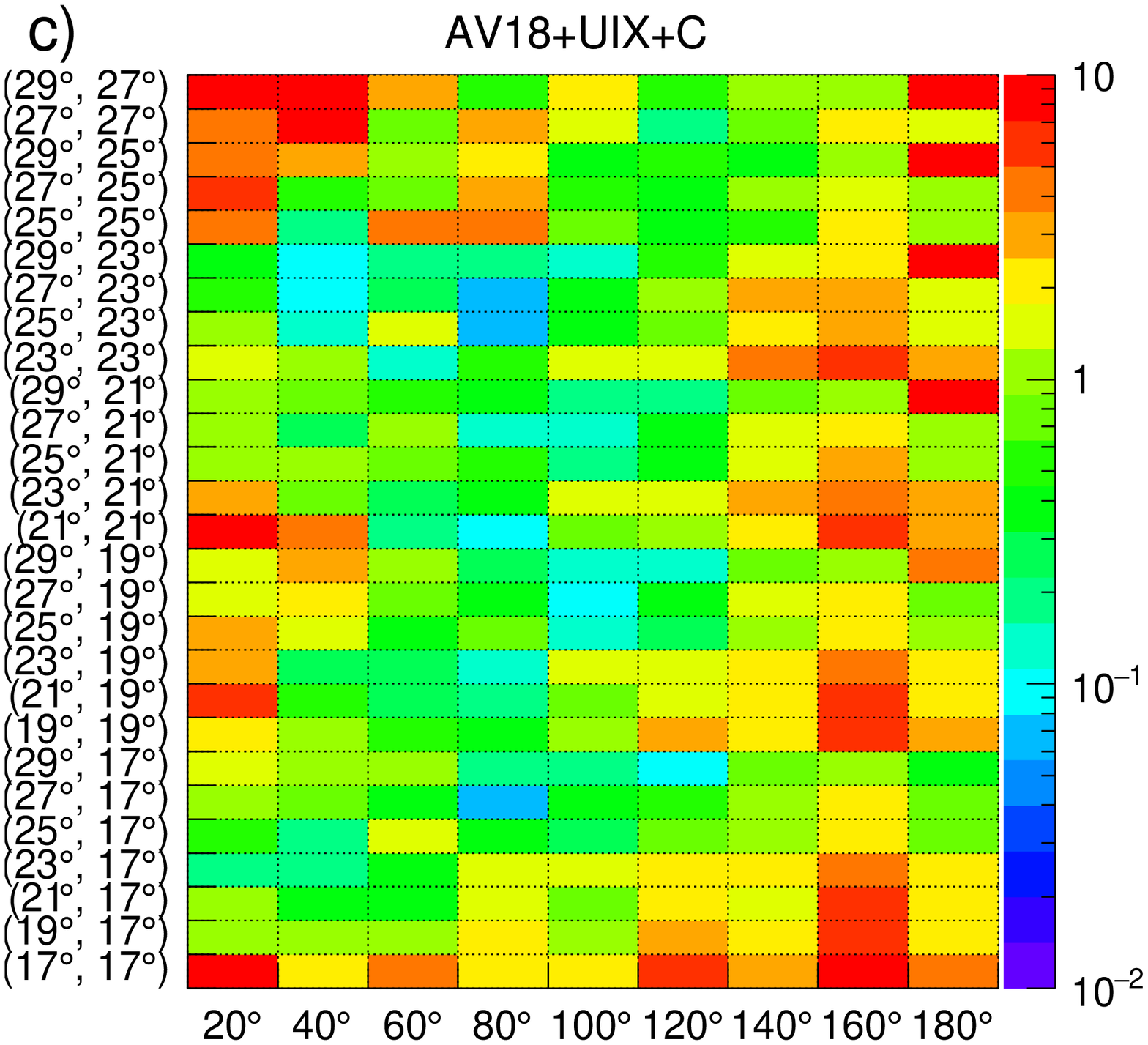}
  \caption{\label{fig:chi2_1} Same as in Fig.~\ref{fig:chi2_0} but for calculations with AV18 potential in combination with Coulomb interaction and/or with Urbana IX force, as specified at the top of each panel.   }
\end{figure}
\begin{figure}
  \includegraphics[width=8.cm]{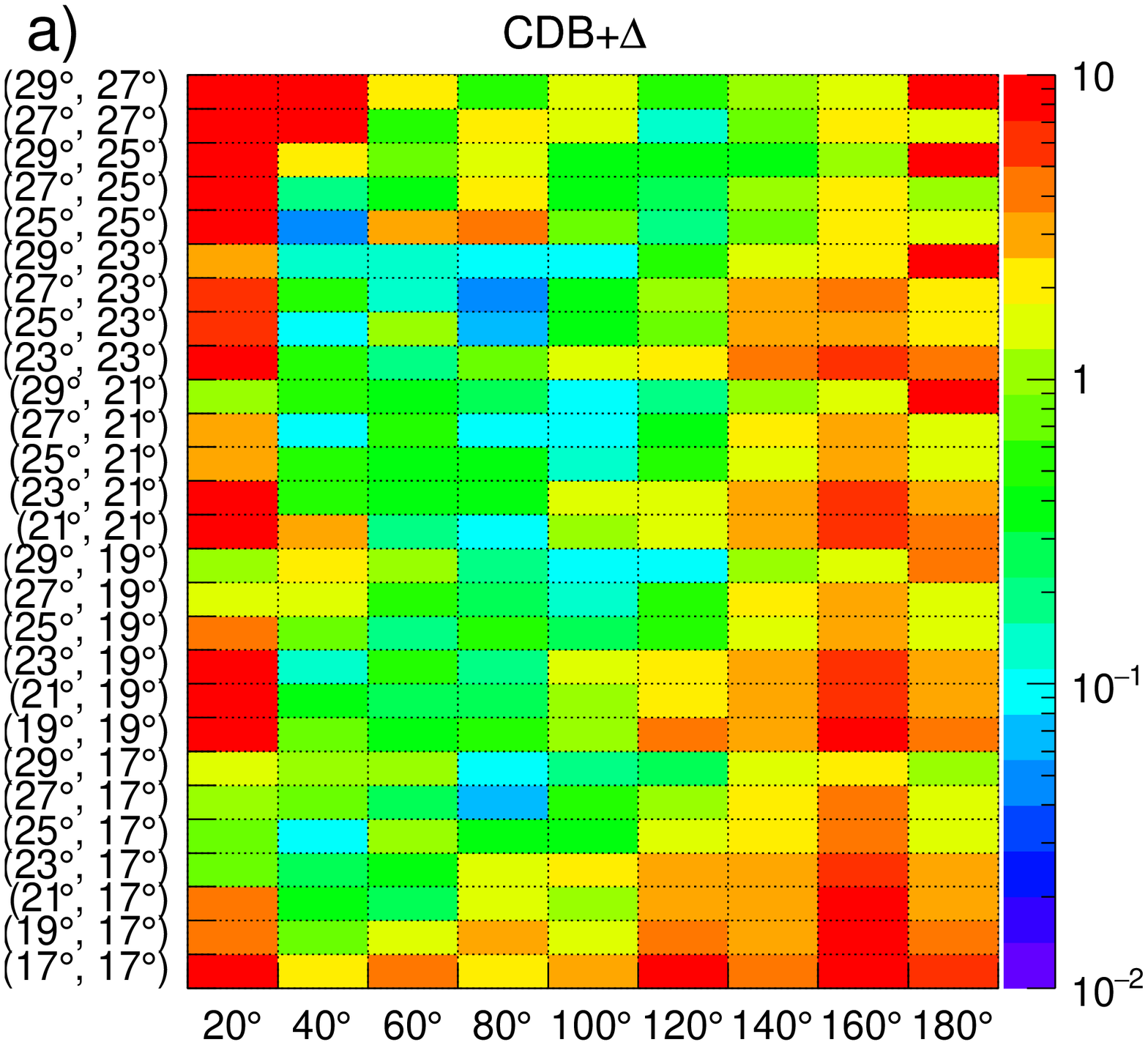}
  \includegraphics[width=8.cm]{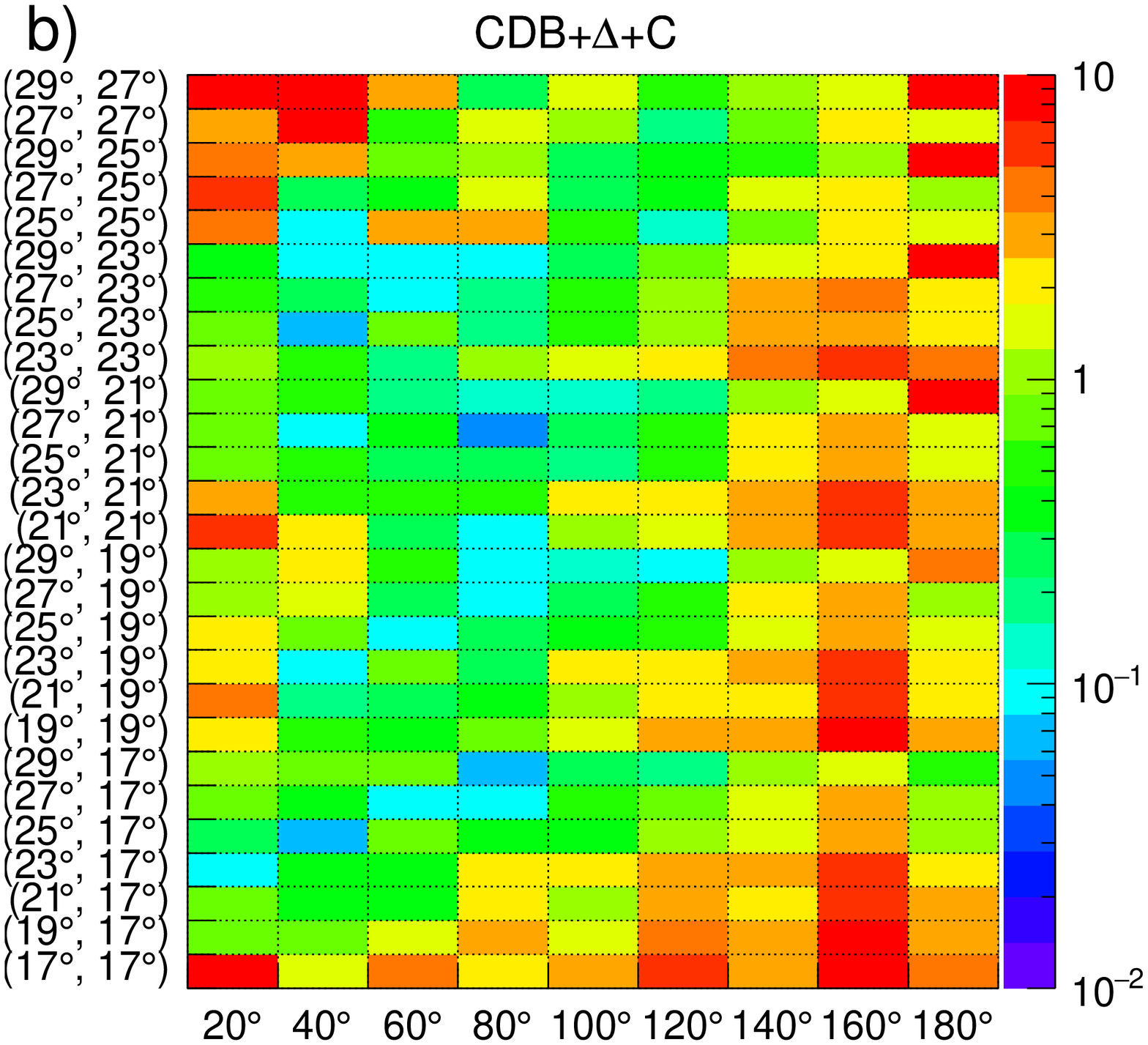}
  \caption{\label{fig:chi2_2} Same as in Fig.~\ref{fig:chi2_0} but for calculations with CDB+$\Delta$ potential, optionally in combination with Coulomb interaction, as specified at the top of each panel.}
\end{figure}

In order to make a quantitative comparison of the data and the theory and to conclude on the compatibility of the theoretical models with the obtained results, the $\chi^{2}$ analysis was carried out. The variable $\chi^{2}$ was calculated for each theoretical model and each geometrical configuration as follows:
\begin{eqnarray} 
\label{eqn:chi2}
\chi^{2}/\text{d.o.f} = \frac{1}{N-1}\sum^{N}_{i=1}\left( \frac{\sigma^{exp}_{i}-\sigma^{th}_{i}}{\Delta\sigma^{exp}_{i}} \right)^2, 
\end{eqnarray}

where $\sigma^{exp}$ corresponds to the measured experimental value of the cross section, $\Delta\sigma^{exp}_{i}$ is the total experimental error including both systematic and statistical uncertainties added in squares. $\sigma^{th}$ is the prediction of the theory being validated. In case of 2N and 2N+TM99 calculations, $\sigma^{th}$ corresponds to the center of the band. Tab.~\ref{tab:chi2} presents global values of $\chi^{2}$ obtained for all presented geometrical configurations.

Data presented for a wide spectrum of geometries allow to select the most reliable subset characterized by high experimental statistics, small value of estimated systematic error and flawless agreement between all  the presented theoretical models predictions. For the data set selected in this way, the agreement between experiment and theory is significantly better (see Figs.~\ref{fig:chi2_0}--\ref{fig:chi2_2}).
Calculations based on the 2N interactions alone (see Fig.~\ref{fig:chi2_0}a) provide very good description of the cross section data in the central part of the studied angular range. The quality of the description deteriorates significantly at low $\varphi_{12}$ values, where final state interactions (FSI) of the proton pair play a more important role. In the FSI region Coulomb interaction between protons should not be neglected and, indeed, calculations including this ingredient (AV18+C in Fig.~\ref{fig:chi2_1}b) provide results, which are much closer to the data in this region. The dominance of Coulomb interaction in the proton-proton FSI region was also observed in the breakup cross section at other energies \cite{ciepal_2015, eslami_phd_2009, mardanpour_phd_2008}.
In extreme cases, like the configuration $\vartheta_{1}=21^{\circ}$, $\vartheta_{2}=21^{\circ}$, $\varphi_{12}=20^{\circ}$ (see Fig.~\ref{fig:cs5}), Coulomb repulsion produces a dip in the middle of $S$ distribution, in the point corresponding to equal proton energies. Although adding of the Coulomb force improves description at low $\varphi_{12}$, it has no positive impact in other regions of discrepancies and even deteriorates the agreement at $\varphi_{12}\geq140^{\circ}$ (see Fig.~\ref{fig:cs1}). The remaining discrepancies can be attributed either to 3NF or relativistic effects.  Calculations including 3NF, like AV18+UIX (Fig.~\ref{fig:chi2_1}a), CDB+$\Delta$ (Fig.~\ref{fig:chi2_2}a) or 2N+TM99 (Fig.~\ref{fig:chi2_0}b) show significant improvement in the whole region of large $\varphi_{12}$, but only calculations including both Coulomb and 3NF, AV18+UIX+C (Fig.~\ref{fig:chi2_1}c) and CDB+$\Delta$+C (Fig.~\ref{fig:chi2_2}b) provide fairly good description for majority of the studied configurations. This success is also reflected in global values of $\chi^{2}$ shown in Table~\ref{tab:chi2}. We can conclude about the importance of 3NF  but, on the other hand, the improvement is not always sufficient. Generally, there are two regions of remaining high $\chi^{2}$ values. In the first one, at $\varphi_{12}\leq100^{\circ}$ and largest studied polar angles, $\vartheta_{1}$, $\vartheta_{2}\geq 25^{\circ}$, all the calculations are above the data and adding 3NF even increases $\chi^{2}$. In the second one, at $\varphi_{12}\geq140^{\circ}$, improvements due to introducing the 3NF are significant, but not sufficient.
\begin{center}
\begin{table}
  \caption{\label{tab:chi2} Calculated global $\chi^{2}$ including all presented breakup geometries.}
  \begin{ruledtabular}
  \begin{tabular}{llc}
    d.o.f.& 2944 &\\
    \hline
    \hline
    2N & 4.55 & H. Wita{\l}a\\
    2N+TM99 & 3.92 & Group\\
    \hline
    AV18+UIX & 3.42 & \\
    AV18+C & 3.89 & A. Deltuva\\
    AV18+UIX+C & 2.60 & \\
    \hline
    CDB+$\Delta$ & 3.51 & \\
    CDB+$\Delta$+C & 2.74 & A. Deltuva
    \end{tabular}
  \end{ruledtabular}
  \end{table}
\end{center}

\section{Summary}
\label{sec:summary}
The measurement of $^{1}H(d, pp)n$ at 80~MeV/nucleon enlarged existing dataset of differential cross sections by \textbf{2944} data points for 243 geometrical configurations creating dense grid in solid angle limited by $\vartheta\in(17^{\circ}, 29^{\circ})$. A set of models including contributions from two--nucleon interaction combined or not with 3NF or Coulomb force dynamics was validated via the $\chi^{2}$--test method referring each model predictions to the cross section distributions. In conclusion, evidence has been found that taking Coulomb and three-nucleon forces into account for modeling effective nuclear interaction globally improves the quality of predictions. Sensitivity of the differential cross section to Coulomb and 3NF effects varies significantly across the studied phase space. There are also configurations where none of the models provides satisfactory description of the data. Underestimation of the cross section data by theoretical calculations was also observed in the corresponding phase space regions in measurements of $^{2}H(p,pp)n$ reaction at 135 and 190~MeV \cite{eslami_phd_2009, mardanpour_phd_2008} and recently for $^{1}H(d,pp)n$ reaction at 170~MeV/nucleon \cite{klos_2019}. This observation  may suggest either important role of relativistic effects or problem with 3NF at higher energies. Further experimental studies, as well as development of fully relativistic calculations with 3NF and Coulomb force included, are important for ultimate understanding of the nature of observed discrepancies.

\begin{widetext}
\begin{acknowledgments}
This work was partially supported by the Polish National Science Center (NCN) from grants DEC-2012/05/B/ST2/02556, 2016/22/M/ST2/00173 and 2016/23/D/ST2/01703 and by the European Commission within the Seventh Framework Programme through IA-ENSAR (contract No. RII3-CT-2010-262010). Thanks also to the guys from AGOR and the source group who delivered a nice beam for the experiment.
\end{acknowledgments}
\begin{center}
\begin{figure}
  \includegraphics[width=17.cm]{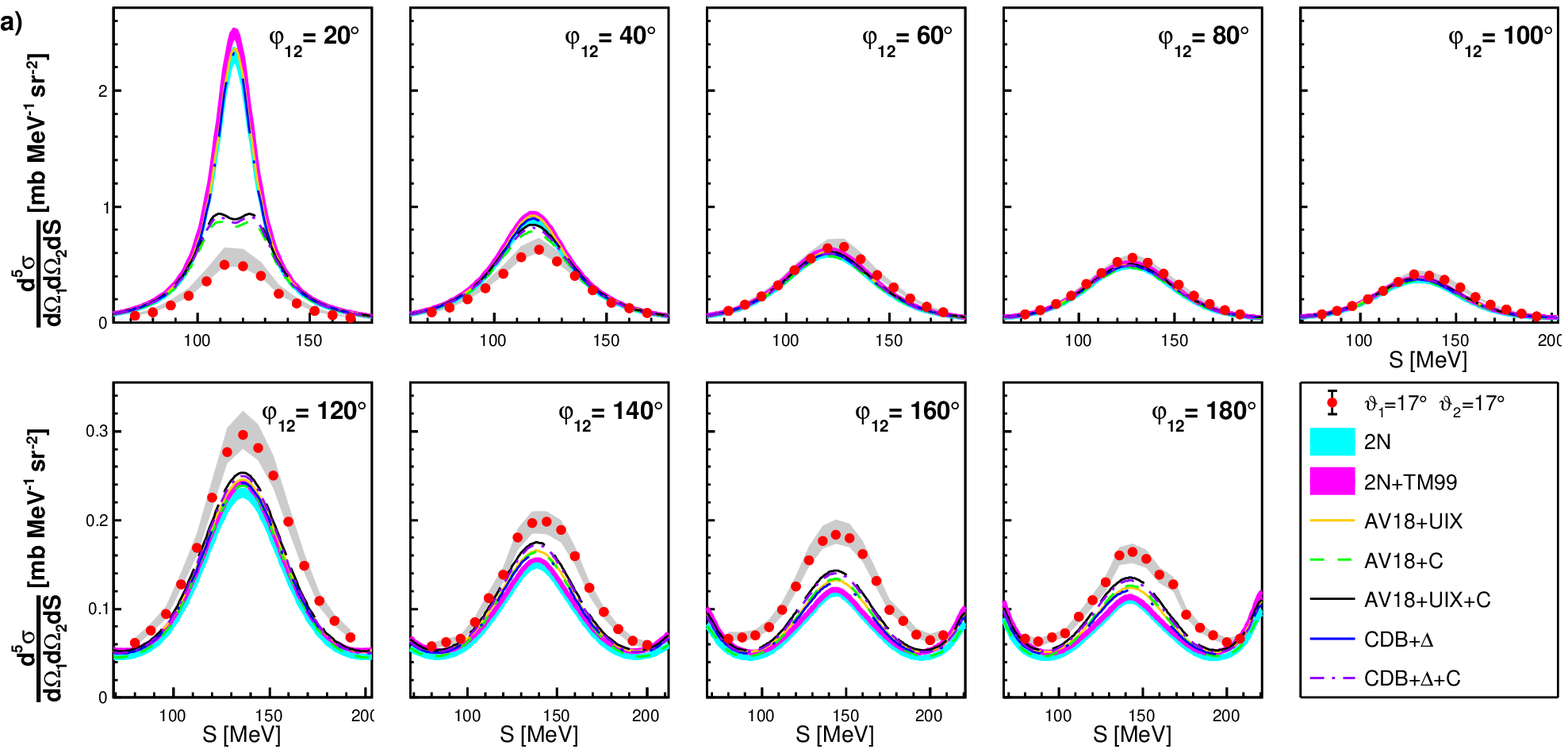}\\
 \noindent\rule[0.5ex]{\linewidth}{1pt}
  \includegraphics[width=17.cm]{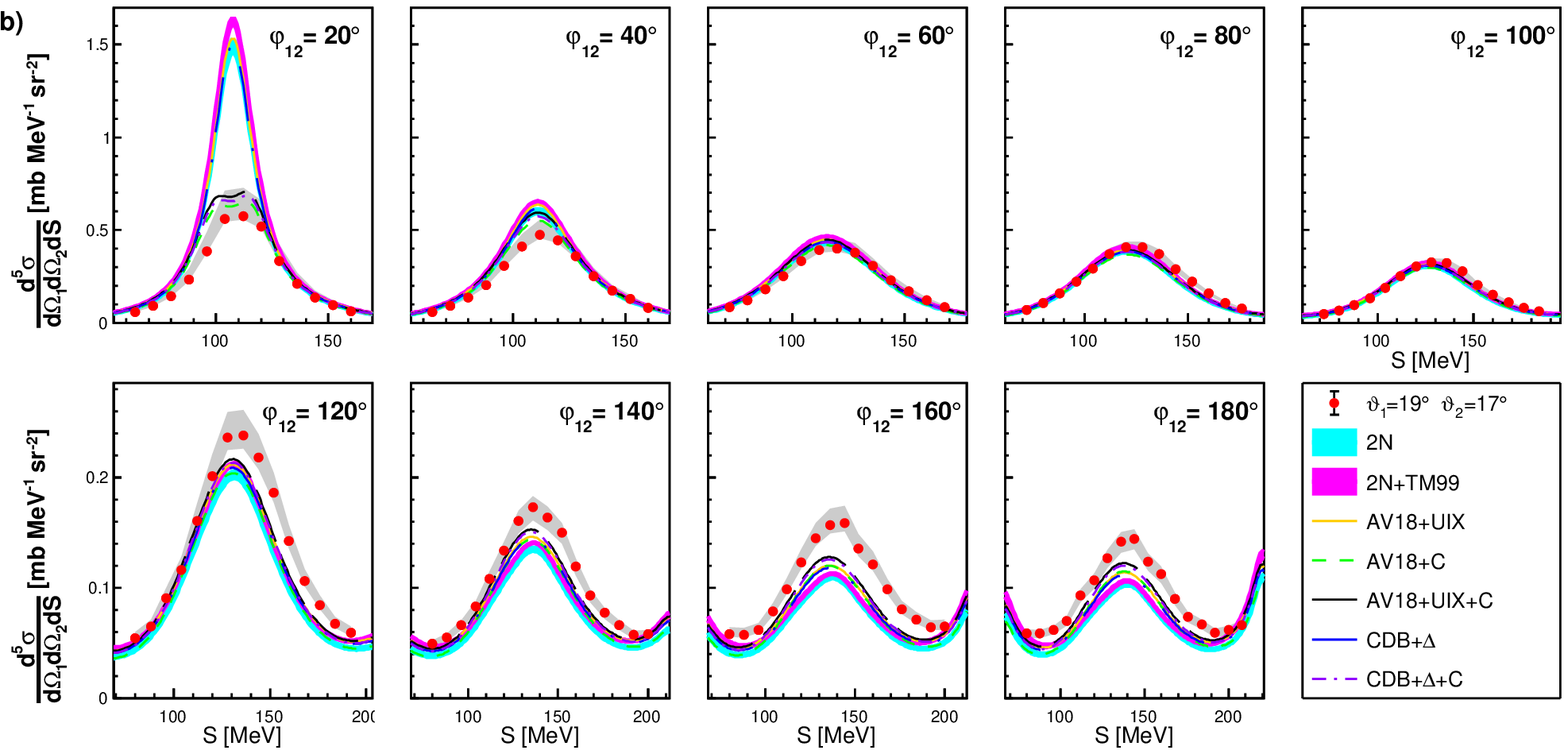}\\
 \noindent\rule[0.5ex]{\linewidth}{1pt}
  \includegraphics[width=17.cm]{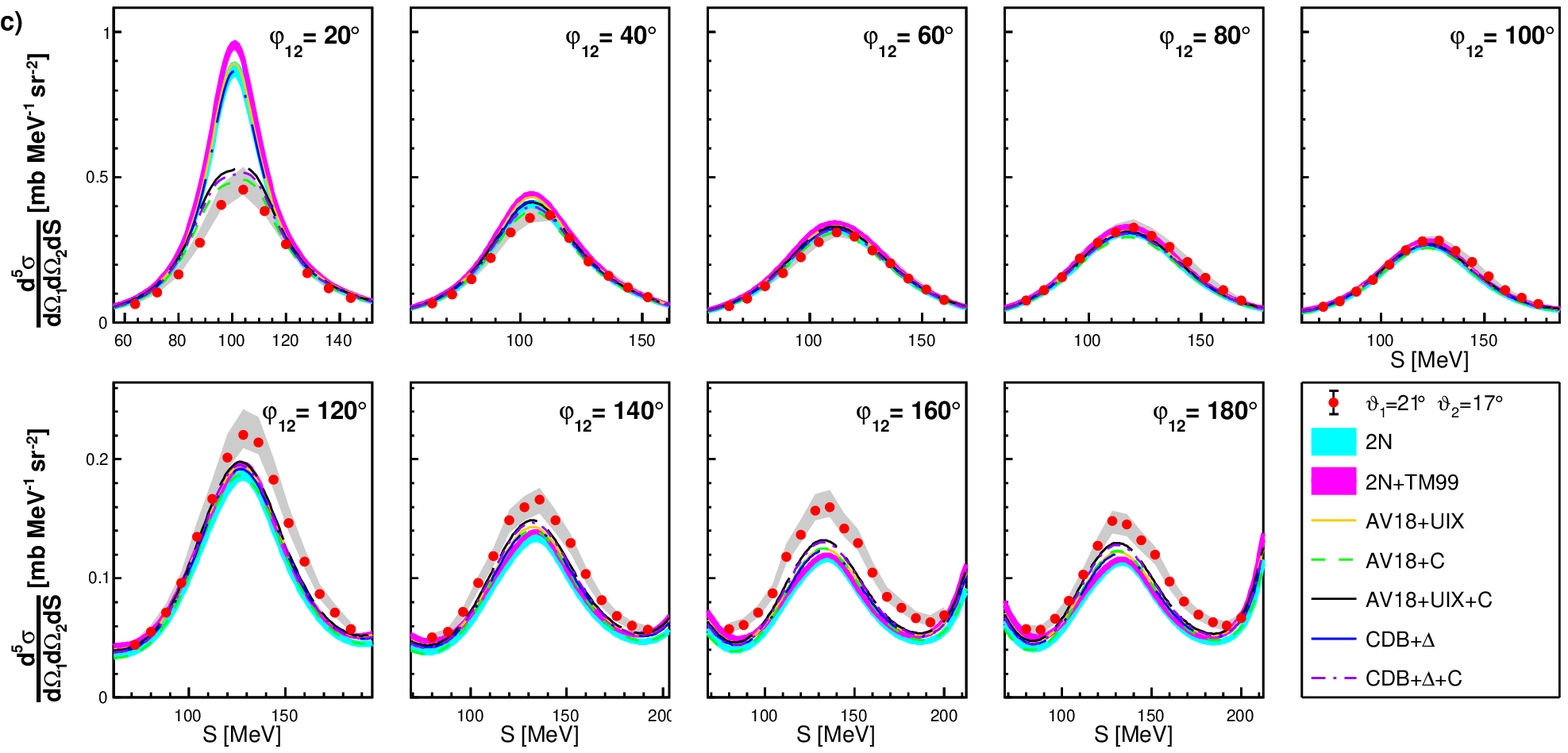}
  \caption{\label{fig:cs1} Differential cross section  for polar angles $\vartheta_{1}$, $\vartheta_{2}$: 17$^{\circ}$,17$^{\circ}$ (a); 19$^{\circ}$,17$^{\circ}$ (b); 21$^{\circ}$,17$^{\circ}$ (c). Details in the text.}
 \end{figure}
\begin{figure}
  \includegraphics[width=17.cm]{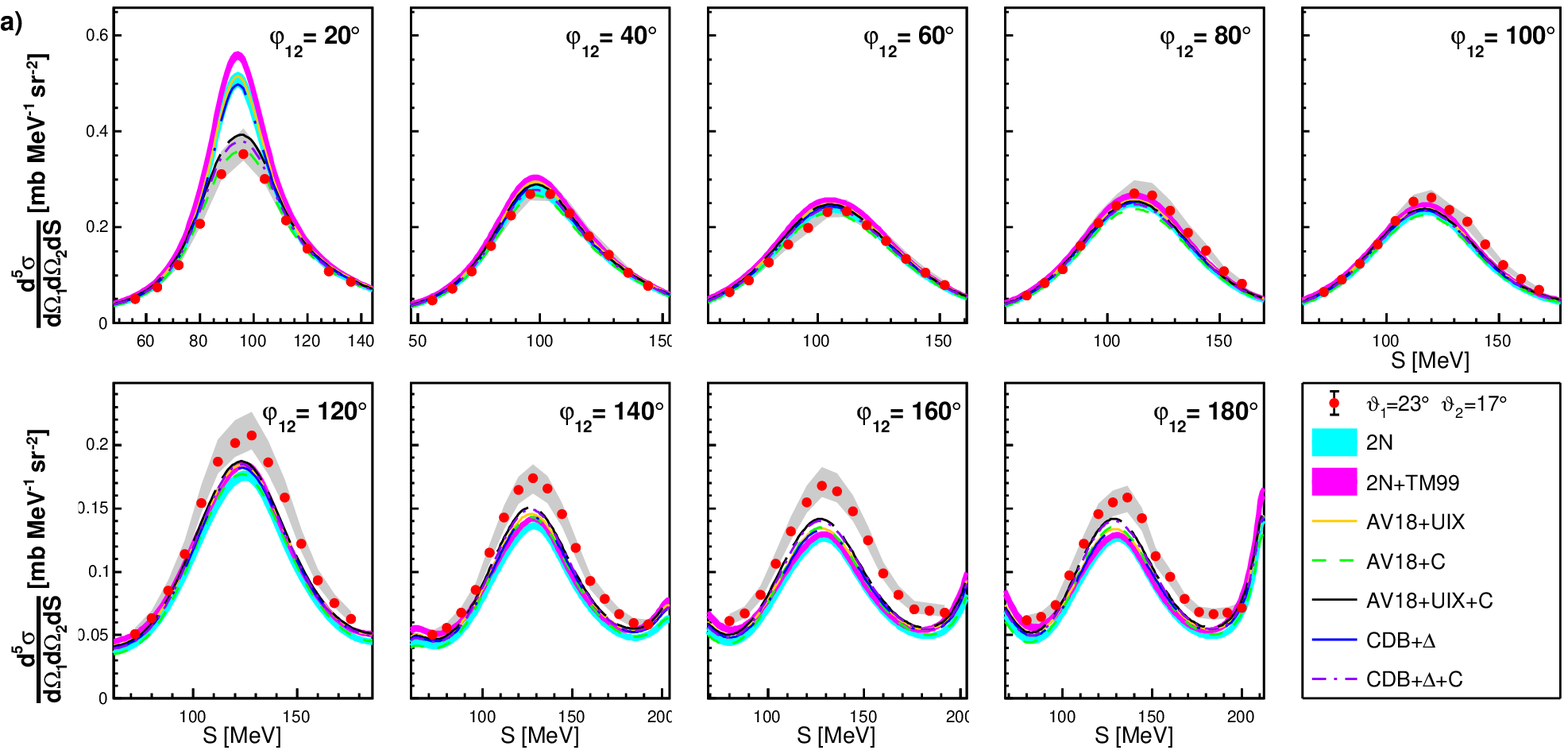}\\
 \noindent\rule[0.5ex]{\linewidth}{1pt}
  \includegraphics[width=17.cm]{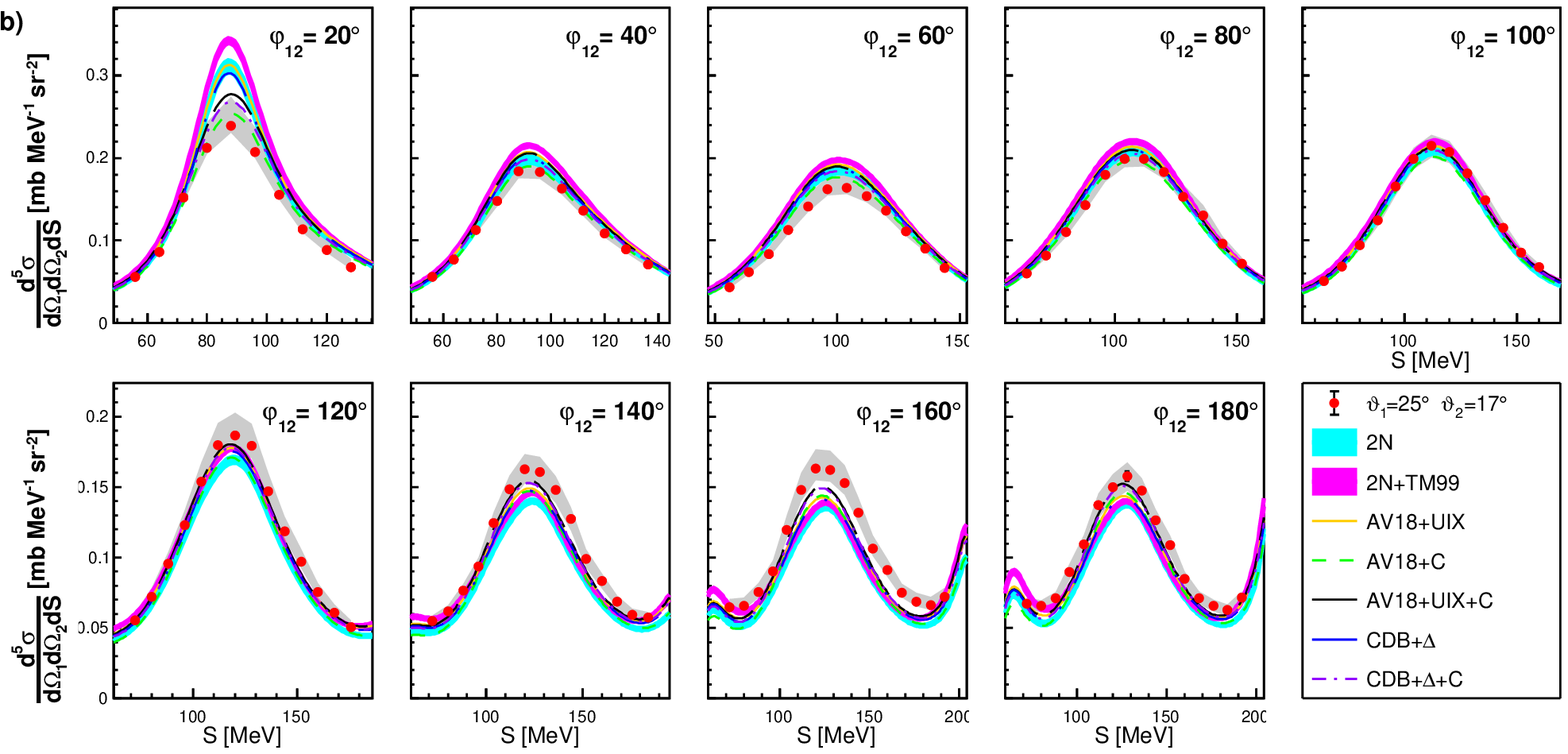}\\
 \noindent\rule[0.5ex]{\linewidth}{1pt}
  \includegraphics[width=17.cm]{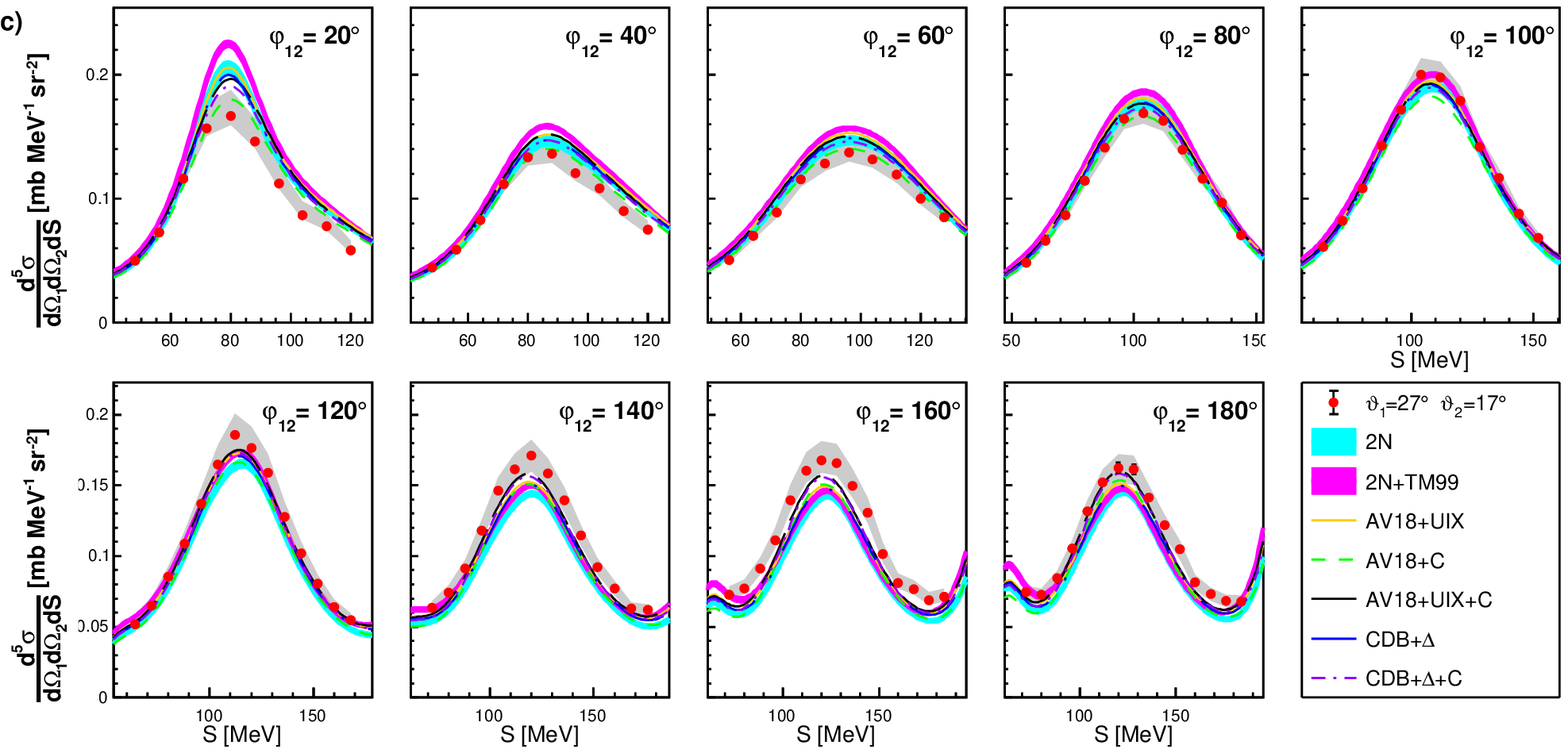}
  \caption{\label{fig:cs2} Differential cross section  for polar angles $\vartheta_{1}$, $\vartheta_{2}$: 23$^{\circ}$,17$^{\circ}$ (a); 25$^{\circ}$,17$^{\circ}$ (b); 27$^{\circ}$,17$^{\circ}$ (c). Details in the text.}
 \end{figure}
\begin{figure}
  \includegraphics[width=17.cm]{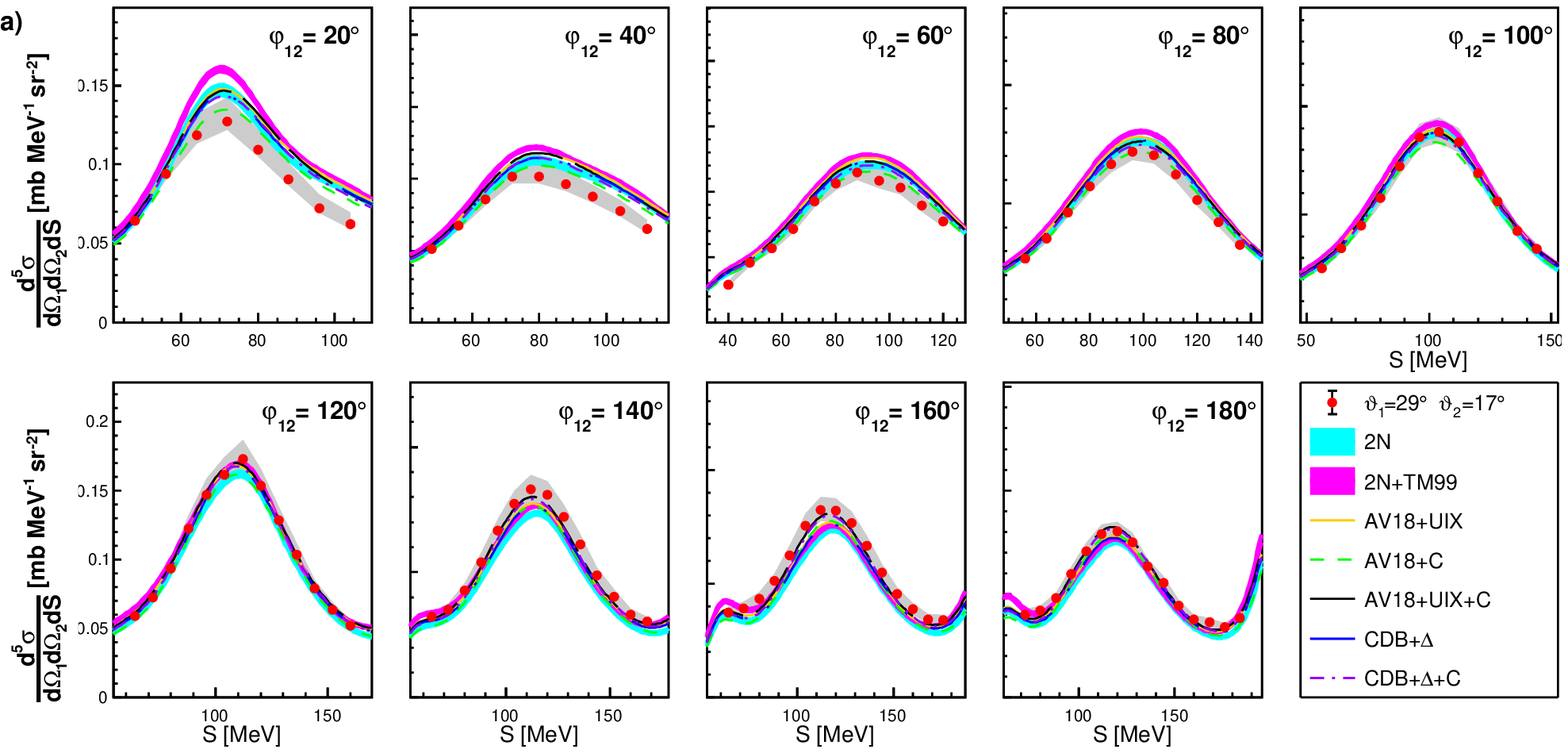}\\
 \noindent\rule[0.5ex]{\linewidth}{1pt}
  \includegraphics[width=17.cm]{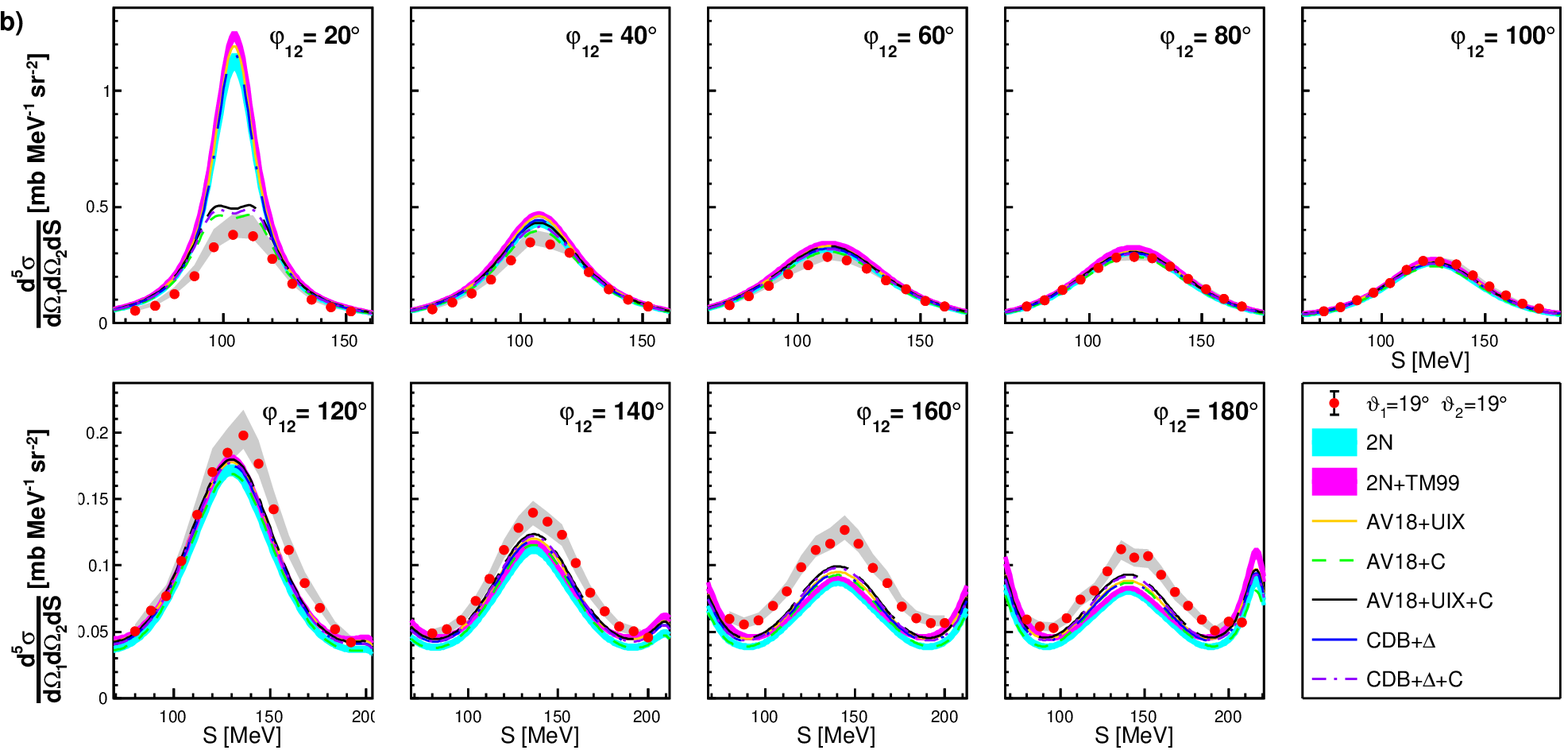}\\
 \noindent\rule[0.5ex]{\linewidth}{1pt}
  \includegraphics[width=17.cm]{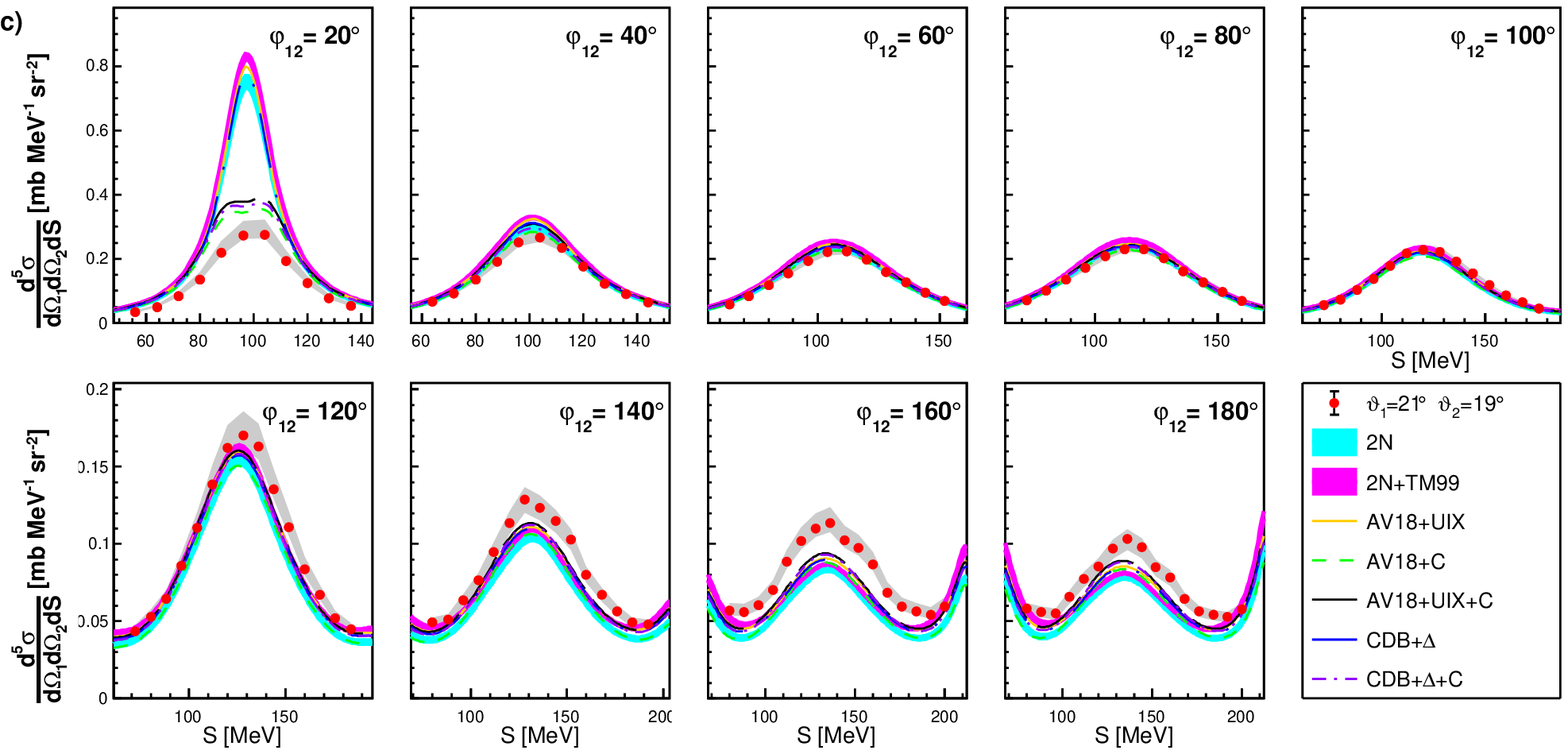}
  \caption{\label{fig:cs3} Differential cross section  for polar angles $\vartheta_{1}$, $\vartheta_{2}$: 29$^{\circ}$,17$^{\circ}$ (a); 19$^{\circ}$,19$^{\circ}$ (b); 21$^{\circ}$,19$^{\circ}$ (c). Details in the text.}
 \end{figure}
 \begin{figure}
  \includegraphics[width=17.cm]{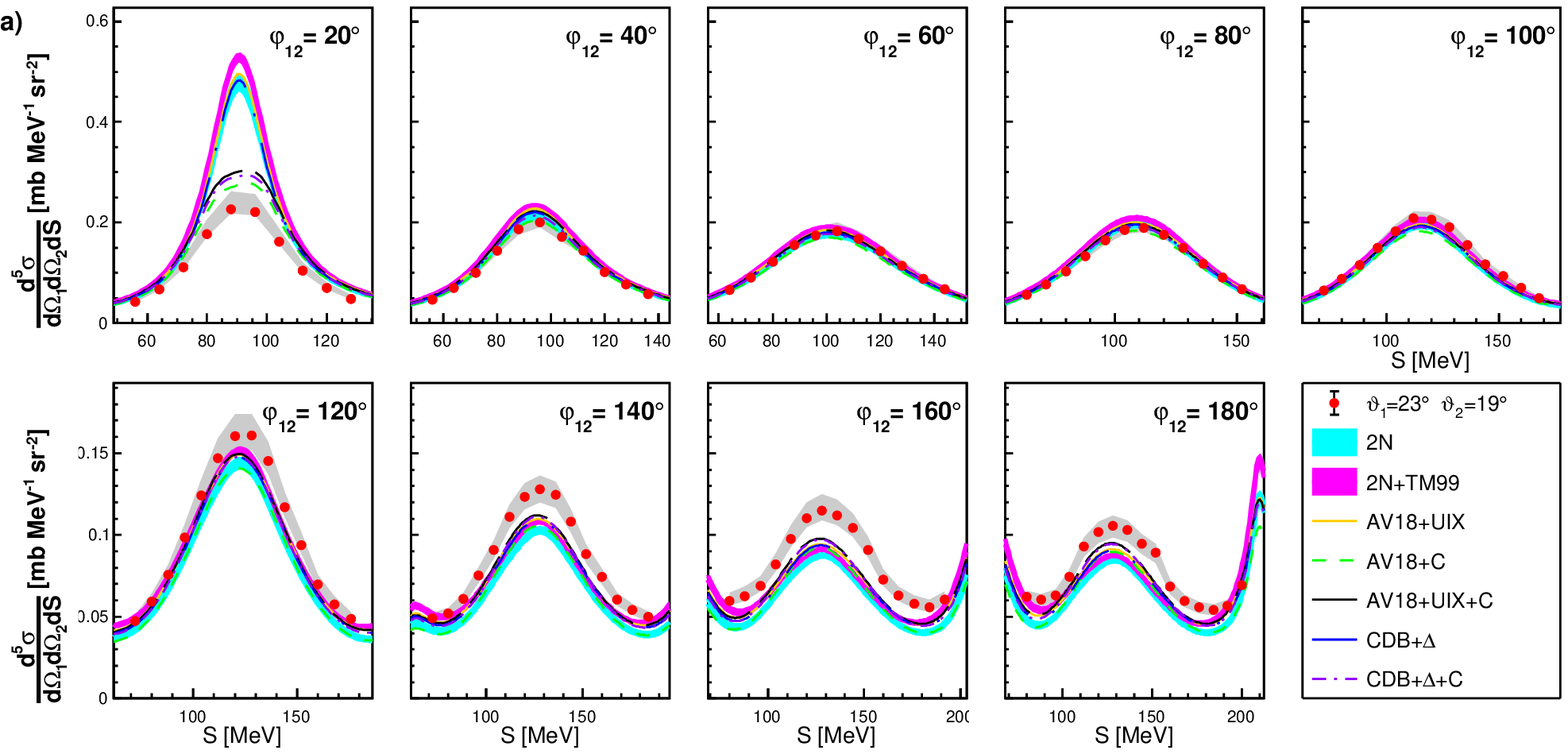}\\
 \noindent\rule[0.5ex]{\linewidth}{1pt}
  \includegraphics[width=17.cm]{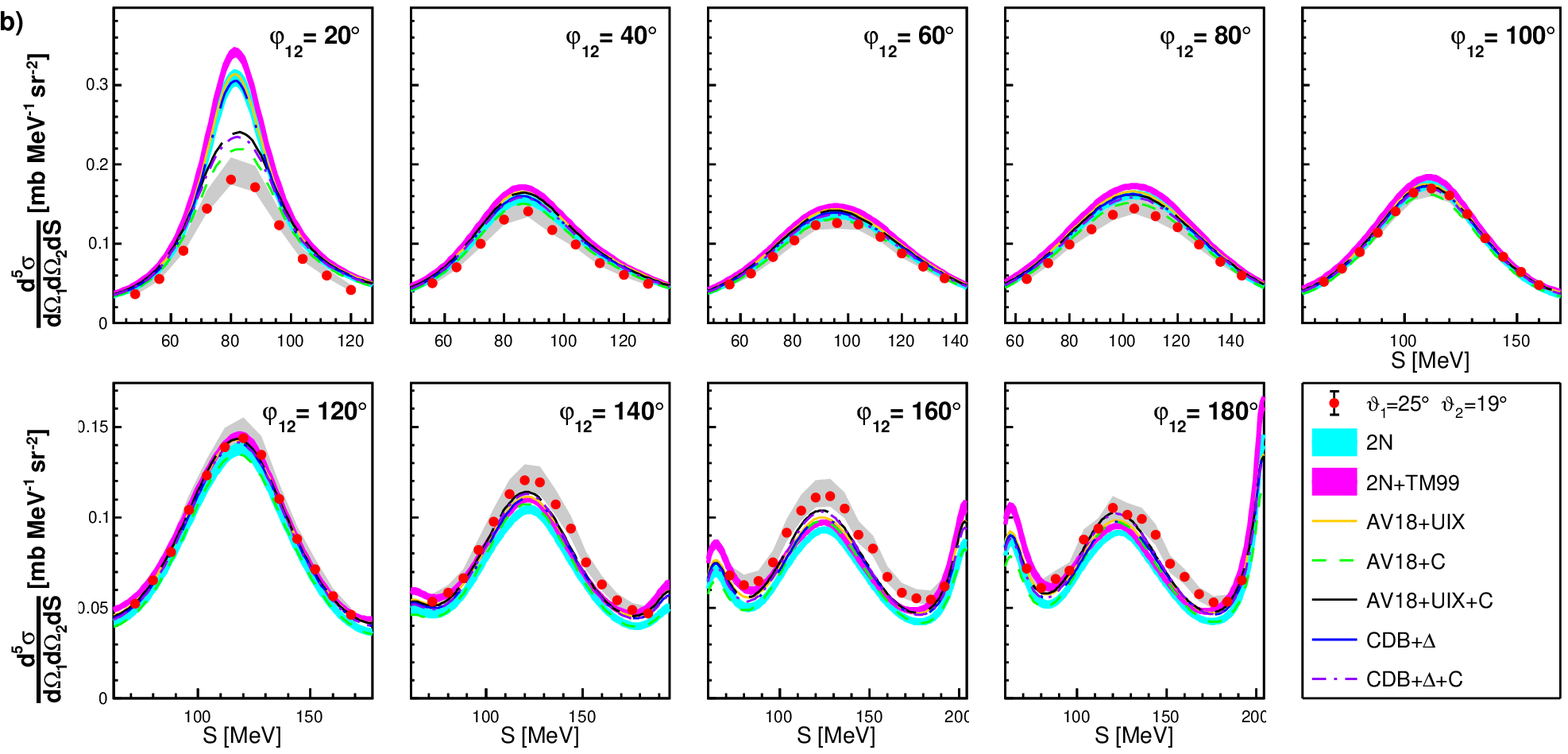}\\
 \noindent\rule[0.5ex]{\linewidth}{1pt}
  \includegraphics[width=17.cm]{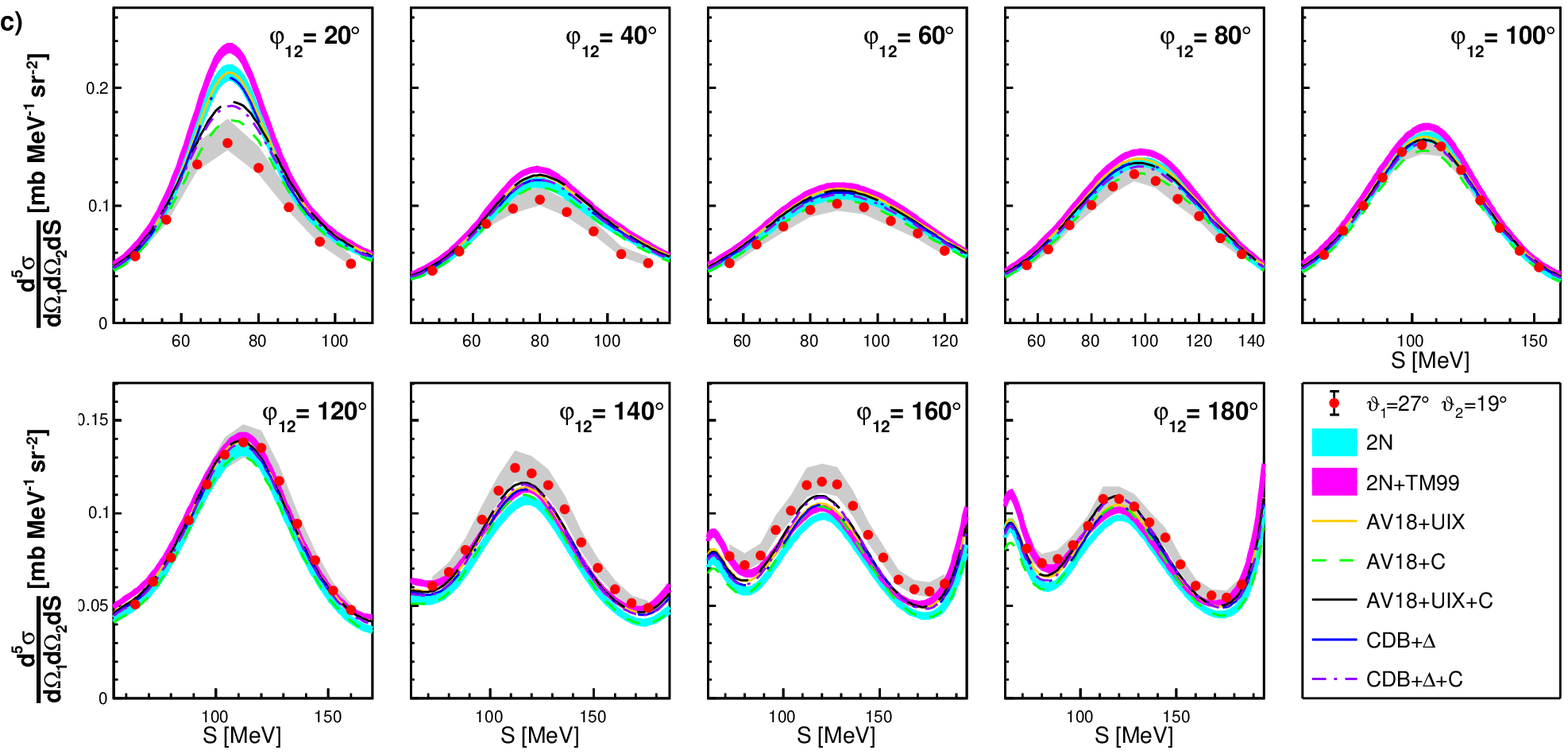}
  \caption{\label{fig:cs4} Differential cross section  for polar angles $\vartheta_{1}$, $\vartheta_{2}$: 23$^{\circ}$,19$^{\circ}$ (a); 25$^{\circ}$,19$^{\circ}$ (b); 27$^{\circ}$,19$^{\circ}$ (c). Details in the text.}
 \end{figure}
 \begin{figure}
  \includegraphics[width=17.cm]{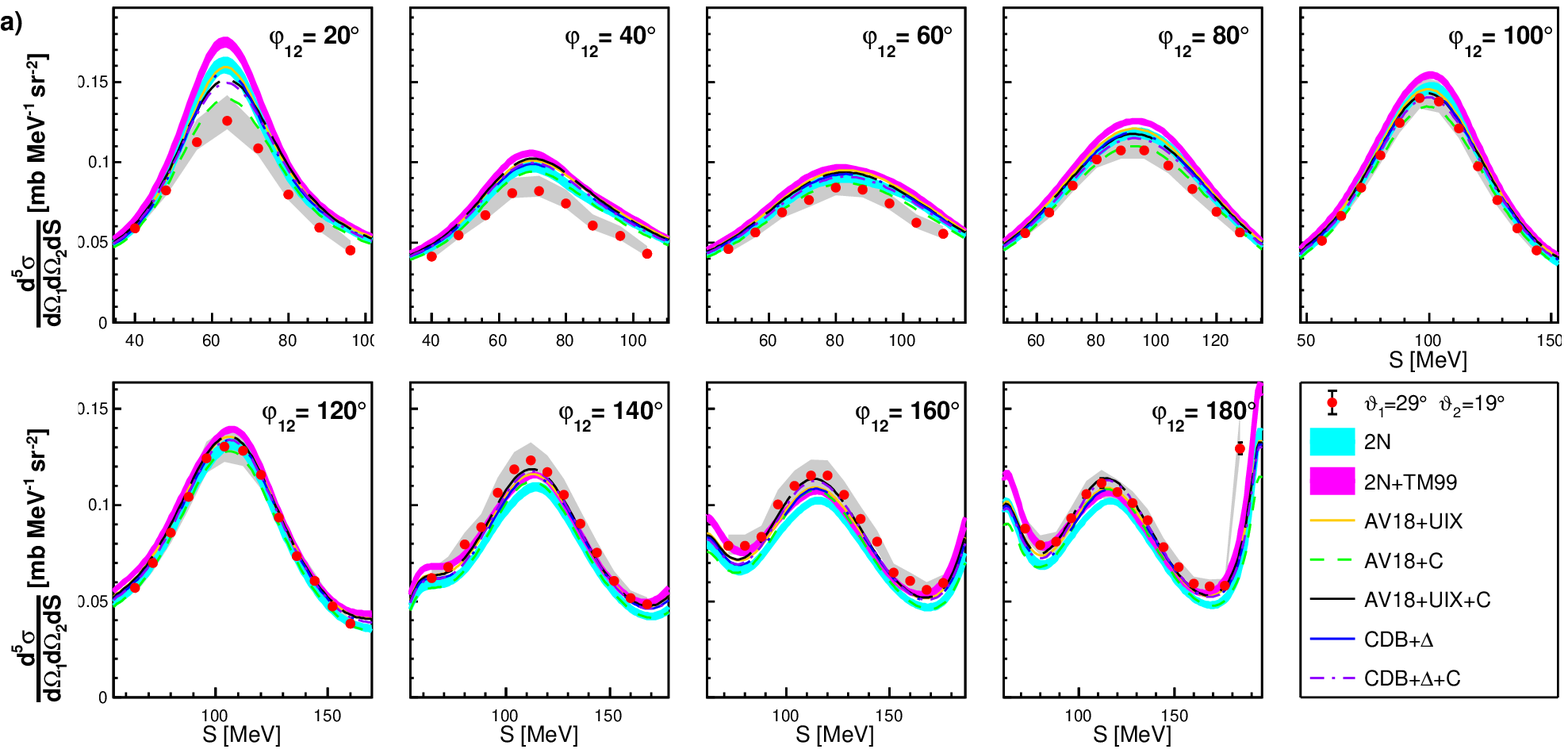}\\
 \noindent\rule[0.5ex]{\linewidth}{1pt}
  \includegraphics[width=17.cm]{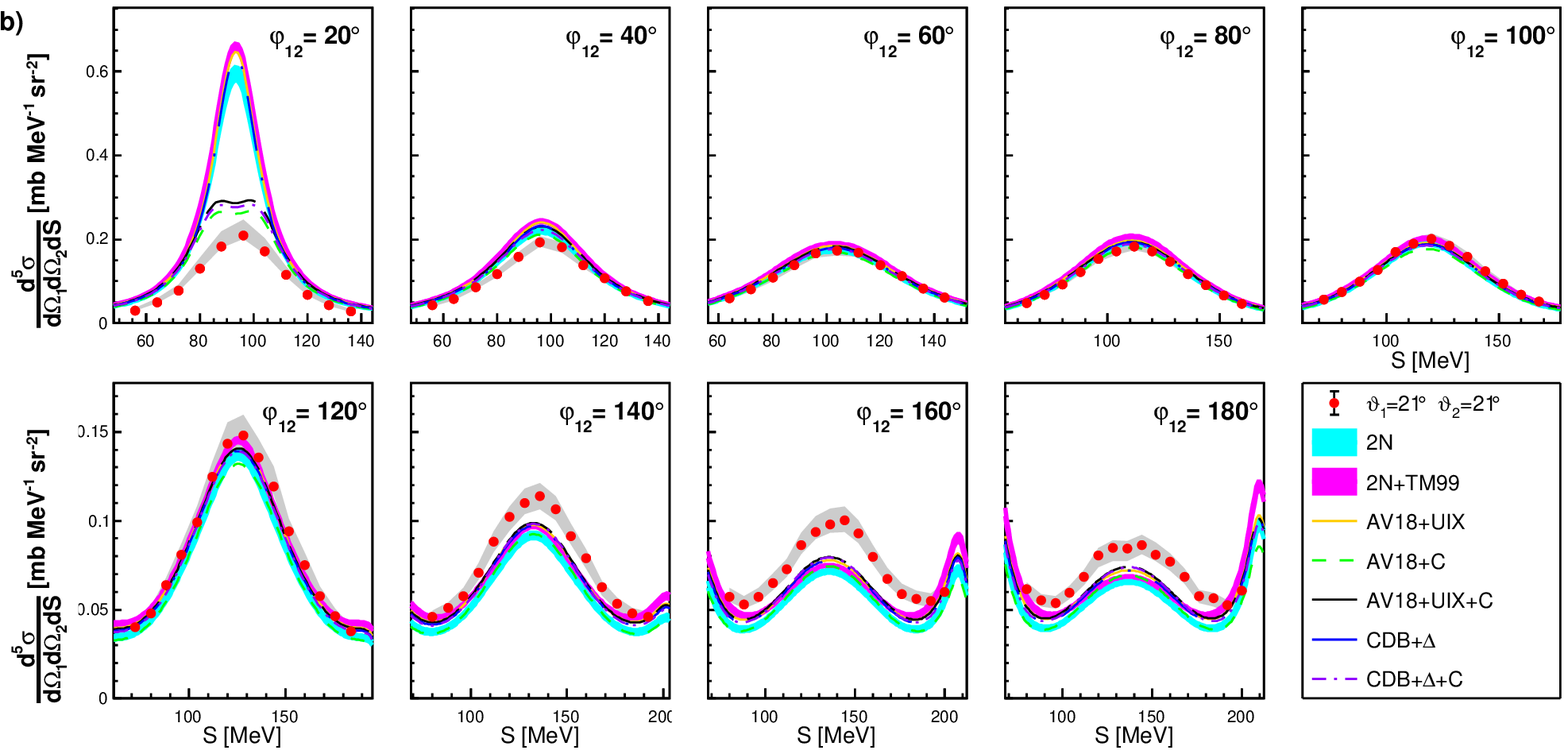}\\
 \noindent\rule[0.5ex]{\linewidth}{1pt}
  \includegraphics[width=17.cm]{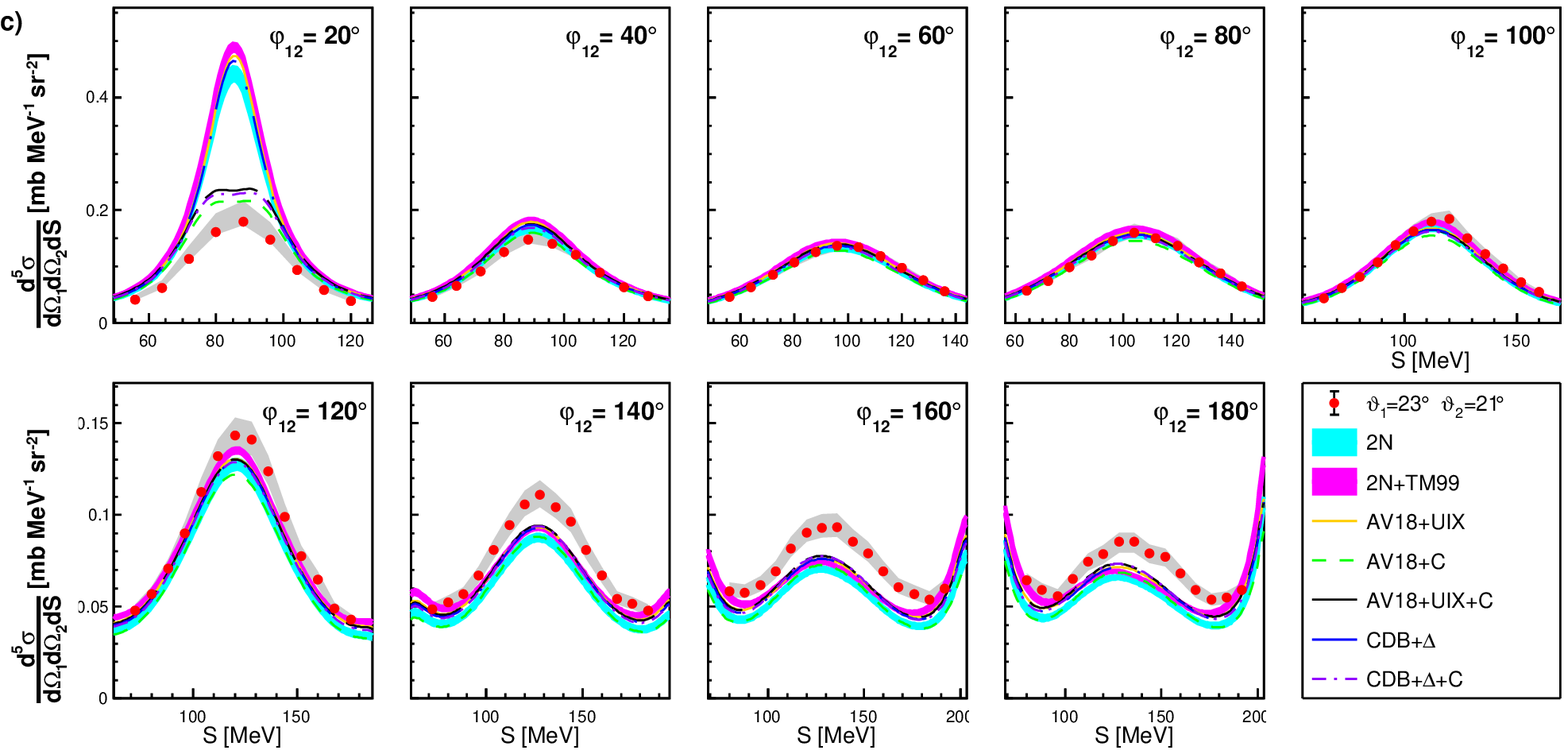}
  \caption{\label{fig:cs5} Differential cross section  for polar angles $\vartheta_{1}$, $\vartheta_{2}$: 29$^{\circ}$,19$^{\circ}$ (a); 21$^{\circ}$,21$^{\circ}$ (b); 23$^{\circ}$,21$^{\circ}$ (c). Details in the text.}
 \end{figure}
 \begin{figure}
  \includegraphics[width=17.cm]{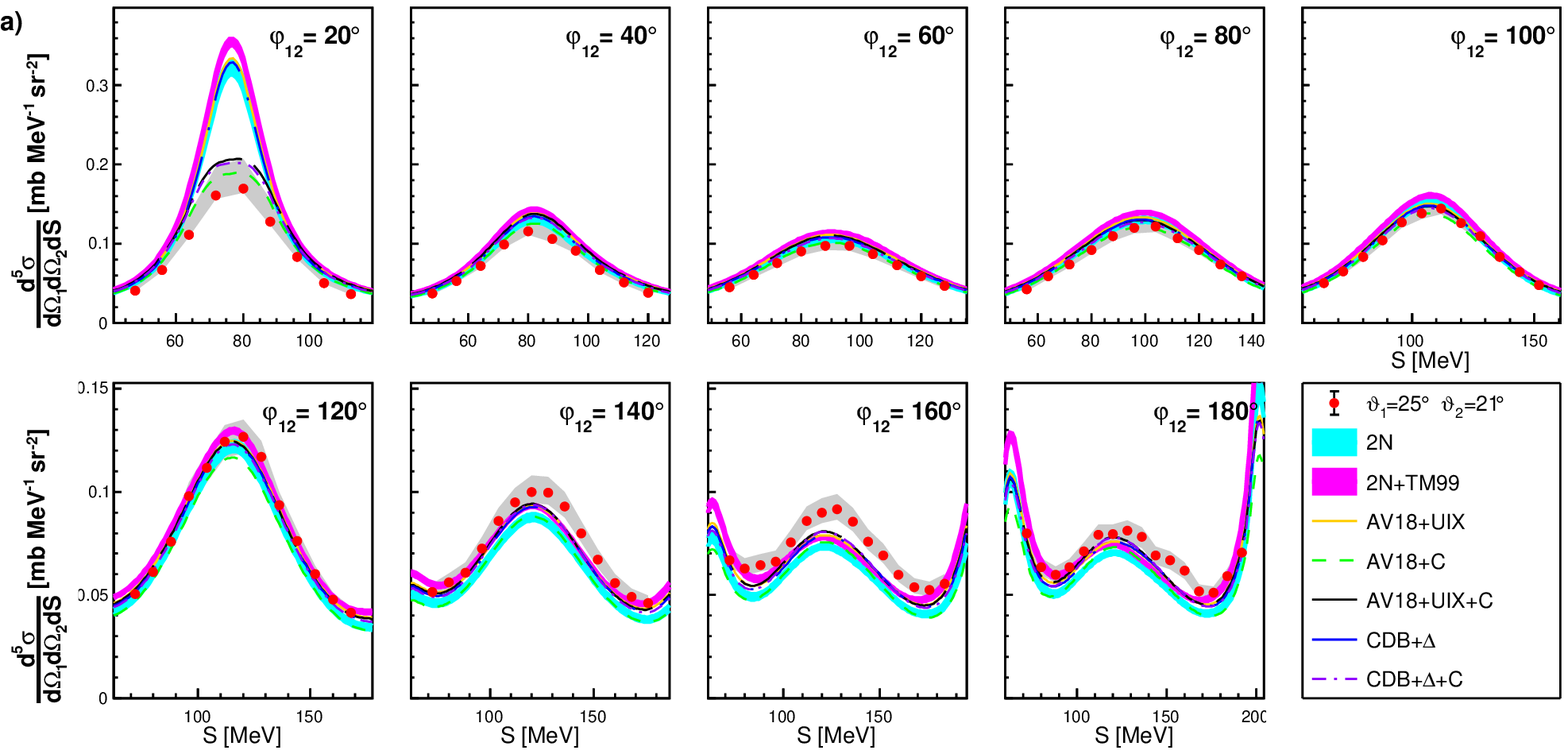}\\
 \noindent\rule[0.5ex]{\linewidth}{1pt}
  \includegraphics[width=17.cm]{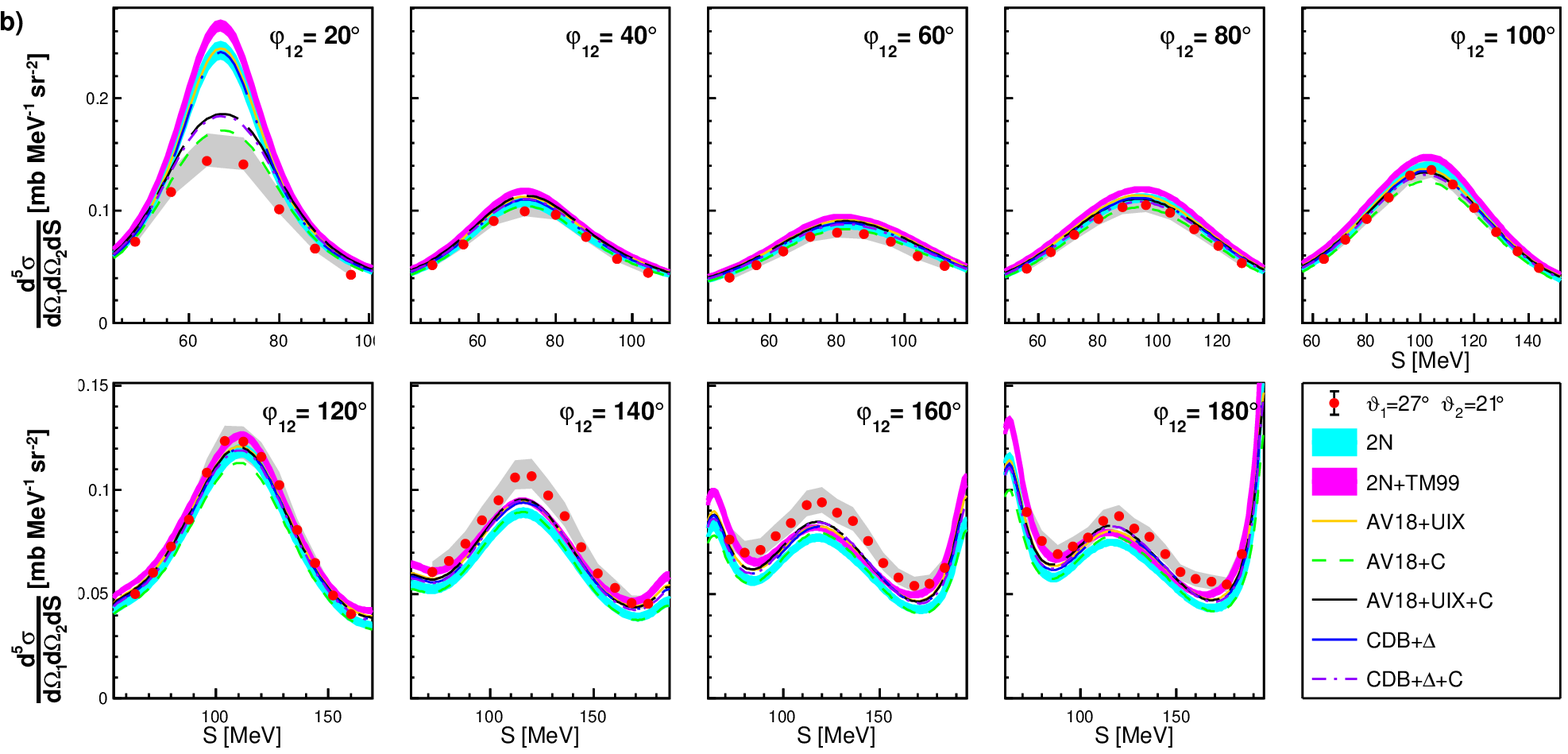}\\
 \noindent\rule[0.5ex]{\linewidth}{1pt}
  \includegraphics[width=17.cm]{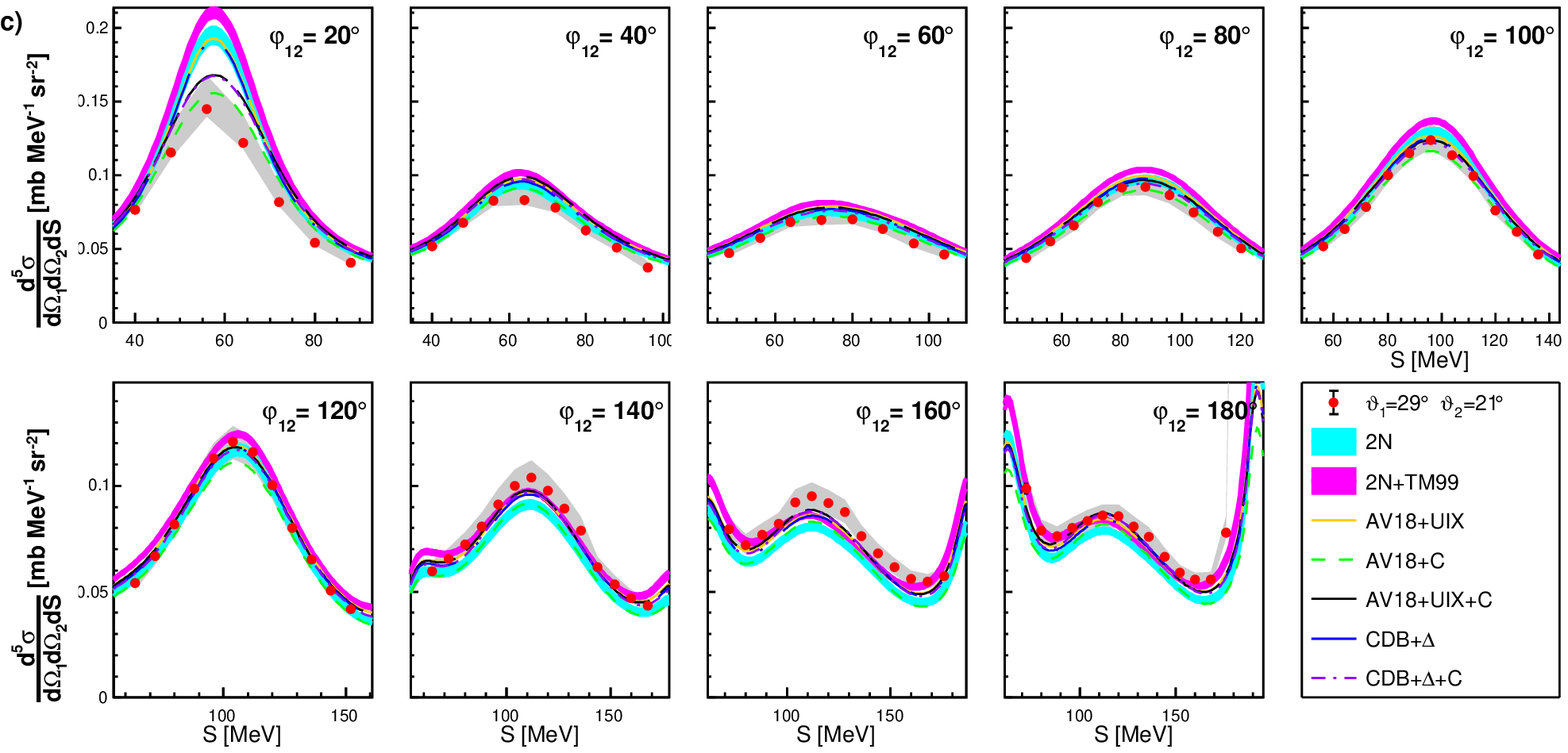}
  \caption{\label{fig:cs6} Differential cross section  for polar angles $\vartheta_{1}$, $\vartheta_{2}$: 25$^{\circ}$,21$^{\circ}$ (a); 27$^{\circ}$,21$^{\circ}$ (b); 29$^{\circ}$,21$^{\circ}$ (c). Details in the text.}
 \end{figure}
  \begin{figure}
  \includegraphics[width=17.cm]{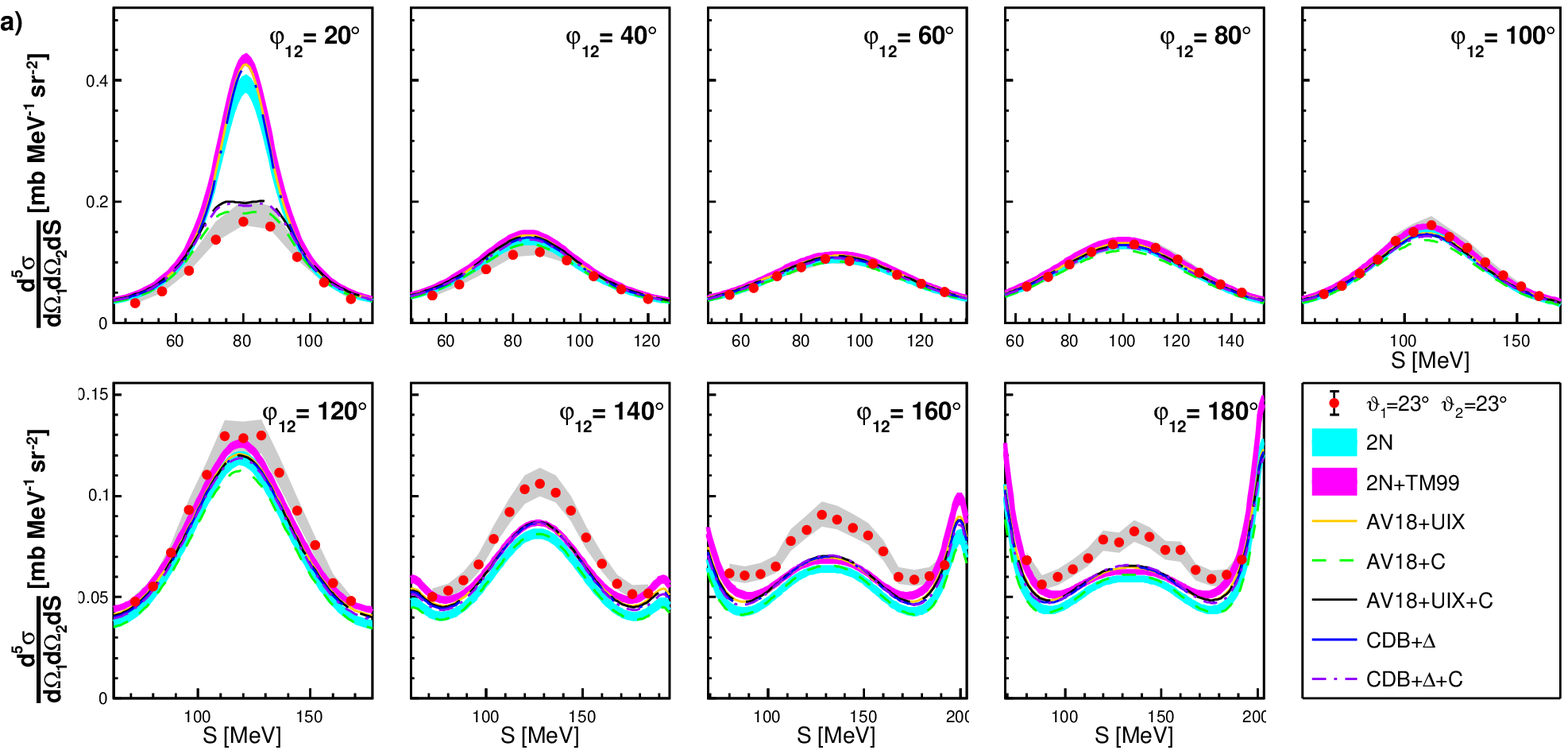}\\
 \noindent\rule[0.5ex]{\linewidth}{1pt}
  \includegraphics[width=17.cm]{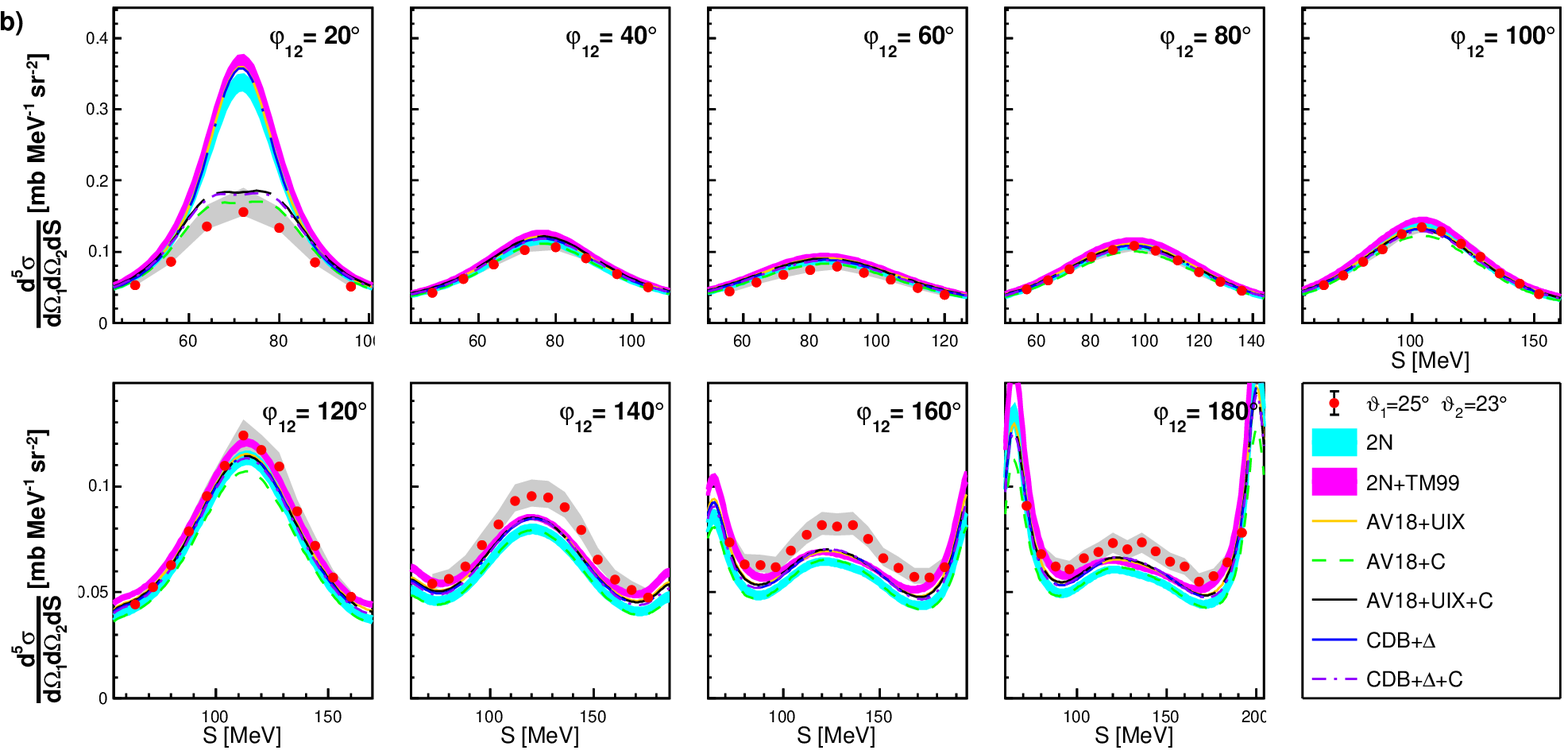}\\
 \noindent\rule[0.5ex]{\linewidth}{1pt}
  \includegraphics[width=17.cm]{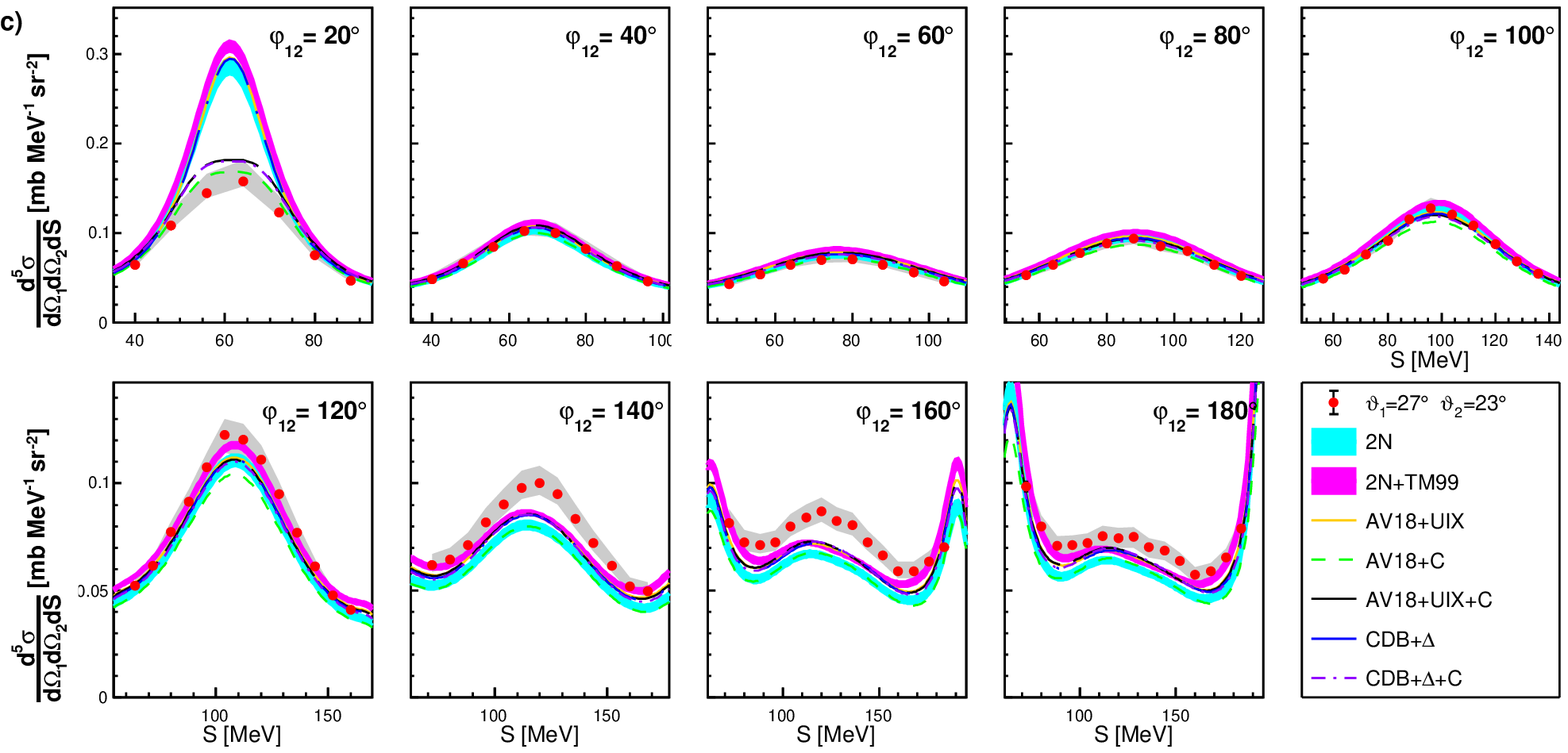}
  \caption{\label{fig:cs7} Differential cross section  for polar angles $\vartheta_{1}$, $\vartheta_{2}$: 23$^{\circ}$,23$^{\circ}$ (a); 25$^{\circ}$,23$^{\circ}$ (b); 27$^{\circ}$,23$^{\circ}$ (c). Details in the text.}
  \end{figure}
  \begin{figure}
  \includegraphics[width=17.cm]{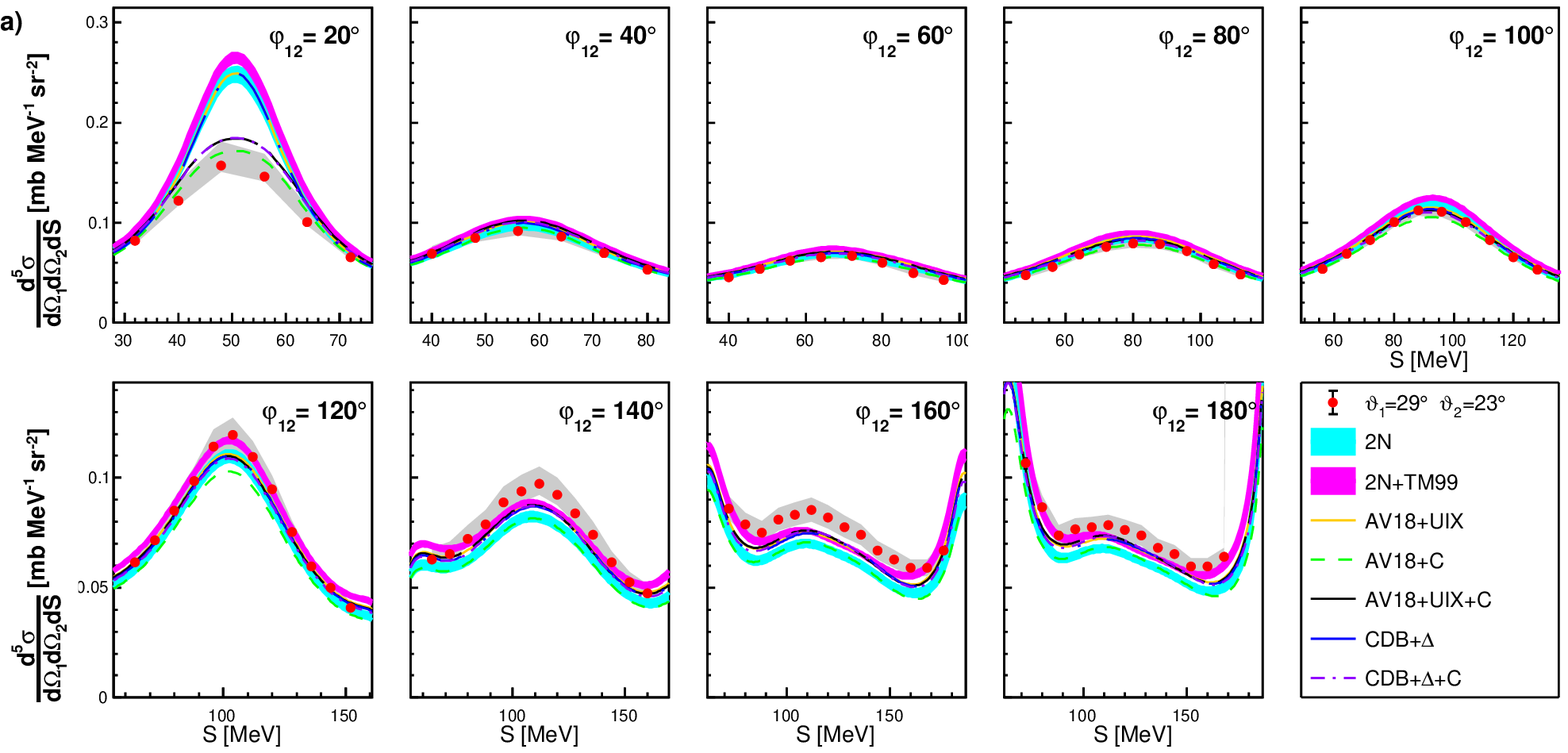}\\
 \noindent\rule[0.5ex]{\linewidth}{1pt}
  \includegraphics[width=17.cm]{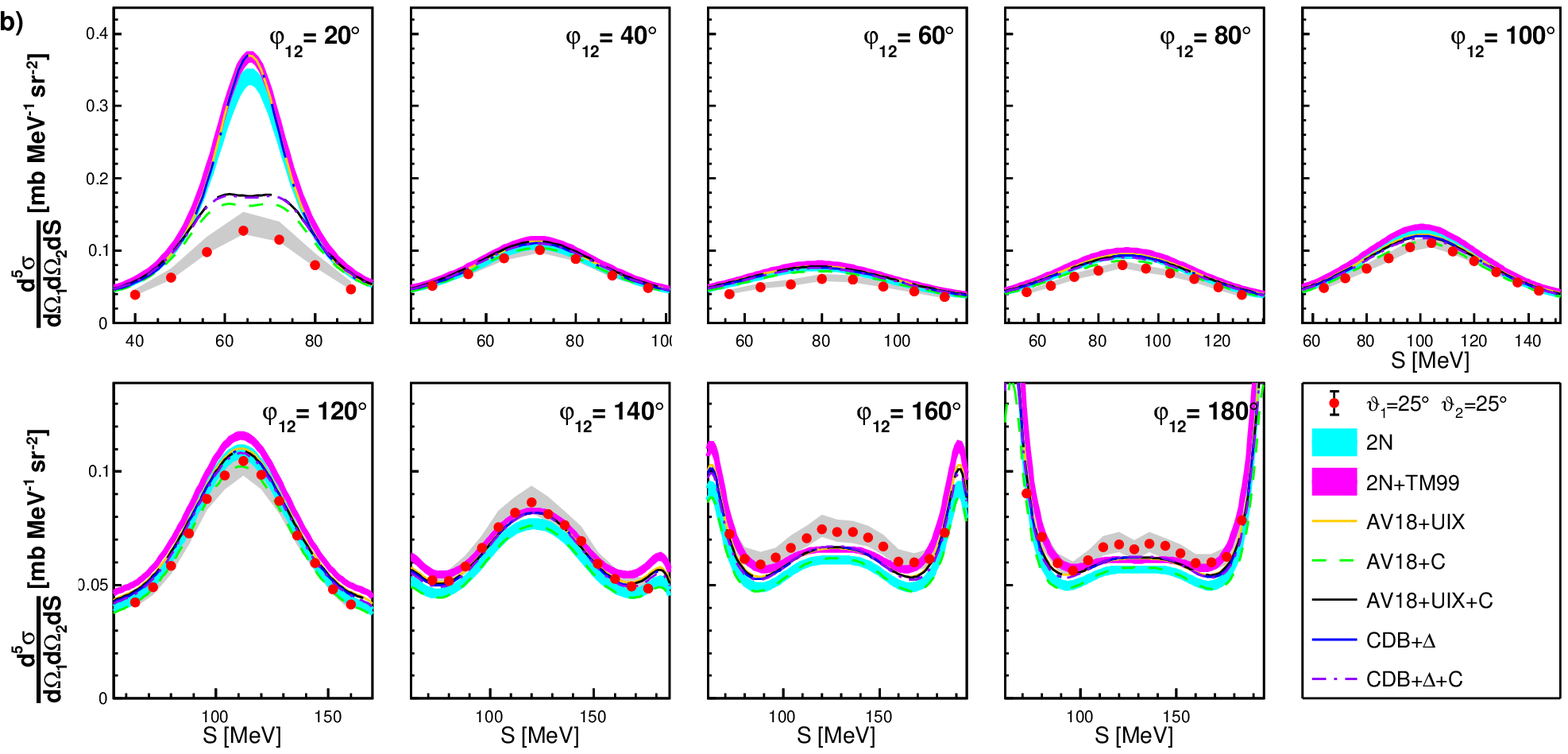}\\
 \noindent\rule[0.5ex]{\linewidth}{1pt}
  \includegraphics[width=17.cm]{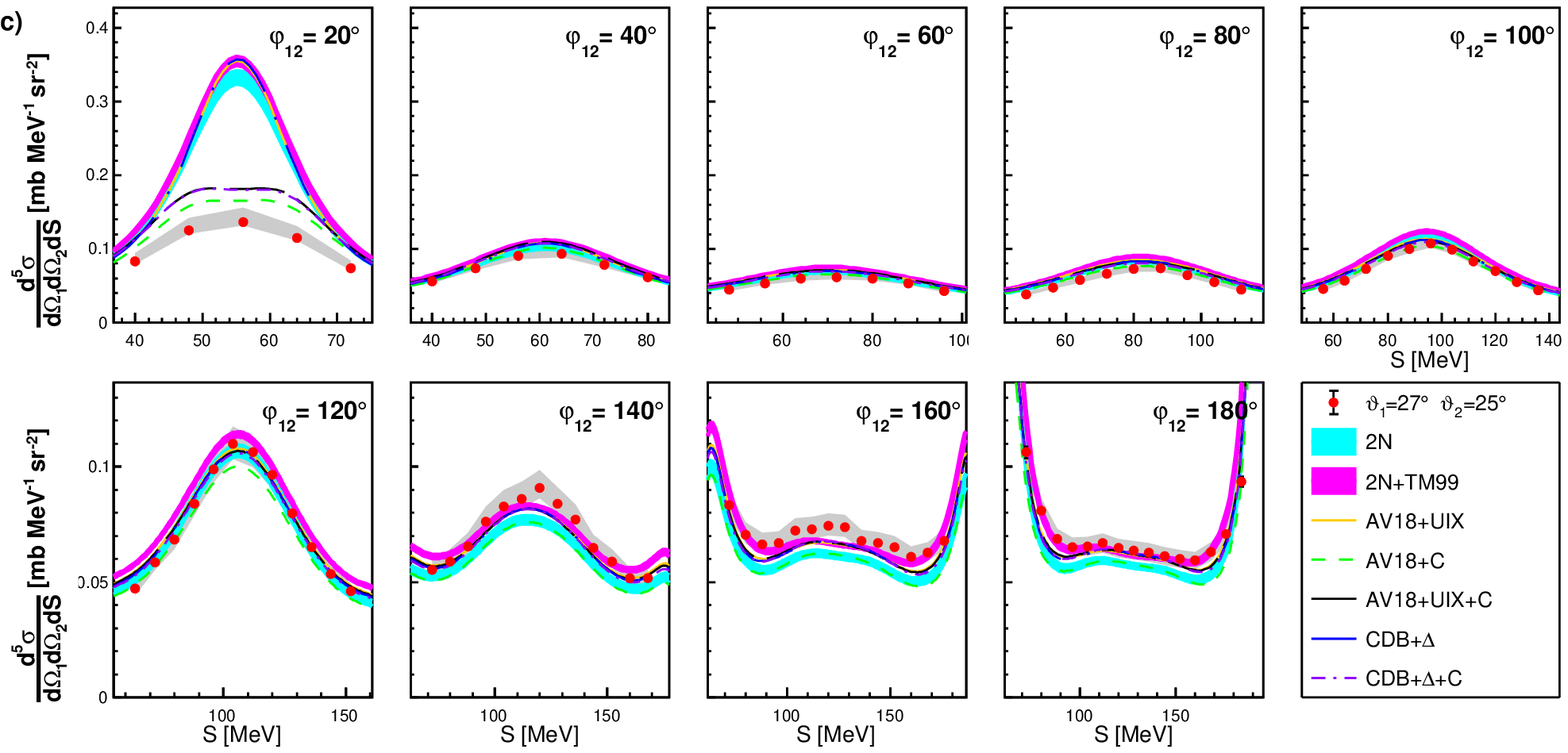}
  \caption{\label{fig:cs8} Differential cross section  for polar angles $\vartheta_{1}$, $\vartheta_{2}$: 29$^{\circ}$,23$^{\circ}$ (a); 25$^{\circ}$,25$^{\circ}$ (b); 27$^{\circ}$,25$^{\circ}$ (c). Details in the text.}
 \end{figure}
  \begin{figure}
  \includegraphics[width=17.cm]{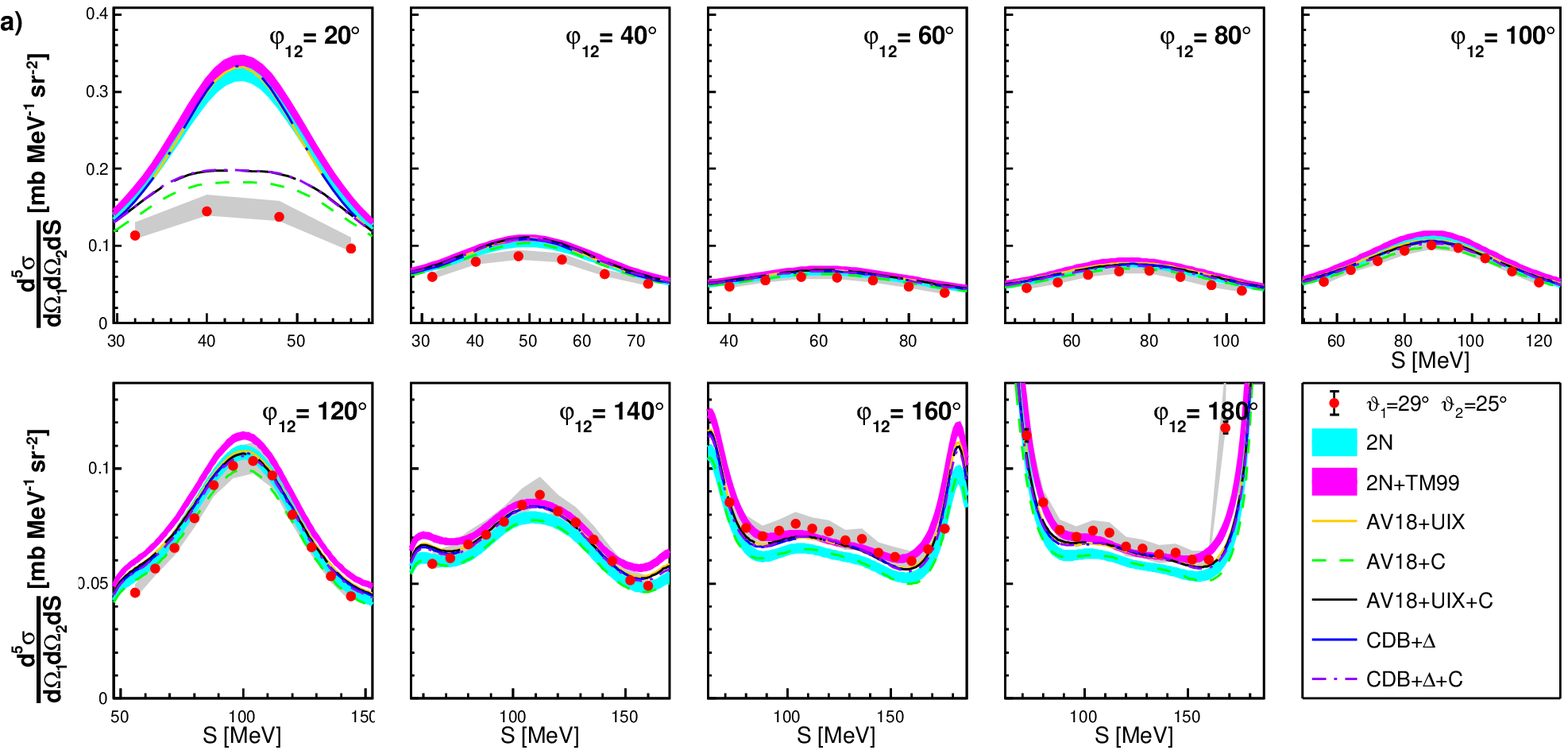}\\
 \noindent\rule[0.5ex]{\linewidth}{1pt}
  \includegraphics[width=17.cm]{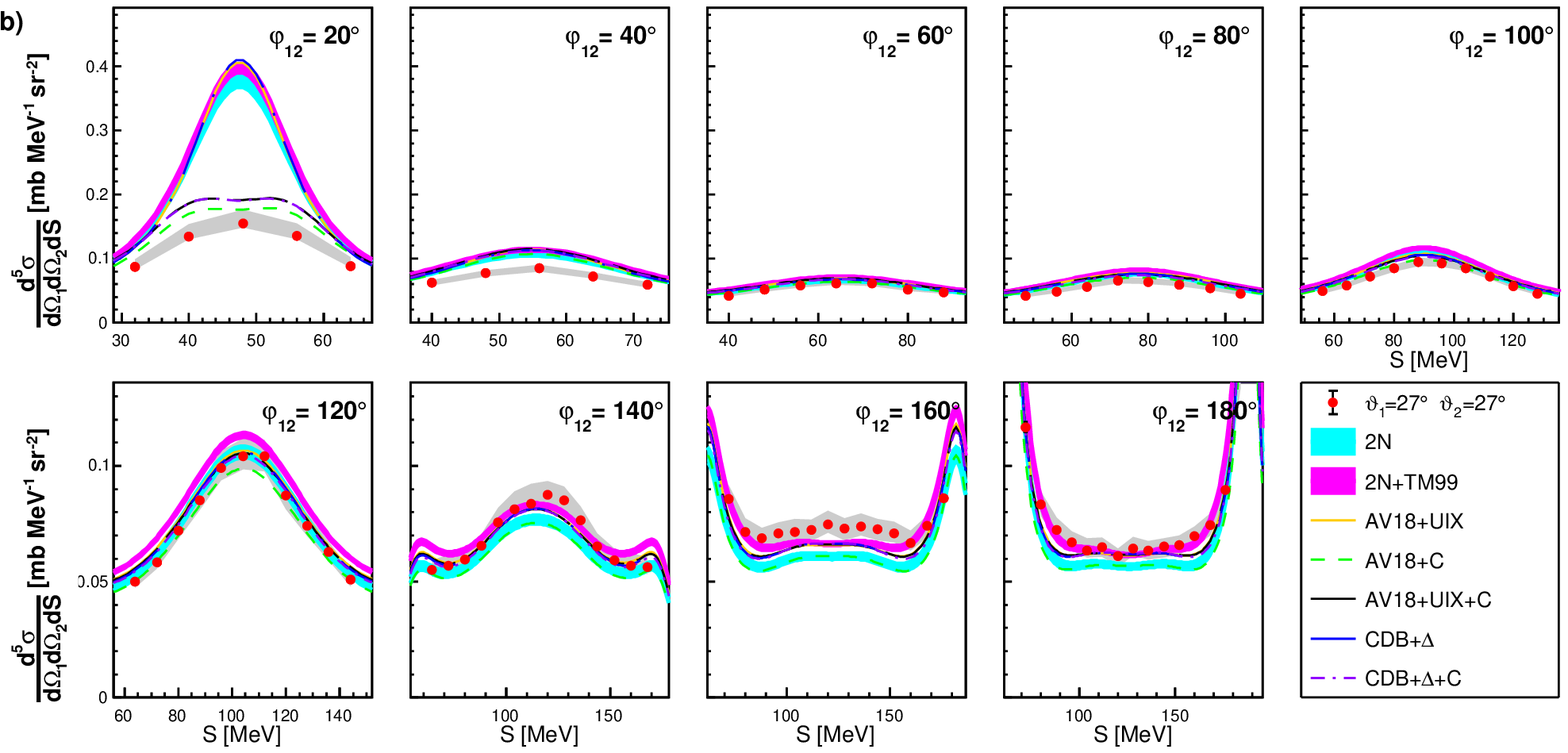}\\
 \noindent\rule[0.5ex]{\linewidth}{1pt}
  \includegraphics[width=17.cm]{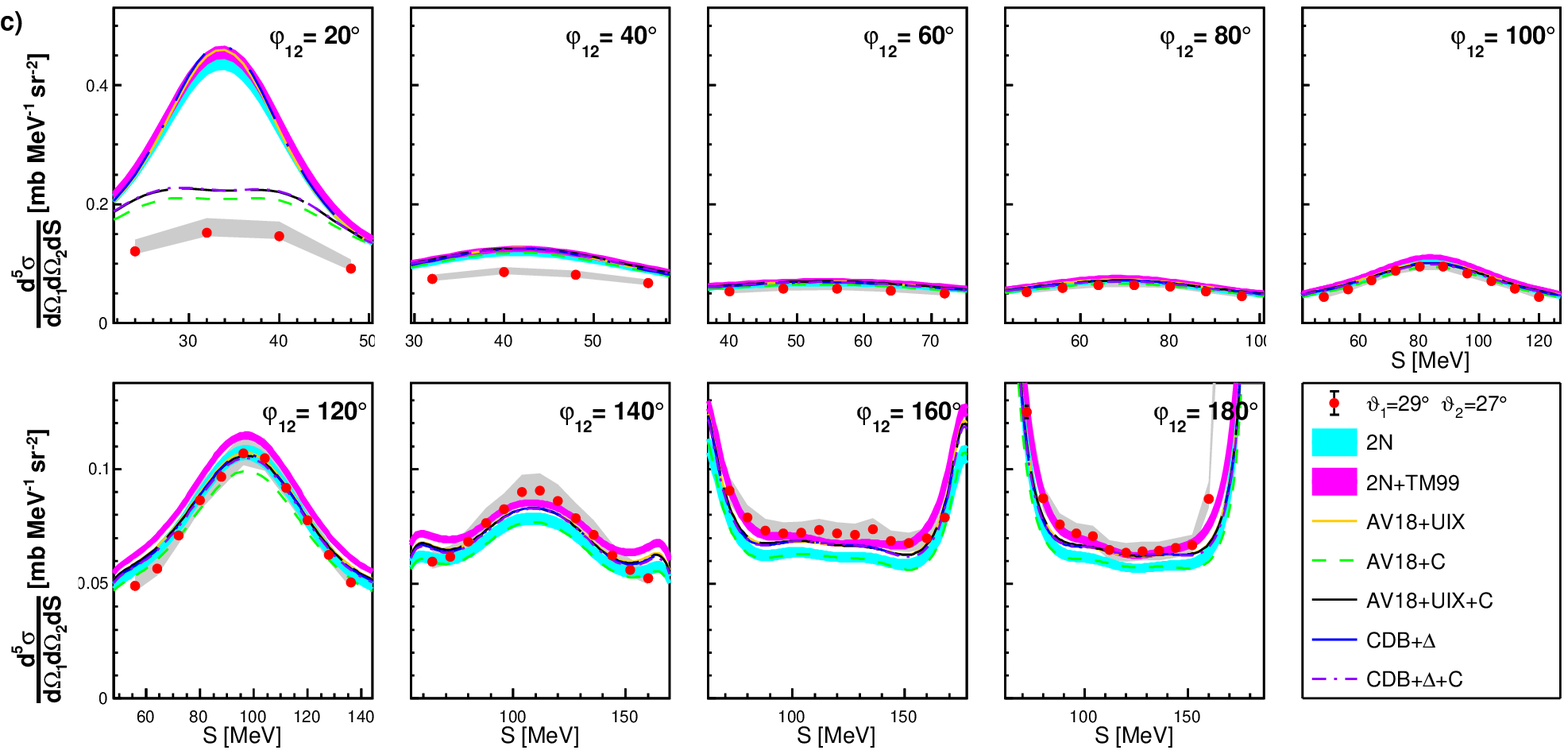}
  \caption{\label{fig:cs9} Differential cross section  for polar angles $\vartheta_{1}$, $\vartheta_{2}$: 29$^{\circ}$,25$^{\circ}$ (a); 27$^{\circ}$,27$^{\circ}$ (b); 29$^{\circ}$,27$^{\circ}$ (c). Details in the text.}
  \end{figure}
  \end{center}
\end{widetext}

\bibliography{apstemplate.bib}

\end{document}